\documentclass[12pt]{article}
\usepackage[affil-it]{authblk}
\usepackage[margin=1in]{geometry}

\usepackage{bm}
\usepackage{multirow}
\usepackage[font=small]{caption}
\usepackage{subcaption}
\usepackage{graphicx, color,times}
\usepackage{natbib}
\usepackage{epstopdf}

\usepackage{amsmath,amsthm,amssymb,amsfonts}
\pdfoutput=1
\theoremstyle{plain}
\newtheorem{assumption}{Assumption}
\newtheorem{corollary}{Corollary}
\newtheorem{theorem}{Theorem}
\newtheorem{remark}{Remark}

\def\bic{\textsc{bic}}

\def\lk{{\Lambda^{(k)}}}
\def\hk{{\mathbf{H}^{(k)} }}
\def\sk{{\Sigma^{(k)} }}
\def\eps{{\varepsilon}}

\def\bZ{{{\bf Z}}}
\def\MN{{\mathcal{N}}}
\def\se{{\rm SE}}
\def\S{\Sigma_0}
\def\St{\tilde{\Sigma}_0}
\def\At{\tilde{A}_0}
\def\St{\tilde{\Sigma}_0}
\def\At{\tilde{\Omega}_0}
\def\Sa{\mathcal{S}^p_{+}}
\def\SE{\mathcal{S}^p_{E}}

\newcommand{\vs}{{\varsigma}}
\newcommand{\mbe}{\mathbb{E}}
\newcommand{\vth}{{\vartheta}}

\def\bfb{\mathbf{b}}
\newcommand{\bX}{{\bf X}}
\newcommand{\bY}{{\bf Y}}
\newcommand{\bA}{{\bf A}}

\newcommand{\bD}{{\bf D}}

\newcommand{\bW}{{\bf W}}
\newcommand{\bP}{\mathbf{P}}
\newcommand{\bIP}{{\bf I}_p}
\newcommand{\bU}{\mathbf{U}}
\newcommand{\bR}{\mathbf{R}}
\newcommand{\bel}{\boldsymbol{\ell}}
\newcommand{\btheta}{\boldsymbol{\theta}}
\newcommand{\bdelta}{\boldsymbol{\delta}}
\newcommand{\beps}{\boldsymbol{\varepsilon}}
\newcommand{\bgamma}{\boldsymbol{\gamma}}
\newcommand{\bbeta}{\boldsymbol{\beta}}
\newcommand{\bmu}{\boldsymbol{\mu}}
\newcommand{\bzero}{\bm{0}}
\newcommand{\mpr}{\mathbb{P}}
\newcommand{\RE}{\mathbf{W}}
\newcommand{\GSA}{GSA}

\newcommand{\argmin}{\operatornamewithlimits{argmin}}
\newcommand{\tr}{\operatornamewithlimits{trace}}
\newcommand{\logdet}{\operatornamewithlimits{logdet}}
\newcommand{\Var}{\operatornamewithlimits{Var}}

\newcommand{\diag}{\operatornamewithlimits{diag}}

\newcommand{\M}[1]{\ensuremath{\boldsymbol{#1}}}

\begin{document}

\title{Network-based Pathway Enrichment Analysis with Incomplete Network Information}
\author[1]{Jing Ma}
\author[2]{Ali Shojaie}
\author[3]{George Michailidis}
\affil[1]{Department of Biostatistics and Epidemiology, University of Pennsylvania Perelman School of Medicine, Philadelphia, 19104, USA}
\affil[2]{Department of Biostatistics, University of Washington, Seattle, 98915, USA}
\affil[3]{Department of Statistics, University of Florida, Gainesville, 32611, USA}

\maketitle

\abstract{\textbf{Motivation:} Pathway enrichment analysis has become a key tool for biomedical researchers to gain insight into the underlying biology of differentially expressed genes, proteins and metabolites. It reduces complexity and provides a system-level view of changes in cellular activity in response to treatments and/or in disease states. Methods that use existing pathway network information have been shown to outperform simpler methods that only take into account pathway membership. However, despite significant progress in understanding the association amongst members of biological pathways, and expansion of data bases containing information about interactions of biomolecules, the existing network information may be incomplete or inaccurate, and is not cell-type or disease condition-specific.  \\

\textbf{Results:} We propose a constrained network estimation framework that combines network estimation based on cell- and condition-specific high-dimensional Omics data with interaction information from existing data bases. The resulting pathway topology information is subsequently used to provide a framework for simultaneous testing of differences in expression levels of pathway members, as well as their interactions. We study the asymptotic properties of the proposed network estimator and the test for pathway enrichment, and investigate its small sample performance in simulated and real data settings.\\

\textbf{Availability:} The proposed method has been implemented in the R-package {\tt netgsa} available on CRAN.\\

\textbf{Contact:} {jinma@upenn.edu}\\

\textbf{Supplementary information:} Supplementary materials including technical details and dataset are available at \textit{Bioinformatics} online.
}

\section{Introduction}\label{intro}

Recent advances in high throughput technologies have transformed biomedical research by 
enabling comprehensive monitoring of complex biological systems. By profiling the activity 
of different molecular compartments (genomic, proteomic, metabolomic), one can delineate 
complex mechanisms that play key roles in biological processes or the development of distinct phenotypes.
These technological advances have thus motivated new methodological developments, most notably the adaptation of systems perspectives to analyze biological systems. Pathway analysis represents a key component in the analysis
process and has been used successfully in generating new biological hypotheses, as well as in determining whether
specific pathways are associated with particular phenotypes. 
Examples include analysis of pathways involved in initiation and progression of cancer and other complex 
diseases \citep{wilson2010epigenetic}, discovering novel transcriptional effects and 
co-regulated genes \citep{green2011signatures}, and understanding the 
basic biological processes in model organisms \citep{houstis2006reactive, gottwein2007viral}. 
See \cite{huang2009systematic} for additional examples of applications.

Pathway analysis methods have evolved since the seminal work by \citet{subramanian2005}.
As pointed out in the review paper by \citet{10.1371/journal.pcbi.1002375}, earlier techniques such as over-representation analysis 
\citep{al2005discovering}, and gene set analysis \citep{subramanian2005, Efron:2007vz} treat each pathway as a set of biomolecules. 
These methods assess whether members of a given pathway have higher than expected levels of activity, either by counting the number of differentially active  members, or by also accounting for the relative rankings of pathway members and/or the magnitude of their associations with the phenotype. 
On the other hand, more recent and statistically powerful methods also account for interactions between biomolecules. These interactions are increasingly available from carefully curated biological
databases, such as Kyoto Encyclopedia of Genes and Genomes \citep[KEGG,][]{Kanehisa01012000},
Reactome \citep{JOSHI-TOPE01012003}, RegulonDB \citep{Huerta01011998} and BioCarta \citep{nishimura2001biocarta}.

A network topology-based method that exhibits superior statistical power in identifying differential activity of pathways was proposed
in \cite{Shojaie:2009lf, Shojaie:2010nz}. The Network-based Gene Set Analysis (NetGSA) method also allows testing for potential changes in
the network structure under different experimental or disease conditions. However, it requires \emph{a priori} knowledge of interactions among pathway members, which, despite rapid progress, remains highly incomplete and occasionally unreliable (see e.g. \cite{23688127} 
and references therein). 
Moreover, existing network information often determine molecular interactions in the normal state of the cell, and do not provide 
any insight into condition/disease-specific alterations in interactions amongst components of biological systems.

The increased availability of large sets of high-dimensional Omics data (e.g. from The Cancer Genome Atlas, http://cancergenome.nih.gov/), coupled with the development of network estimation 
techniques based on graphical models \citep{lauritzen1996graphical} offers the possibility to validate and complement existing network information, and to obtain condition-specific estimates of molecular interactions. Such an approach for leveraging existing knowledge to enhance the analysis of low signal-to-noise biological datasets was advocated in \cite{ideker2011boosting}.

The first contribution of this paper is the development of a method for constrained network estimation from high-dimensional data, together with establishing the consistency of the resulting estimate.
Estimation of high-dimensional networks subject to hard (or soft) constraints on conditional dependence relationships amongst random variables represents a canonical problem in the context of graphical models, and the proposed method for addressing this problem is of independent interest. 
By incorporating the condition-specific network estimates from the proposed method into the NetGSA framework, we also provide a rigorous statistical framework for assessing alterations in biological pathways, referred to as \emph{differential network biology} \citep{ideker2012differential}. 

The proposed framework accounts for two sources of uncertainty: the first concerns the reliability of the external information used for constructing the network estimate from data. The second is the variability of the network estimate, which can impact the pathway enrichment testing procedure. We establish that, under certain regularity conditions, consistent estimates of the network can be obtained, leading, in turn, to an asymptotically most power unbiased test for pathway enrichment analysis.
Our theoretical analysis also sheds light into the potential improvements in accuracy and power by directly accounting for the amount of reliable external network information.

A second objective of this study is to scale up the NetGSA estimation algorithm to very large size networks. 
The main bottleneck in applying the NetGSA methodology arises from the estimation of mixed effects linear parameters---specifically the variance components---for thousands of variables. We develop efficient and stable computational methods for 
estimation of these parameters based on a profile likelihood approach. In particular, we employ a Cholesky factorization of 
the covariance matrices to speed up matrix inversions, and use it to develop a stable algorithm based on Newton's method 
with backtracking line search \citep[][page 487]{Boyd:2004gs} for step size selection. To supply reliable starting points for this 
algorithm, we further develop an approximate method-of-moment-type estimator.

The proposed methods are illustrated on both metabolomics and gene expression data. For mass spectrometry metabolomics profiling 
one can obtain good quality measurements for a few hundred metabolites that do not provide complete coverage of the underlying biochemical pathways. The small number of metabolites in each pathway and the incomplete coverage of the metabolites particularly hinder the  application of over-representation and gene set analysis methods in this setting. In our experience, only topology-based pathway enrichment analysis methods, such as NetGSA, are capable of reliably delineating pathway activity, as illustrated in Section~\ref{realdata}. Further, our investigation of previously analyzed gene expression data set on lung and breast cancer provides new useful insights.

The remainder of the paper is organized as follows. Section~\ref{NetEst} presents the new method for network estimation under external information constraints and establishes its consistency.
Section~\ref{speed} outlines the new computational algorithm for scaling up NetGSA, as well as the inference procedure for both pathway enrichment and differential network analysis.
The performance of the developed methodology is evaluated in Section~\ref{simulation} and is examined on real data sets in Section~\ref{realdata}.

%\begin{methods}
\section{Methods}

Gaussian graphical models \citep[][Chapter 5]{lauritzen1996graphical} are widely used in biological applications to model the interactions among components of biological systems \citep[][chapter 6]{dehmer2008analysis}. 
Specifically, partial correlation networks are commonly used to model interactions in molecular networks; these networks are represented by an undirected graph $G=(V, E)$ with node set $V$ and edge set $E$ corresponding to biomolecules interactions among them, respectively. The edge set $E$ corresponds to the $p\times p$ precision, or inverse covariance, matrix $\Omega$, whose nonzero elements $\omega_{ii'}$ refer to edges between nodes $i$ and $i'$, and indicate that $i$ and $i'$ are conditionally dependent given all other nodes in the network. The magnitude of the partial correlation $\bA_{ii'}=-\omega_{ii'}/\sqrt{\omega_{ii}\omega_{i'i'}}$ determines the strength (positive or negative) of the conditional association between the respective nodes. In the sequel, the matrix $\bA$ will also be called the weighted adjacency matrix, with $\bA_{ii'}$ being the association weight between $i$ and $i'$. 

\subsection{Network Estimation Under External Information Constraints}\label{NetEst}
As discussed in Section~\ref{intro}, the availability of large collections of samples for different disease states and biological processes together with carefully curated information of biomolecular interactions enables the estimation of network structures within the setting of Gaussian graphical models. However, the availability of  external network information provides a novel and unexplored modification of the corresponding network estimation problem. 
Denote by $E^c$ the set of node pairs not connected in the network, i.e. $\omega_{ii'}=0$.
Then, the external information can be represented by the following two subsets
\begin{align*}
E_1 & = \{ (i,i') \in E: i \ne i' , \omega_{ii'} \neq 0 \}, \\
 E_0 &= \{ (i,i') \in E^c: i \ne i', \omega_{ii'} = 0 \}.
\end{align*}
In words, $E_1$ contains known edges, while $E_0$
contains node pairs where it is known that no interaction exists between them. Note that $E_1 \subseteq E$ and $E_0 \subseteq E^c$. 
The external information available in $E_1$ does not imply exact knowledge of the magnitude of $\omega_{ii'}$ nor $\bA_{ii'}$.

Suppose we observe an $m\times p$ data matrix $\bZ=(\bZ_1, \ldots, \bZ_p)$, where each row represents one sample from a $p$-variate Gaussian distribution $\MN(0, \Omega^{-1})$ for a given biological condition (e.g., cancer or normal).
Our goal is then to estimate the network structure, or equivalently the precision matrix $\Omega$, subject to external information encoded in $E_1$ and $E_0$. Let $\bD=\diag(\Omega)$ represent the diagonal matrix whose diagonal entries are the same as $\Omega$ and $\bIP$ be the $p$-identity matrix. Then, $\bA=\bIP - \bD^{-1/2}\Omega \bD^{-1/2}$ is the partial correlation matrix. When $E_1 = E$ and $E_0 = E^c$, the problem becomes that of covariance selection \citep{dempster1972covariance}, which has been studied extensively in the literature. However, to the best of our knowledge, the problem of estimating $\Omega$ (and the partial correlation matrix $\bA$) when $E_1$ and $E_0$ only contain partial information ($E_1 \subsetneq E$ and $E_0 \subsetneq E^c$) has not been investigated before. 

In this section, we assume that the $m$ observations used for estimating condition-specific networks are separate from those used for pathway enrichment analysis (highlighted by the use of $\bZ_i$'s and $m$ to denote the random variables and sample size, respectively). 
The framework introduced in this section reduces the potential bias in small sample settings, and takes advantage of the additional publicly available samples, in lieu of reliable network information. 
With large enough samples, network estimation and pathway enrichment can be performed using the same set of samples by incorporating sample splitting strategies. 
While the problem considered in this section is seemingly similar to matrix completion \citep{Candes:2009mk}, the two problems are fundamentally different in nature. In particular, in this setting, matrix completion corresponds to completing the remaining entries from the partially observed $p\times p$ matrix $\bA$, under some structural assumptions on $\bA$, such as low-rankness. 
On the other hand, in the setting of graphical models, the entries of the weighted adjacency matrix are estimated based on data on the nodes of the graph.

In biological settings, both the structure of the network, as well as strengths of associations may be condition-specific. Therefore, we need to accurately estimate the nonzero entries in $\Omega$ to recover both the structure of the network and the strength of associations between nodes. 
In the absence of any external information, the $\ell_1$-penalized negative log-likelihood estimate of $\Omega$ is obtained by solving
\begin{equation}\label{abs_info}
\argmin_{\Omega \succ 0} \left\{ \tr(\Omega\hat{\Sigma}) -\logdet \Omega + \lambda \|\Omega\|_1 \right\},
\end{equation}
wherein $\hat{\Sigma} = \bZ^{T}\bZ/m$ is the empirical covariance matrix of the data,
$\|\Omega\|_1 = \sum_{i\ne i'}|\omega_{ii'}|$ denotes the $\ell_1$ norm of the parameters, and $\lambda$ is the regularization parameter.
In the presence of external information, the problem can be cast as the following constrained optimization one
\begin{align}\label{e:constraints}
\min_{\Omega \succ 0} & ~  \left\{ \tr (\Omega \hat{\Sigma}) -\logdet \Omega \right\},
\end{align}
subject to $\omega_{ii'} = 0$ for $(i,i') \in E_0$, $\omega_{ii'} \ne 0$ for $(i,i') \in E_1$, and $\sum_{i\ne i', ~(i,i') \notin E_0 \cup E_1} |\omega_{ii'}| \le t.$

In the following, we present a two-step procedure to solve the constrained optimization problem~\eqref{e:constraints}. The proposed approach combines the neighborhood selection technique \citep{Meinshausen:2006rw} with constrained maximum likelihood estimation.
It exploits the fact that the estimated neighbors of each node using neighborhood selection coincide with the nonzero entries of the inverse covariance matrix \citep{Friedman:2008lt}.
Specifically, in neighborhood selection the network structure is estimated by finding the optimal set of predictors when regressing the random variable $\bZ_i$ corresponding to node $i \in V$ on all other variables, using an $l_1$-penalized linear regression. The coefficients for this optimal prediction $\btheta^i$ are closely related to the entries of the inverse covariance matrix: for all $i' \ne i$, $\theta^i_{i'} = -\omega_{ii'}/\omega_{ii}$. The set of nonzero coefficients of $\btheta^i$ is thus the same as the set of nonzero entries in the row vector of $\omega_{ii'} \ (i' \ne i)$, which defines the set of neighbors of node $i$. 

Let $J_1^i$ and ${J}_0^i$ denote the subsets $V \backslash i$ for which external information is available: $J_1^i$ is the set of nodes which are known to be in the neighborhood of $i$, and ${J}_0^i$ is the set of nodes which are known to be not connected to $i$. Let $\bZ_{-i}$ denote the submatrix obtained by removing the $i$th column of $\bZ$. Assume all columns of $\bZ$ are centered and scaled to have norm 1. 
Denote by $\Sa$ the set of all $p\times p$ positive definite matrices and $\SE = \{\Omega \in \mathbb{R}^{p\times p}: \omega_{ii'}=0, \mbox{ for all } (i,{i'}) \notin {E} \mbox{ where } i\ne {i'} \}$. 
The proposed algorithm proceeds in two steps.

\begin{itemize}
\item [(i)] Estimate the network structure $\hat{E}$. For every node $i$, find $\hat{\btheta}^i$ via the following steps.
\begin{itemize}
\item [(a)] For $i' \in J_0^i$, set $\hat{\theta}_{i'}^{i} = 0 $.
\item [(b)] For $i' \in J_1^i$, find $\hat{\theta}_{i'}^{i} $ using linear regression
\begin{equation}\label{opt0-0}
\hat{\btheta}_{J_1^i}^{i} = \argmin_{\btheta \in \mathbb{R}^{|J_1^i|}}   \frac{1} {m} \|\bZ_i - \bZ_{J_1^i}\btheta\|^2_2.
\end{equation}
\item [(c)] For $i'\in \tilde J \equiv V\backslash \{J_1^i \cup J_0^i \cup \{i\}\}$, find $\hat{\theta}_{i'}^{i} $ using lasso
\begin{equation}\label{opt0-1}
\hat{\btheta}_{\tilde J}^{i} = \argmin_{\btheta \in \mathbb{R}^{|\tilde J|}} \frac{1} {m} \|\RE_i - \bZ_{\tilde J}\btheta\|^2_2  + 2\lambda \sum_{i'\in \tilde J } |\theta_{i'}|,
\end{equation}
where $\RE_i = \bZ_i - \bZ_{J_1^i}\hat \btheta_{J_1^i}^i$ is the residual vector after regressing $\bZ_i$ on the known connections.
\end{itemize}
The edge set $\hat{E}$ is estimated to be $\{ (i, i'): \hat{\theta}^i _{i'} \neq 0 \mbox{ OR } \hat{\theta}^{i'}_i \neq 0\}$.
\item
[(ii)] Given the structure $\hat{E}$, estimate the inverse covariance matrix $\hat{\Omega}$ by
\begin{equation}\label{opt3}
\hat{\Omega} = \argmin_{\Omega \in \Sa \cap \mathcal{S}^p_{\hat{E}}   } \left\{ \tr(\hat{\Sigma}\Omega) -\logdet \Omega \right\}.
\end{equation}
\end{itemize}

\begin{remark}\label{remark:uncertainty}
In step (i-b) of the algorithm, the coefficients $\btheta^i$ for known edges have not been penalized in \eqref{opt0-0}. In settings where the external information may be unreliable, we can augment \eqref{opt0-0} with a lasso penalty $\lambda \sum_{i'\in  J_1^i } t_{i'}|\theta_{i'}|$, where the penalty weights $t_{i'}\ (i' \in  J^i_1)$ allow for different penalization depending on the reliability of existing information. 

The second step focuses on estimation of the magnitude of nonzero entries in the precision matrix $\Omega$, given the estimated network topology $\hat E$. The optimization problems in both steps are convex and can be solved efficiently using existing software (e.g., \verb=glmnet= and \verb=glasso= in \verb=R=).
\end{remark}

The proposed estimator enjoys nice theoretical properties under certain regularity conditions. Before presenting the main result, we introduce some additional notations. Let $\Sigma_0$ be the true covariance matrix and $\Omega_0 = \Sigma_0^{-1}$. For $i = 1, \ldots, p$, 
denote by $\|\btheta^i\|_0 = \# \{i': ~\theta^i_{i'} \ne 0\}$ the $l_0$ norm of $\btheta^i$. 
Write $s=\underset{i=1,\ldots,p}{\max} \|\btheta^i\|_0$ and $S_0=\sum_{i=1}^p \|\btheta^i\|_0$.
For a subset $J\subset \{1,\ldots,p\}$, let $\bZ_J$ be the submatrix obtained by removing the columns whose indices are not in $J$. We make the following assumptions.
\begin{assumption}\label{minmaxeig}
There exist $\phi_1, \phi_2>0$ such that the eigenvalues of $\Sigma_0$ are bounded, i.e.
$
0<  \phi_2  \le  \phi_{\min}(\Sigma_0) \le \phi_{\max}(\Sigma_0) \le {1/ \phi_1} < \infty.
$
\end{assumption}
\begin{assumption}\label{res_assmp2}
There exists $\kappa(s)>0$ such that 
\begin{align}\label{res_eq2}
 \min_{\substack{|{J} |\le s} }\min_{\substack{\bdelta \in \mathbb{R}^{p}\\  \| \bdelta_{{J}^c}\|_1 \le 3\|\bdelta_{{J}}\|_1} } \frac{1}{\sqrt{m} } \frac{\|\bZ\bdelta\|_2}{ \|\bdelta_{{J}}\|_2}  \ge \kappa(s) .
\end{align}
\end{assumption}
Assumption~\ref{minmaxeig} is standard in high-dimensional settings. Assumption~\ref{res_assmp2} corresponds to the restricted eigenvalue assumption introduced in \cite{Bickel:2009vp}, which is presented here for completeness.

Denote by $|E|$ the cardinality of the edge set $E$. Let $r \equiv ( |E_0| + |E_1| ) / \{ p(p-1)/2 \}$ represent the percentage of external network information available. Clearly, $0\le r < 1$. We are now ready to state our first result. 
\begin{theorem}\label{thm1e}
Suppose Assumption~\ref{minmaxeig} holds and Assumption~\ref{res_assmp2} is satisfied with $\kappa(2s)$. For constants $c_1>4$ and $0<k_1<1$, assume also that the sample size satisfies
\begin{equation}\label{c1k1}
m \ge \left\{\frac{16c_1}{ k_1 \phi_1 \kappa^2(2s)}\right\}^2 (1-r) S_0 \log(p-rp),
\end{equation} 
where $S_0$ is the total number of nonzero parameters excluding the diagonal. Consider $\hat{\Omega}$ defined in \eqref{opt3}. Then, with probability at least $1-2p^{2-c_1^2/8}$, under appropriately chosen $\lambda$, we have
\begin{equation}\label{e:conv:rate}
\| \hat{\Omega} - \Omega_0 \|_2 \le \| \hat{\Omega} - \Omega_0 \|_F = O\left( \sqrt{\frac{S_0\log (p-rp) }{ m}}\right).
\end{equation}
\end{theorem}
\noindent
\begin{remark}
In addition to the improved sample complexity \eqref{c1k1}, the convergence rate in \eqref{e:conv:rate} indicates an improvement of the order of $\sqrt{S_0\log(1-r)^{-1}/m}$ in the presence of external information. This improvement is particularly important for our analysis of power properties of NetGSA in Section~\ref{speed-2}, which requires norm consistency of adjacency matrix estimation. While consistency can be established using a theoretical analysis similar to graphical lasso \citep{Rothman:2008ce}, our proofs in Section A of the Supplementary Materials utilize the techniques from \cite{Bickel:2009vp} and \citet{Zhou:2010on} to characterize the improvement in rates resulting from the external information.
\end{remark}

Let $\bA_0$ be the true partial correlation matrix, i.e. $\bA_0 = \bIP - \bD_0^{-1/2} \Omega_0 \bD_0^{-1/2}$, where $\bD_0 = \diag(\Omega_0)$. The following corollary is an immediate result of Theorem~\ref{thm1e}.
\begin{corollary}\label{cor1}
Let assumptions in Theorem~\ref{thm1e} be satisfied. Assume further that  $S_0=o(m/\log (p-rp))$. For $\hat{\Omega}$ defined in \eqref{opt3}, let $\hat{\bA}$ be the corresponding partial correlation matrix. Then, with probability at least $1-2p^{2-c_1^2/8}$, under appropriately chosen $\lambda$, we have
\begin{equation*}
\| \hat{\bA} - \bA_0 \|_2 = o(1).
\end{equation*}
\end{corollary}
\begin{remark}
Corollary~\ref{cor1} implies that, under certain regularity conditions, the error in the condition-specific network estimate $\hat \bA$ is negligible. This proves essential for establishing power properties of NetGSA with estimated network information, as shown in the next section. The proof of Corollary~\ref{cor1} is available in Section A of the Supplementary Materials. 
\end{remark}

The tuning parameter $\lambda$ in the first step of the proposed algorithm is important for selecting the correct structure of the network, which further affects the magnitude of the network interactions in the second step. Accurate estimation of these magnitudes is crucial for topology-based pathway enrichment methods. We propose to select $\lambda$ using the Bayesian Information Criterion (\bic). Specifically, for a given $\lambda$, we define 
\begin{equation}\label{eq:BIC}
\bic(\lambda) =\tr ( \hat{\Sigma}\hat \Omega_{\lambda}) - \log \det (\hat \Omega_{\lambda} ) + \frac{\log(m)}{m} |\hat E_{\lambda}|,
\end{equation}
where $\hat \Omega_{\lambda}$ is the estimated precision matrix from the data and $\hat E_{\lambda}$ is the estimated edge set. The optimal tuning parameter is thus $\lambda^*=\argmin_{\lambda} \bic(\lambda)$.

\subsection{NetGSA with Estimated Network Information}\label{speed}
Next, we discuss how (condition-specific) estimates of bimolecular interactions from Section~\ref{NetEst} can be incorporated into the NetGSA framework to obtain a rigorous inference procedure for both pathway enrichment and differential network analysis. 
To this end, we formally define the NetGSA methodology based on undirected Gaussian graphical models in Section~\ref{speed-1}. In Section~\ref{speed-2}, we discuss how the constrained-network estimation procedure of Section~\ref{NetEst} can be combined with NetGSA to rigorously infer differential activities of biological pathways, as well as changes in their network structures. 

\subsubsection{The latent variable model}\label{speed-1}

Consider $p$ genes (proteins/metabolites) whose activity levels across $n$ samples are organized in a $p \times n$ matrix $\mathcal{D}$. 
In the framework of NetGSA, the effect of genes (proteins/metabolites) in the network is captured using a latent variable model \citep{Shojaie:2009lf,Shojaie:2010nz}. Denote by $\bY$ an arbitrary column of the data matrix $\mathcal{D}$. %Consider the simple network in Figure \ref{simple:net}. 
Suppose the observed data can be decomposed into signal, $\bX$, plus noise $\beps\sim \MN_p(\bzero,\sigma_\eps^2 \bIP)$, i.e. $\bY = \bX + \beps$.
The latent variable model assumes that the signal $\bX$ follows a multivariate normal distribution with partial correlation matrix $\bA$. Based on the connection between linear recursive equations and covariance selection proposed in \citet{wermuth1980linear}, there exists a lower triangular matrix $\Lambda$ such that $\Lambda^{-1}\bX = \bgamma$, where $\bgamma \sim \MN_p(\bmu,\sigma_{\bgamma}^2 \bIP)$ and $\Lambda \Lambda^T = (\bIP - \bA)^{-1} $. Note that the current version of the NetGSA model differs from the original model in \citet{Shojaie:2009lf,Shojaie:2010nz}. This difference is primarily manifested through the definition of $\Lambda$ in the two models: $\Lambda$ is defined here based on the {\em undirected partial correlation network} $\bA$, whereas it was previously defined based on directed (physical) interactions among genes (proteins/metabolites) in \citet{Shojaie:2009lf,Shojaie:2010nz}.

Assuming that $\bgamma$ and $\beps$ are independent, the NetGSA model can then be summarized as 
\begin{equation}\label{TSmodel}
  \bY = \Lambda\bgamma + \beps.
\end{equation}

The NetGSA methodology allows for more complex models, including time course observations. For expositional clarity, we present the methodology in the setting of two experimental conditions and consider the general case where $\bA^{(1)} \ne \bA^{(2)}$. Details of NetGSA under multiple conditions can be found in \citet{Shojaie:2010nz} and are applicable for the undirected networks presented in this work. Let $\bY_j^{(k)}\ (j=1,\ldots, n; k=1, 2)$ be the $j$-th sample in the expression data under condition $k$ ($j$th column of data matrix $\mathcal{D}$), with the first $n_1$ columns of $\mathcal{D}$ corresponding to condition $1$ (control) and the remaining $n_2 = n-n_1$ columns to condition $2$ (treatment). Denote by $\lk $ the \emph{influence matrix} and $\bmu^{(k)}$ the mean vector under condition $k$. The NetGSA framework considers a latent variable model of the form
\begin{align*}
\bY_j^{(1)} &= \Lambda^{(1)} \bmu^{(1)} + \Lambda^{(1)} \bgamma_j + \beps_j, \quad (j=1, \ldots, n_1),\\
\bY_j^{(2)} &= \Lambda^{(2)} \bmu^{(2)} + \Lambda^{(2)} \bgamma_j + \beps_j, \quad (j=n_1+1, \ldots, n).
\end{align*}
Here, $\bgamma_j$ is the vector of (unknown) random effects, and $\beps_j$ is the vector of random errors. They are independent and normally distributed with mean $\bzero$ and variances $\sigma^2_{\bgamma} \bIP$ and $\sigma^2_{\beps} \bIP$, respectively.

Inference in NetGSA requires estimation of the mean parameters $\bmu^{(1)}$ and $\bmu^{(2)}$ and variance components ${\sigma}_{\bgamma}^2$ and ${\sigma}_{\beps}^2$. The variance components can be estimated via maximum likelihood or restricted maximum likelihood, which can be computationally demanding for large networks. To extend the applicability of the NetGSA, we consider using Newton's method for estimating the variance parameters based on the profile log-likelihood to improve the computational stability. See Section B of the Supplementary Materials for more details.

%%%%%%%%%%%%%%%%%%%%
\subsubsection{Joint pathway enrichment and differential network analysis using NetGSA}\label{speed-2}
%%%%%%%%%%%%%%%%%%%%

To test for enrichment of a pre-specified pathway $P$, \citet{Shojaie:2009lf} propose the contrast vector \citep{searle1971linear} 
$\bel=(- {\bfb} \Lambda^{(1)} \odot {\bfb}, {\bfb} \Lambda^{(2)} \odot {\bfb}) $, where ${\bfb}$ is a row binary vector determining the membership of genes in a pre-specified pathway $P$ and $\odot$ denotes the Hadamard product. 
The advantage of this contrast vector is that it isolates influences from nodes outside the pathways of interest.
Let $\bbeta = ({\bmu^{(1)}}^T, {\bmu^{(2)}}^T)^T$ be the concatenated vector of means.
The null hypothesis of no pathway activity vs the alternative of pathway activation then becomes
  \begin{equation}\label{TStest}
     \ H_0: \bel {\bbeta} =0, \quad H_1: \bel {\bbeta} \ne 0.
  \end{equation}
The significance of individual contrast vectors in \eqref{TStest} can be tested using the following Wald test statistic
\begin{equation}\label{TSt-test}
  TS = \frac{ \bel\hat{{\bbeta}} }{ \se(\bel\hat{{\bbeta}} )},
\end{equation}
where $ \se(\bel\hat{{\bbeta}} )$ represents the standard error of $ \bel\hat{{\bbeta}} $ and $\hat \bbeta$ is the estimate of $\bbeta$. Both $\bel$ and $ \se(\bel\hat{{\bbeta}} )$ depend on the underlying networks, which are estimated using data from the two experimental conditions. Under the null hypothesis, $TS$ follows approximately a $t$-distribution whose degrees of freedom can be estimated using the Satterthwaite approximation method \citep{Shojaie:2010nz}.

The above general framework allows for test of pathway enrichment in arbitrary subnetworks, while automatically adjusting for overlap among pathways. In addition, the above choice of contrast vector $\bel$ accommodates changes in the network structure. Such changes have been found to play a significant role in development and initiation of complex diseases \citep{chuang2012subnetwork}, and NetGSA is currently the only method that systematically combines the changes in expression levels and network structures, when testing for pathway enrichment.
However, the applicability of the existing NetGSA framework \citep{Shojaie:2009lf, Shojaie:2010nz} is limited by the assumption of known network structure (namely $\Lambda^{(k)}, k=1,2$). In the current framework, we estimate $\Lambda^{(k)}\ (k=1,2)$ from data as discussed in Section~\ref{NetEst}. 
We next show that NetGSA with estimated network information provides valid inference for pathway enrichment and differential network analysis.

For $k=1, 2$, let $\bZ^{(k)}$ of dimension $m_k \times p$ be the data matrix used to separately estimate the partial correlation matrix  under condition $k$. 
Denote by $S_{k}$ the number of nonzero off-diagonal entries in the true partial correlation matrix $\bA^{(k)}_0$ and by $r_k$ the percentage of available external information. We obtain the following result.
\begin{theorem}\label{thm:ump}
Let assumptions in Theorem~\ref{thm1e} be satisfied and $S_{k}=o(m_k/\log (p-r_kp))$ under each condition $k\ (k=1, 2)$. Consider the inverse covariance matrices $\hat{\Omega}^{(k)}$ estimated from~\eqref{opt3} of Section~\ref{NetEst}. Then the test statistic in \eqref{TSt-test} based on the corresponding networks $ \hat{\bA}^{(k)}$ is an asymptotically most powerful unbiased test for \eqref{TStest}.
\end{theorem}
\begin{remark}
Theorem 2.1 of \cite{Shojaie:2010nz} says that NetGSA is robust to uncertainty in network information. Specifically, \cite{Shojaie:2010nz} show that if the error in network information $\Delta_{\bA_0^{(k)}} = \hat{\bA}^{(k)} - \bA_0^{(k)}$ satisfies  $\| \Delta_{\bA_0^{(k)}} \|_2 = o_{\mpr}(1)$, then NetGSA is an asymptotically most powerful unbiased test for \eqref{TStest}. The result in Theorem~\ref{thm:ump} establishes this property for (partially) estimated networks using the consistency of our proposed network estimation procedure in Theorem~\ref{thm1e} and Corollary~\ref{cor1}. A detailed proof can be found in Section A of the Supplementary Materials. 
\end{remark}

%\end{methods}

\section{Simulation Results}\label{simulation}

We present two simulation studies to assess the performance of the proposed network estimation procedure, as well as its impact on NetGSA. We refer readers to Section C of the Supplementary Materials for additional simulation scenarios ---including validation of Type I errors and settings with a large number of variables $p$--- and subsequent discussion.

Our first experiment is based on an undirected network of size $p=100$. The network structure is extracted from the DREAM3 challenge \citep{DREAM32010} corresponding to the Ecoli network (labeled Ecoli 1). The pathways of interest are determined through a community detection algorithm based on the leading nonnegative eigenvector of the modularity matrix of the network \citep{igraph2006}. Under the null hypothesis, all nodes have the same mean expression values of 1. Under the alternative hypothesis,  the mean expression levels of 0\%, 30\%, 40\% and 60\% of nodes in subnetworks 1, 3, 5 and 7 are increased by 0.5, respectively.  
%
%the proportion of nodes that have positive mean changes of magnitude 0.5 is 0\%, 30\%, 40\% and 60\% for subnetworks 1, 3, 5 and 7, respectively. The same pattern applies to subnetworks 2, 4, 6 and 8.

Our second experiment considers a network of size $p=160$ with 8 subnetworks of equal sizes, all of which are generated from the same scale-free graph of size 20. To allow for interactions across subnetworks, there is 20\% chance for the hub node in each subnetwork to connect to the hub node in another subnetwork. Mean expression values for all nodes are the same under the null hypothesis. Under the alternative hypothesis, we allow, respectively, 0\%, 40\%, 60\% and 80\% of the nodes to have positive mean changes of magnitude 0.5 for subnetworks 1-4. Subnetworks 5-8 follow the same pattern. 

\begin{figure}[ht] 
%\captionsetup{width=0.9\textwidth}  
\centering
\begin{minipage}{\linewidth}
\centerline{\includegraphics[scale= 0.3]{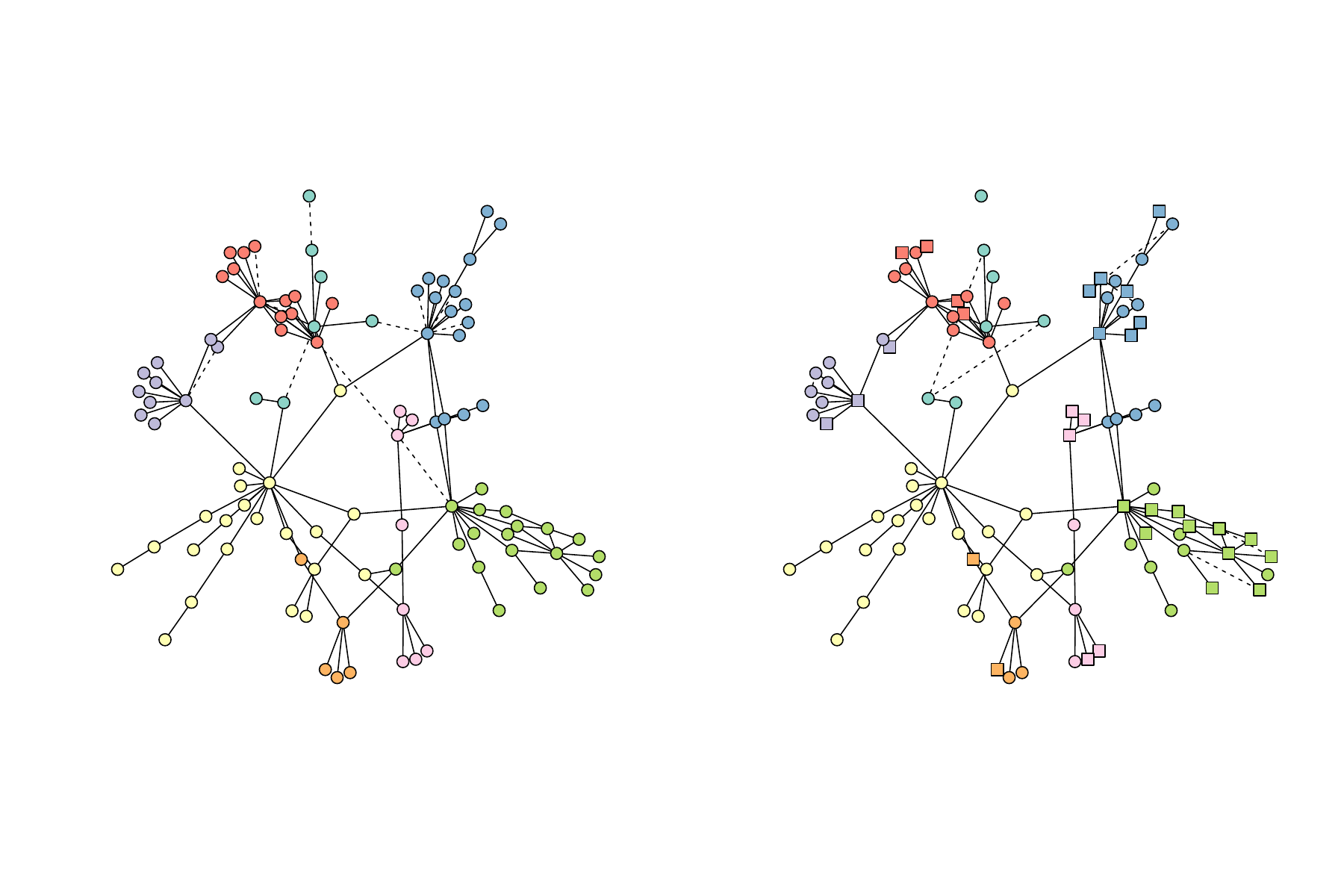}}
%\centerline{\includegraphics[scale= 0.4,bb=0 0 900 450]{NetVis_p100_20160504_v3_100.eps}}
\centerline{\includegraphics[scale= 0.3]{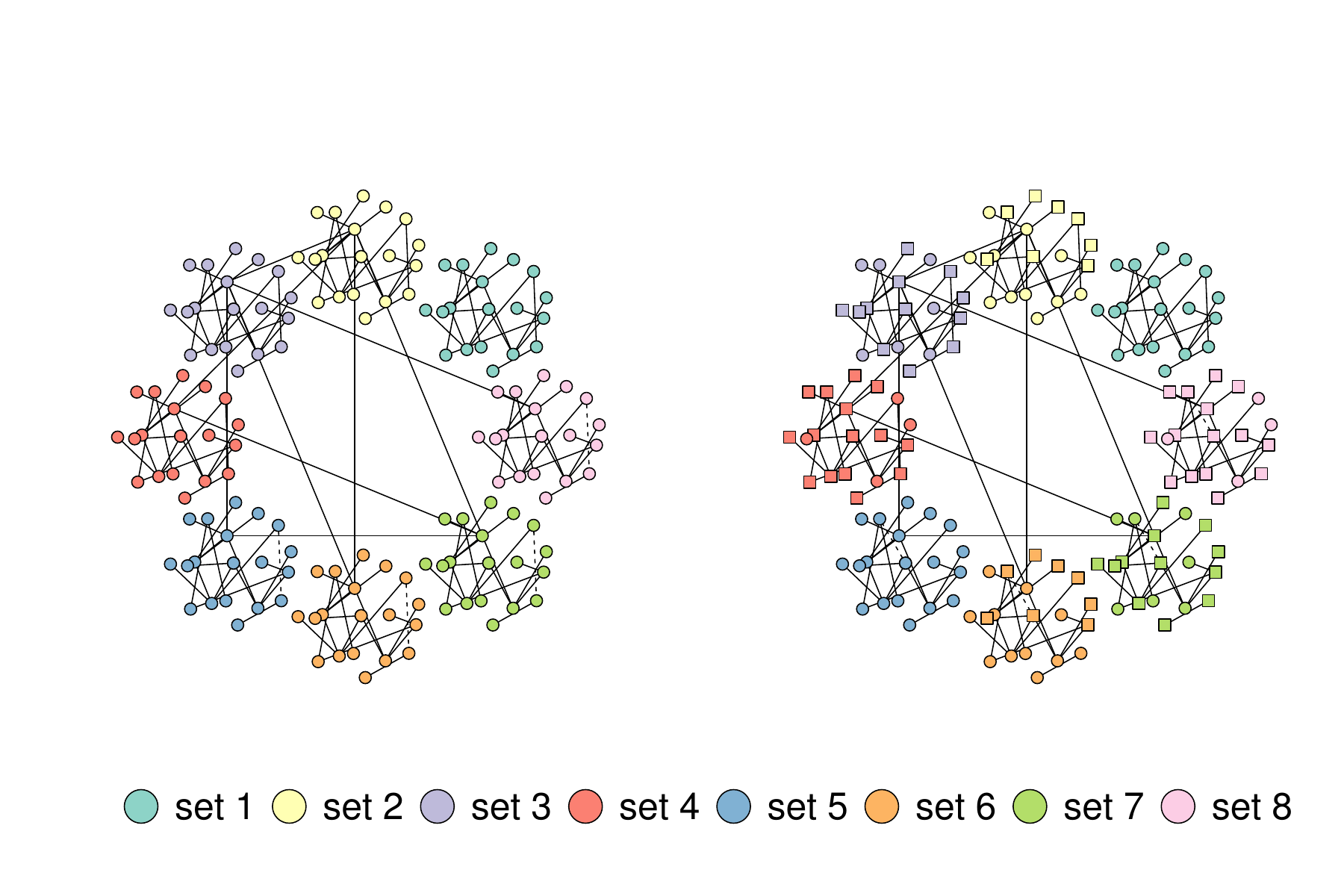}}
   \caption{The network and subnetwork topology in experiment 1 under the null (top left) and alternative (top right), and experiment 2 under the null (bottom left) and alternative (bottom right). Dashed lines represent edges that are present in only one condition. Nodes in square are associated with mean changes.}
\label{fig:p100:net}
\end{minipage}
\end{figure} 

In both experiments, we also allow the structures in four subnetworks under the alternative hypothesis to differ from their null equivalent by a small amount, in order to simultaneously test pathway enrichment and differential networks. Fig.~\ref{fig:p100:net} shows the network topologies as well as the structural changes for the chosen subnetworks from the null to the alternative hypothesis in the two experiments. 
Further, we study the robustness of NetGSA to model misspecification by including scenarios where a proportion (50\% for $r=0.2$ and 20\% for $r=0.8$ in experiment 1, and 60\% in experiment 2) of the supplied structural information is incorrectly specified, i.e. they are not present in the true model. 

%%%%%%%%%%%%%%%%%%%

To illustrate how external network information affects the estimation accuracy, we vary the percentage of information $r$ from 0 to 1. When $r$ is less than 1, we estimate the adjacency matrices using the proposed two-step procedure and fill in the nonzero edges with the estimated weights. When full knowledge of the network topology is given ($r=1$), we only apply the second step to estimate the edge weights. When there exist misspecified edges in the external information, we use two tuning parameters for network estimation, one for controlling the overall sparsity of the network and the other for correcting the misspecified edges. The optimal tuning parameters are selected over a grid of values using \bic~defined in \eqref{eq:BIC}. 

Table~\ref{p160_err_sparse} compares the estimated networks with the true model under several deviance measures based on 100 simulation replications; in both experiments, the sample size for both null and alternative hypotheses is $m = 100$. The Matthews correlation coefficients improve significantly as the percentage of external information $r$ increases from 20\% to 80\%, and the Frobenius norm loss shows a clear decreasing trend, both indicating the improvement in estimation accuracy when more external information is available. In cases where the information is misspecified (denoted by 0.2(m) and 0.8(m)), one can see that the performance of network estimation is not compromised by much after properly selecting the tuning parameters. 

\begin{table}[ht]
\centering
\caption{False positive rate (FPR in percentage), false negative rate (FNR in percentage), Matthews correlation coefficient (MCC) and Frobenius norm loss (Fnorm) for network estimation in experiments 1 and 2}\label{p160_err_sparse}
\resizebox{0.85\linewidth}{!}{% 
\begin{tabular}{llccccccccc}
\hline\hline
\multicolumn{2}{c}{} &\multicolumn{4}{c}{$p=100$} & &\multicolumn{4}{c}{$p=160$} \\
 &$r$ & FPR(\%) & FNR(\%) & MCC & Fnorm & & FPR(\%) & FNR(\%) & MCC & Fnorm \\ 
 \hline
\multirow{5}{*}{Null}
&0.0 & 9.46 & 2.78 & 0.43  & 0.48   & & 2.94 & 0.84 & 0.54 & 0.36 \\  
&0.2 & 7.64 & 5.83 & 0.45  & 0.46   & & 2.77 & 1.03 & 0.55 & 0.34 \\ 
&0.8 & 1.81 & 1.22 & 0.75  & 0.28   & & 1.18 & 0.02 & 0.72 & 0.24 \\ 
&0.2(m) & 7.91 & 4.85 & 0.45 & 0.46 & & 2.76 & 0.95 & 0.55 & 0.34 \\
&0.8(m) & 2.29 & 3.82 & 0.70 & 0.31 & & 1.22 & 0.02 & 0.71 & 0.25 \\
&  &  & &  &  & &  &  &  &  \\ 
\multirow{5}{*}{Alt}
&0.0 & 8.71 & 1.52 & 0.44 & 0.45    & & 2.90 & 0.88 & 0.54 & 0.36 \\  
&0.2 & 7.09 & 3.82 & 0.47 & 0.42    & & 2.73 & 0.88 & 0.55 & 0.35 \\ 
&0.8 & 1.80 & 1.19 & 0.75 & 0.25    & & 1.19 & 1.89 & 0.71 & 0.26 \\ 
&0.2(m) &7.29 & 2.62 & 0.47 & 0.42  & & 2.72 & 0.78 & 0.55 & 0.34  \\
&0.8(m) &2.17 & 5.50 & 0.69 & 0.29  & & 1.22 & 1.93 & 0.70 & 0.27\\
\hline
\end{tabular}%
}
\end{table}

Next, we examine the performance of NetGSA in detecting pathway enrichment by comparing it with Gene Set Analysis \citep[GSA,][]{Efron:2007vz}. GSA tests a competitive null hypothesis and compares the set of genes in the pathway with its complement in terms of association with the phenotype. The underlying model consists of both randomization of the genes and permutation of the samples, which are combined into the idea of `restandardization'. This method is later denoted by GSA-c. In addition, we consider GSA with permutation of the samples only, later denoted by GSA-s, since this version of GSA compares the set of genes in the pathway with itself. 

Tables~\ref{GSANetGSA:MLM} and \ref{GSANetGSA:exp2} present the estimated powers for each subnetwork in the two experiments from 100 simulation replicates, respectively. Here we use $n_1=n_2=25$ samples for each condition in experiment 1 and $n_1=n_2=40$ in experiment 2, which are different from the datasets used for network estimation. The powers are calculated as the proportion of replicates that show differential changes, based on the false discovery rate (FDR) controlling procedure of \cite{Benjamini:1995jh}. {To facilitate comparison, different FDR cutoffs are used for GSA and NetGSA to ensure consistent type I error for the first pathway in both experiments}. For NetGSA, we look at scenarios when there is 20\% and 80\% external structural information (with and without misspecification) and use the estimated networks to test enrichment for each subnetwork. We also include the scenario where the exact networks with correct edge weights are provided, in which case only the variance components and mean expression values are estimated from the mixed linear model. True powers for each subnetwork are calculated by replacing all unknown parameters with their corresponding known values. 

\begin{table}[ht]
\centering
\caption{{Powers in experiment 1. False discovery rate cutoffs are $q^*=0.01$ for 0.2, 0.8, 0.2(m) and 0.8(m), 0.05 for GSA-s and 0.10 for E and GSA-c. Here 0.2/0.8 refer to NetGSA with 20\%/80\% external information; E refers to NetGSA with the exact networks; T refers to the true power; GSA-c/GSA-s refer to Gene Set Analysis with/without randomization of the genes based on 1000 permutations; 0.2(m)/0.8(m)  refer to NetGSA with 20\%/80\% misspecified external information.} }\label{GSANetGSA:MLM}
%\resizebox{0.9\linewidth}{!}{%
\begin{tabular}{ccccccccc}
\hline\hline
\multicolumn{1}{c}{}& \multicolumn{8}{c}{$p=100$}  \\
Pathway & 0.2 & 0.8 & E & T & GSA-s & GSA-c &0.2(m) & 0.8(m)\\
\hline
  1 & 0.03 & 0.03 & 0.08 & 0.06 & 0.15 & 0.04 & 0.03 & 0.02 \\ 
  2 & 0.08 & 0.08 & 0.08 & 0.06 & 0.09 & 0.00 & 0.09 & 0.09 \\ 
  3 & 0.36 & 0.33 & 0.43 & 0.46 & 0.24 & 0.00 & 0.40 & 0.38 \\ 
  4 & 0.38 & 0.26 & 0.09 & 0.07 & 0.26 & 0.05 & 0.37 & 0.24 \\ 
  5 & 0.91 & 0.91 & 0.95 & 0.97 & 0.95 & 0.00 & 0.92 & 0.89 \\ 
  6 & 0.27 & 0.24 & 0.24 & 0.26 & 0.37 & 0.00 & 0.26 & 0.25 \\ 
  7 & 0.72 & 0.80 & 0.99 & 0.99 & 0.98 & 0.14 & 0.69 & 0.86 \\ 
  8 & 0.45 & 0.61 & 0.63 & 0.57 & 0.87 & 0.00 & 0.51 & 0.58 \\
\hline
\end{tabular}
\end{table}

\begin{table}[ht]
\centering
\caption{Powers in experiment 2. False discovery rate cutoffs are $q^*=0.01$ for 0.2, 0.8, 0.2(m) and 0.8(m), 0.05 for GSA-s and 0.10 for E and GSA-c. Here 0.2/0.8 refer to NetGSA with 20\%/80\% external information; E refers to NetGSA with the exact networks; T refers to the true power; GSA-c/GSA-s refer to Gene Set Analysis with/without randomization of the genes based on 1000 permutations; 0.2(m)/0.8(m)  refer to NetGSA with 20\%/80\% misspecified external information.}\label{GSANetGSA:exp2}
%\resizebox{0.9\linewidth}{!}{%
{\begin{tabular}{ccccccccc}
\hline\hline
\multicolumn{1}{c}{}& \multicolumn{8}{c}{$p=160$}  \\
Pathway & 0.2 & 0.8 & E & T & GSA-s & GSA-c &0.2(m) & 0.8(m) \\
\hline 
  1 & 0.04 & 0.06 & 0.02 & 0.05 & 0.06 & 0.02 & 0.04 & 0.06 \\ 
  2 & 0.37 & 0.36 & 0.30 & 0.36 & 0.41 & 0.00 & 0.36 & 0.36 \\ 
  3 & 0.88 & 0.94 & 0.96 & 0.99 & 0.99 & 0.00 & 0.89 & 0.94 \\ 
  4 & 0.97 & 0.99 & 1.00 & 1.00 & 1.00 & 0.23 & 0.97 & 0.99 \\ 
  5 & 0.36 & 0.25 & 0.11 & 0.11 & 0.13 & 0.15 & 0.35 & 0.25 \\ 
  6 & 0.38 & 0.27 & 0.03 & 0.07 & 0.26 & 0.01 & 0.35 & 0.27 \\ 
  7 & 0.66 & 0.72 & 0.92 & 0.92 & 1.00 & 0.00 & 0.67 & 0.72 \\ 
  8 & 0.90 & 0.95 & 1.00 & 1.00 & 1.00 & 0.13 & 0.91 & 0.94 \\
  \hline
\end{tabular}%
}
\end{table} 

For $p=100$, the results from NetGSA with the exact networks agree with the true powers, indicating low powers for subnetworks 1, 2 and 4, slightly higher powers for 3, 6 and 8, high powers for 5 and 7 due to significant changes in mean expression levels and structures. When the exact networks are unknown, we see clear improvement in the estimated powers for subnetworks 4, 7 and 8 as the percentage of external information increases from 20\% to 80\%. GSA-s does reasonably well with overestimated powers for subnetwork 8. The last two columns in Table~\ref{GSANetGSA:MLM} show the estimated powers from NetGSA when the external information is misspecified. For both cases ($r=0.2$ and $r=0.8$), the results bear high similarity to those in the first two columns, which suggests that the proposed framework is robust to inaccuracy in network information.

For $p=160$, the NetGSA estimated powers when 20\% external information is available match the true powers reasonably well, with a small underestimation of powers for subnetworks 3, 7 and 8. We note marked improvement in the three corresponding values when the external information increases to 80\%. 
Moreover, NetGSA is able to distinguish subnetworks 5-8 that have both changes in mean values and subnetwork topology from their corresponding counterparts 1-4. When the external information is misspecified, the last two columns indicate that NetGSA still returns valid powers that are comparable to those obtained with correctly specified structural information. \GSA-s yields a small overestimation of powers for subnetwork 7. 

In both experiments, \GSA~with randomization of the genes (GSA-c) fails to identify any of the differential subnetworks.
%, which can be due to the small number of pathways. 

%%%%%%%%%%%%%%%%%%%%

\section{Applications to Metabolomics and Genomics Data}\label{realdata}
We apply NetGSA to three Omics data sets to demonstrate its potential in revealing biological insights. In all three studies, the $p$-values were corrected for multiple comparisons using the false discovery rate control procedure proposed in \citet{benjamini2001control} to account for the dependency among KEGG pathways.

\begin{figure}[ht]   
\centering
\begin{minipage}{\linewidth}
\centerline{\includegraphics[scale=0.45]{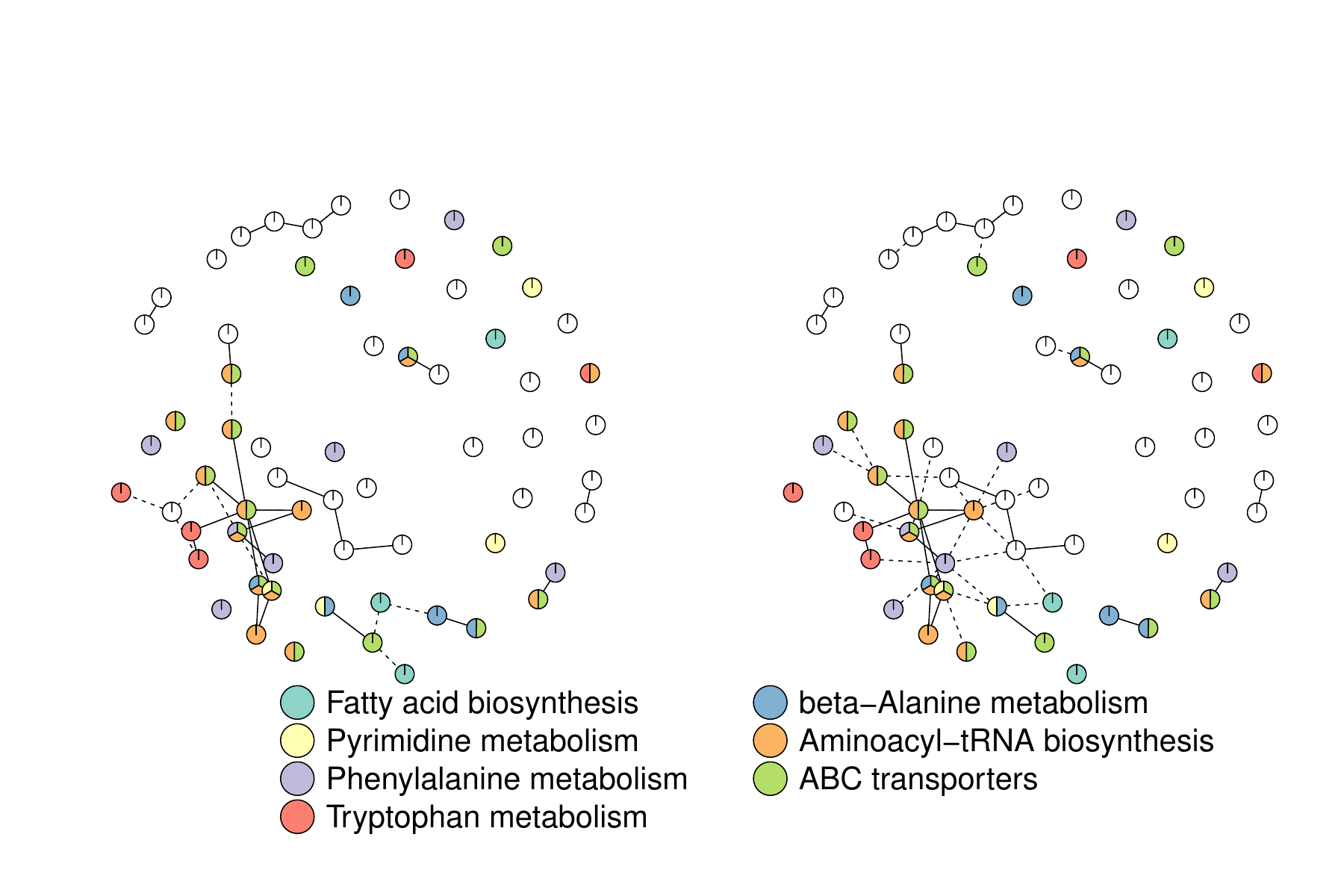}}
\end{minipage}
   \caption{The estimated network topology and enriched pathways in the metabolomics study for the benign class (left) and cancer class (right). Dashed lines represent edges that are present in only one class. Nodes in multiple colors are present in multiple pathways.}
\label{metabs:net}
\end{figure}

Our first application is based on the metabolomics data set from \citep{Putluri:2011ap} to examine changes in metabolic profiles associated with bladder cancer using untargeted mass spectrometry data acquisition strategy. 
The data consists of 31 cancer and 27 benign tissue samples and  63 detected metabolites. 
Here we focused on estimating the network of metabolic interactions, enhanced by information gleaned from KEGG \citep{Kanehisa01012000}. 
For each condition, we used the \bic~ criterion to select the tuning parameter $\lambda$. At the optimal $\lambda$, we applied the proposed network estimation procedure to identify the metabolic network; see Fig.~\ref{metabs:net} for an illustration of the estimated networks for the cancer and benign classes, respectively. It can be seen that there are numerous interactions between pathways that describe energy metabolism in the cancer state, due to the greater need of cancer cells for energy.

We tested for differential activity of biochemical pathways extracted from KEGG using the same set of data. 
Shown in Table \ref{metabs} are estimated $p$-values after false discovery rate correction with a $q$-value of 0.01 for the significant pathways selected from NetGSA. 
These identified pathways include those that describe altered utilization of amino acids and their aromatic counterparts, as well as metabolism of fatty acids and intermediates of tricarboxylic acid cycle (TCA) which were followed up for biological insights in the original study by \citet{Putluri:2011ap}.
Among the selected pathways, fatty acid biosynthesis is not identified by \GSA-s. 
Interestingly, \GSA-c fails to report any pathway as being significantly enriched. This again confirms our hypothesis that incorporating pathway topology information allows sophisticated enrichment methods in detecting important regulatory pathways. 

\begin{table}[ht]
\centering
\caption{$p$-values for the pathways in the metabolomics study, with false discovery rate correction at $q^*=0.01$. Here $0.00$ represents a zero $p$-value produced out of finite permutations. \GSA-c/\GSA-s refer to Gene Set Analysis with/without randomization of the genes based on 3000 permutations.
}\label{metabs}
{
\begin{tabular}{lrrr}
\hline\hline
Pathway & NetGSA & GSA-s &  GSA-c   \\\hline
  Tryptophan metabolism         & $3e^{-5}$ & 0.00 & 1.00 \\ 
  beta-Alanine metabolism       & $3e^{-5}$ & 0.00 & 1.00 \\ 
  Aminoacyl-tRNA biosynthesis & $2e^{-4}$ & 0.00 & 1.00 \\ 
  ABC transporters                  & $4e^{-4}$ & 0.00 & 1.00 \\ 
  Fatty acid biosynthesis          & $2e^{-3}$ & 1.00 & 1.00 \\ 
  Pyrimidine metabolism         & $2e^{-3}$ & 0.00 & 1.00 \\ 
  Phenylalanine metabolism    & $4e^{-3}$ & 0.00 & 1.00 \\
\hline
\end{tabular}
}
\end{table}

The second data set \citep{subramanian2005} consists of gene expression profiles of 5217 genes for 62 normal and 24 lung cancer patients. We considered 47 KEGG pathways of size at least 5 that describe signaling and biochemical mechanisms and excluded genes that eitrher are not present in the 47 pathways, or without recorded network information. The number of genes that remain for pathway enrichment analysis is 303. Based on the external topology information from the BioGRID database, we applied the proposed network estimation procedure coupled with \bic~to estimate the underlying interaction networks for both normal and cancer conditions.
We then explored whether \GSA~and NetGSA with the estimated networks are able to detect enriched pathways using the same data set. After correcting for multiple comparisons, using a  false discovery rate of $q^*= 0.01$, none of the three methods identifies any pathway as being significantly differential
enriched. The lack of statistical power in obtaining differential pathways was also noted in the original paper of \citep{subramanian2005}; see Table 9 of the Supplementary Materials for the complete list of FDR adjusted $p$-values. 
Nevertheless, using NetGSA and a relaxed FDR cutoff threshold of 0.30 (similar to the strategy adopted in \citep{subramanian2005}), we obtain the following top three ranked signaling pathways: Jak-STAT, p53 and Wnt. All three are implicated in lung cancer, although the latter two are also implicated in multiple other types of human malignancies. However, the Jak/STAT pathway has been recently shown to play a key role in non small cell lung cancer cells \citep{song2011}.

\begin{figure}[ht]   
\centering
\begin{subfigure}{0.45\linewidth}
\centerline{\includegraphics[scale=0.35]{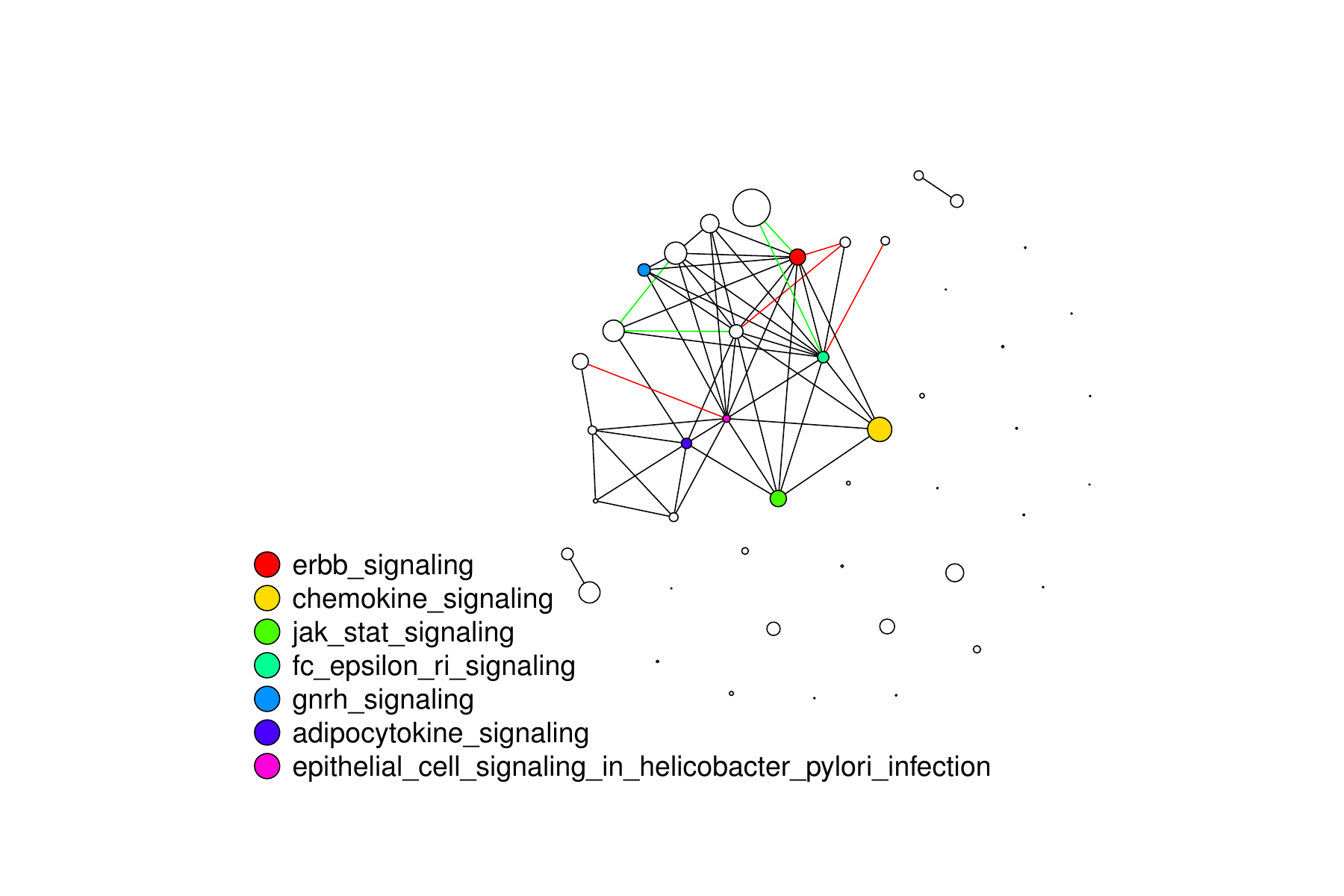}}
\end{subfigure}
\begin{subfigure}{0.45\linewidth}
\centerline{\includegraphics[scale=0.35]{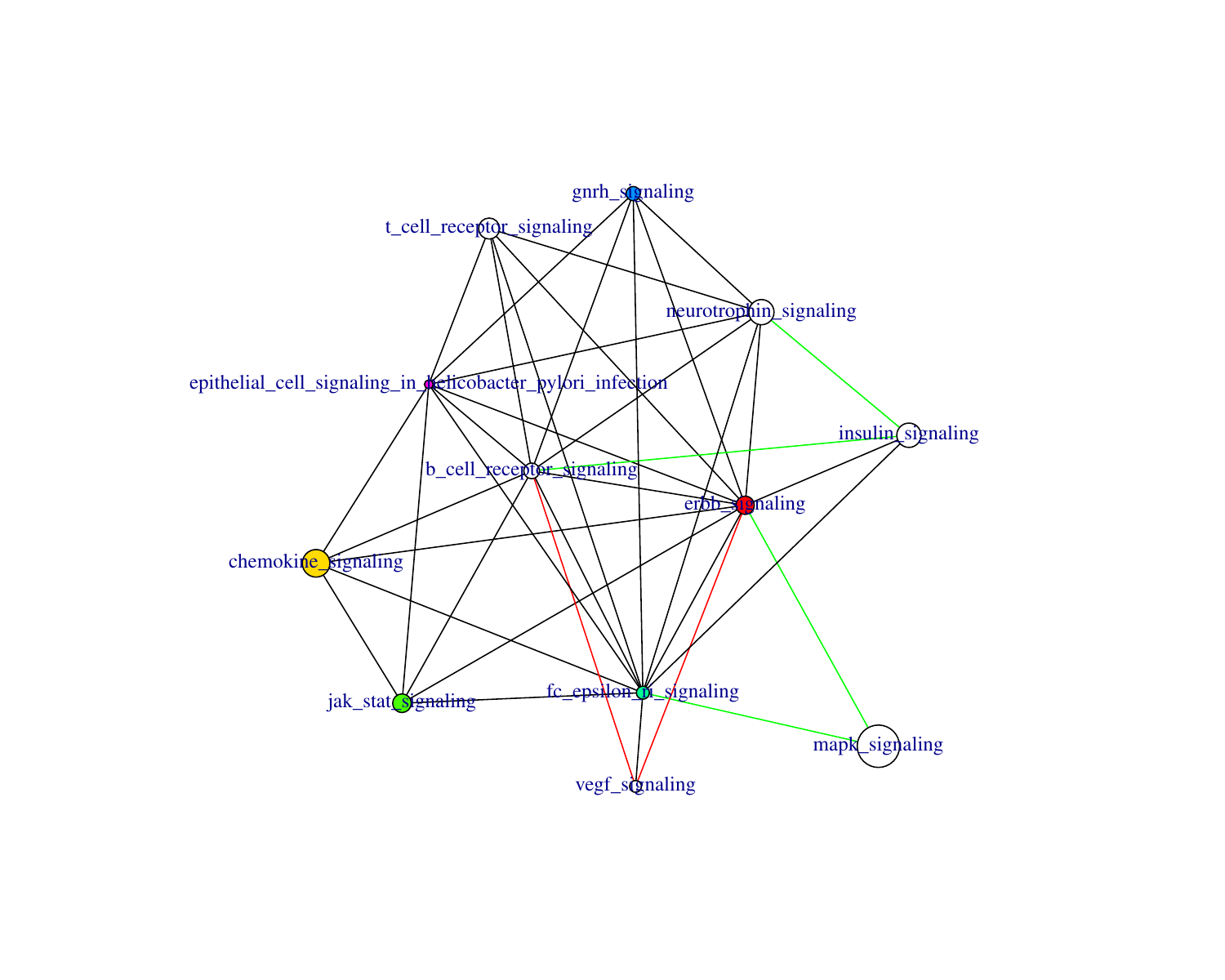}}
\end{subfigure}
\caption{The estimated pathway topology (left) and the subnetwork surrounding the ErbB pathway (right) in the TCGA cancer study.
Edges in black are present in both classes, whereas red and green  edges are only present in ER positive and ER negative class, respectively. Node size is proportional to the size (number of genes) of the corresponding pathway.}
\label{tcga:net}
\end{figure}

Our third and final application is based on a data set from The Cancer Genome Atlas \citep{cancer2012comprehensive}. The data set contains RNA-seq measurements for 17296 genes from 1033 breast cancer specimens, including ER positive, ER negative and other unevaluated cases. As in the previous gene microarray study, the external network information is extracted from the BioGRID database. We focused on a subset of the genes that have recorded network information and are present in KEGG pathways with at least 5 members. This leaves for further consideration 800 genes with 403 samples from the ER positive and 117 from the ER negative classes, spanning over 45 KEGG pathways. We then applied the constrained network estimation procedure with the tuning parameter selected via  \bic~in \eqref{eq:BIC} to obtain the partial correlation networks for the ER positive and ER negative classes, respectively. Due to the large number of variables, visualization of the estimated networks at the individual gene level is challenging. Instead, we examine the interactions among pathways in Fig.~\ref{tcga:net} to gain insight into their co-regulation behavior. The weighted pathway level network is defined as follows. Let each node in the network represent one pathway, with size proportional to the size of the corresponding pathway. A weighted edge between two pathways $P_1$ and $P_2$ is defined as the number of nonzero partial correlations between genes in $P_1$ and those in $P_2$ (normalized by the sizes of the two pathways). Links visualized in Fig.~\ref{tcga:net} are the top 5\% of the weighted edges, where ranking is based on edge weights.  

Table~\ref{TCGA} presents the false discovery rate corrected $p$-values for the selected differential pathways using NetGSA based on the estimated partial correlation networks, as well as \GSA-c and \GSA-s.
 The complete table is presented in Section D of the Supplementary Materials.
At $q^*=0.01$, NetGSA reports 25 out of the 45 KEGG pathways as significantly enriched, whereas \GSA~either rejects the null for all pathways (\GSA-c) or fails to reject any pathway (\GSA-s). Selected differential pathways identified by NetGSA are also highlighted in Fig.~\ref{tcga:net}.
Of particular interest is the set of connected, enriched pathways centered around the ErbB pathway in the right penal of Fig.~\ref{tcga:net}. This pathway contains receptors that signal through various pathways to regulate cell proliferation, migration, differentiation, apoptosis, and cell motility and play a key role in breast cancer \citep{howe2011targeting}, although its role in breast carcinogenesis not very well understood. Note that the Jak-STAT pathway is downstream of the ErbB one and can be activated by key epidermal growth factor receptors in the former to create signaling cascades \citep{henson2006surviving}. Further, the GnRH signaling pathway has been reported to interact with the ErbB pathway receptors \citep{morgan2011gnrh}. All these interconnected pathways are related to receptors that have been implicated in various studies with over-expression in the ER negative class and hence faster tumor growth and poorer clinical outcomes. 

\begin{table}[ht]
\centering
\caption{$p$-values for the differential pathways in the TCGA data, with false discovery rate correction at $q^*=0.01$. Here $0.00$ represents a zero $p$-value produced out of finite permutations. \GSA-c/\GSA-s refer to Gene Set Analysis with/without randomization of the genes based on 3000 permutations.
}\label{TCGA}
\resizebox{0.85\linewidth}{!}{%
{
\begin{tabular}{lrrr}
\hline\hline
Pathway & NetGSA & GSA-s &  GSA-c   \\\hline
  Epithelial cell signaling in Helicobacter pylori infection & $5e^{-95}$ & 0.00 & 1.00 \\ 
  Cell cycle & $2e^{-47}$ & 0.00 & 1.00 \\ 
  Galactose metabolism & $3e^{-31}$ & 0.00 & 1.00 \\ 
  Glutathione metabolism & $1e^{-27}$ & 0.00 & 1.00 \\ 
  NOD-like receptor signaling pathway & $1e^{-24}$ & 0.00 & 1.00 \\ 
   Pyrimidine metabolism & $4e^{-23}$ & 0.00 & 1.00 \\ 
  Cysteine and methionine metabolism & $1e^{-22}$ & 0.00 & 1.00 \\ 
  Starch and sucrose metabolism & $1e^{-18}$ & 0.00 & 1.00 \\ 
  Toll-like receptor signaling pathway & $1e^{-18}$ & 0.00 & 1.00 \\ 
  Glycolysis / Gluconeogenesis & $3e^{-17}$ & 0.00 & 1.00 \\ 
  Jak-STAT signaling pathway & $9e^{-15}$ & 0.00 & 1.00 \\ 
  Chemokine signaling pathway & $3e^{-14}$ & 0.00 & 1.00 \\ 
  ErbB signaling pathway & $7e^{-13}$ & 0.00 & 1.00 \\ 
  p53 signaling pathway & $7e^{-12}$ & 0.00 & 1.00 \\ 
  Hedgehog signaling pathway & $5e^{-10}$ & 0.00 & 1.00 \\ 
  beta-Alanine metabolism & $1e^{-7}$ & 0.00 & 1.00 \\ 
  Fc epsilon RI signaling pathway & $5e^{-7}$ & 0.00 & 1.00 \\ 
  Fructose and mannose metabolism & $2e^{-6}$ & 0.00 & 1.00 \\ 
  Pentose phosphate pathway & $2e^{-6}$ & 0.00 & 1.00 \\ 
  PPAR signaling pathway & $5e^{-6}$ & 0.00 & 1.00 \\ 
  Adipocytokine signaling pathway & $4e^{-5}$ & 0.00 & 1.00 \\ 
  Purine metabolism & $6e^{-5}$ & 0.00 & 1.00 \\ 
  Valine, leucine and isoleucine degradation & $5e^{-4}$ & $1e^{-3}$ & 1.00 \\ 
  GnRH signaling pathway & $2e^{-3}$ & 0.00 & 1.00 \\ 
  TGF-beta signaling pathway & $3e^{-3}$ & 0.00 & 1.00\\
 \hline
\end{tabular}
}
}%
\end{table}

%%%%%%%%%%%%%%%%%%%%
\section{Discussion}\label{sec:conc}
%%%%%%%%%%%%%%%%%%%%

This paper introduces a constrained partial correlation network estimation method that seamlessly incorporates  externally available interaction information for genes and other biomolecules. The end product is a reliable condition-specific estimate of the underlying networks. The resulting estimated network structures are then used for network-based pathway enrichment analysis. 
For the purpose of constrained network estimation, one might also try the one-step constrained maximum likelihood estimation (a functionality offered in the \verb=R=-package \verb=glasso=) to recover the underlying partial correlation network. 
However, this one-step approach requires sophisticated specification of the tuning parameters at positions for which structural information is available, and can be challenging to implement in practice. 

Two sources of uncertainty can be identified in the proposed framework: one from the reliability of the external database information in the network estimation procedure and the other from the uncertainty regarding the estimated network itself, as well as how it propagates into the NetGSA testing procedure. As discussed in Remark~\ref{remark:uncertainty}, the proposed method can conveniently accommodate the first source of uncertainty by incorporating a non-zero penalty on parameters that are uncertain. Further, as shown in Theorem~\ref{thm:ump}, the proposed test via the extended NetGSA framework is asymptotically unbiased and most powerful, given the consistency of the estimated network, and hence accounts for the second source of uncertainty. 
Nevertheless, in finite samples as the numerical work in Section C of the Supplementary Materials illustrates, Type I errors may be slightly off in the presence of numerous errors in the estimated network (either due to misspecification of the external information or lack of samples for accurate estimation). The topic of dealing with network estimation errors and possible ways to address it is discussed in \citet{narayan2015mixed}.

Finally, the current framework of NetGSA uses the Cholesky decomposition of the covariance matrix of the underlying network. It is natural to ask  whether the order of the variables affects the result of enrichment analysis.  In simulations and the real data analyses, we find that the estimated powers/$p$-values from NetGSA are comparable after permutation of the variables.

\section*{Acknowledgements}

The authors would like to thank the reviewers for many constructive comments and suggestions.
\vspace*{-12pt}

\bibliographystyle{natbib}
\bibliography{Reference}

\begin{thebibliography}{}

\bibitem[Al-Shahrour {\em et~al.}(2005)Al-Shahrour, D{\'\i}az-Uriarte, and
  Dopazo]{al2005discovering}
Al-Shahrour, F., D{\'\i}az-Uriarte, R., and Dopazo, J. (2005).
\newblock Discovering molecular functions significantly related to phenotypes
  by combining gene expression data and biological information.
\newblock {\em Bioinformatics\/}, {\bf 21}(13), 2988--2993.

\bibitem[Benjamini and Hochberg(1995)Benjamini and Hochberg]{Benjamini:1995jh}
Benjamini, Y. and Hochberg, Y. (1995).
\newblock Controlling the false discovery rate: A practical and powerful
  approach to multiple testing.
\newblock {\em Journal of the Royal Statistical Society. Series B
  (Methodological)\/}, {\bf 57}(1), 289--300.

\bibitem[Benjamini and Yekutieli(2001)Benjamini and
  Yekutieli]{benjamini2001control}
Benjamini, Y. and Yekutieli, D. (2001).
\newblock The control of the false discovery rate in multiple testing under
  dependency.
\newblock {\em Annals of statistics\/}, pages 1165--1188.

\bibitem[Bickel and Levina(2008)Bickel and Levina]{Bickel:2008fk}
Bickel, P.~J. and Levina, E. (2008).
\newblock Regularized estimation of large covariance matrices.
\newblock {\em Annals of Statistics\/}, {\bf 36}(1), 199--227.

\bibitem[Bickel {\em et~al.}(2009)Bickel, Ritov, and Tsybakov]{Bickel:2009vp}
Bickel, P.~J., Ritov, Y., and Tsybakov, A.~B. (2009).
\newblock Simultaneous analysis of lasso and dantzig selector.
\newblock {\em Annals of Statistics\/}, {\bf 37}(4), 1705--1732.

\bibitem[Boyd and Vandenberghe(2004)Boyd and Vandenberghe]{Boyd:2004gs}
Boyd, S. and Vandenberghe, L. (2004).
\newblock {\em Convex Optimization\/}.
\newblock Cambridge University Press.

\bibitem[Byrd {\em et~al.}(1995)Byrd, Lu, Nocedal, and Zhu]{byrd1995limited}
Byrd, R.~H., Lu, P., Nocedal, J., and Zhu, C. (1995).
\newblock A limited memory algorithm for bound constrained optimization.
\newblock {\em SIAM Journal on Scientific Computing\/}, {\bf 16}(5),
  1190--1208.

\bibitem[Candes and Recht(2009)Candes and Recht]{Candes:2009mk}
Candes, E.~J. and Recht, B. (2009).
\newblock Exact matrix completion via convex optimization.
\newblock {\em Foundations of Computational Mathematics\/}, {\bf 9}, 717--772.

\bibitem[Chuang {\em et~al.}(2012)Chuang, Rassenti, Salcedo, Licon, Kohlmann,
  Haferlach, Fo{\`a}, Ideker, and Kipps]{chuang2012subnetwork}
Chuang, H.-Y., Rassenti, L., Salcedo, M., Licon, K., Kohlmann, A., Haferlach,
  T., Fo{\`a}, R., Ideker, T., and Kipps, T.~J. (2012).
\newblock Subnetwork-based analysis of chronic lymphocytic leukemia identifies
  pathways that associate with disease progression.
\newblock {\em Blood\/}, {\bf 120}(13), 2639--2649.

\bibitem[Csardi and Nepusz(2006)Csardi and Nepusz]{igraph2006}
Csardi, G. and Nepusz, T. (2006).
\newblock The igraph software package for complex network research.
\newblock {\em InterJournal\/}, {\bf Complex Systems}, 1695.

\bibitem[Dehmer and Emmert-Streib(2008)Dehmer and
  Emmert-Streib]{dehmer2008analysis}
Dehmer, M. and Emmert-Streib, F. (2008).
\newblock {\em Analysis of Microarray Data: a Network-Based Approach\/}.
\newblock John Wiley \& Sons.

\bibitem[Dempster(1972)Dempster]{dempster1972covariance}
Dempster, A.~P. (1972).
\newblock Covariance selection.
\newblock {\em Biometrics\/}, {\bf 28}(1), 157--175.

\bibitem[Efron and Tibshirani(2007)Efron and Tibshirani]{Efron:2007vz}
Efron, B. and Tibshirani, R. (2007).
\newblock On testing the significance of sets of genes.
\newblock {\em Annals of Applied Statistics\/}, {\bf 1}(1), 107--129.

\bibitem[Friedman {\em et~al.}(2008)Friedman, Hastie, and
  Tibshirani]{Friedman:2008lt}
Friedman, J., Hastie, T., and Tibshirani, R. (2008).
\newblock Sparse inverse covariance estimation with the graphical lasso.
\newblock {\em Biostatistics\/}, {\bf 9}(3), 432--441.

\bibitem[Gottwein {\em et~al.}(2007)Gottwein, Mukherjee, Sachse, Frenzel,
  Majoros, Chi, Braich, Manoharan, Soutschek, Ohler, {\em
  et~al.}]{gottwein2007viral}
Gottwein, E., Mukherjee, N., Sachse, C., Frenzel, C., Majoros, W.~H., Chi,
  J.-T.~A., Braich, R., Manoharan, M., Soutschek, J., Ohler, U., {\em et~al.}
  (2007).
\newblock A viral microrna functions as an orthologue of cellular mir-155.
\newblock {\em Nature\/}, {\bf 450}(7172), 1096--1099.

\bibitem[Green {\em et~al.}(2011)Green, Monti, Dalla-Favera, Pasqualucci,
  Walsh, Schmidt-Supprian, Kutok, Rodig, Neuberg, Rajewsky, {\em
  et~al.}]{green2011signatures}
Green, M.~R., Monti, S., Dalla-Favera, R., Pasqualucci, L., Walsh, N.~C.,
  Schmidt-Supprian, M., Kutok, J.~L., Rodig, S.~J., Neuberg, D.~S., Rajewsky,
  K., {\em et~al.} (2011).
\newblock Signatures of murine b-cell development implicate yy1 as a regulator
  of the germinal center-specific program.
\newblock {\em Proceedings of the National Academy of Sciences\/}, {\bf
  108}(7), 2873--2878.

\bibitem[Haberman(1989)Haberman]{haberman1989concavity}
Haberman, S.~J. (1989).
\newblock Concavity and estimation.
\newblock {\em Annals of Statistics\/}, {\bf 17}(4), 1631--1661.

\bibitem[Henson and Gibson(2006)Henson and Gibson]{henson2006surviving}
Henson, E.~S. and Gibson, S.~B. (2006).
\newblock Surviving cell death through epidermal growth factor (egf) signal
  transduction pathways: implications for cancer therapy.
\newblock {\em Cellular signalling\/}, {\bf 18}(12), 2089--2097.

\bibitem[Houstis {\em et~al.}(2006)Houstis, Rosen, and
  Lander]{houstis2006reactive}
Houstis, N., Rosen, E.~D., and Lander, E.~S. (2006).
\newblock Reactive oxygen species have a causal role in multiple forms of
  insulin resistance.
\newblock {\em Nature\/}, {\bf 440}(7086), 944--948.

\bibitem[Howe and Brown(2011)Howe and Brown]{howe2011targeting}
Howe, L.~R. and Brown, P.~H. (2011).
\newblock Targeting the her/egfr/erbb family to prevent breast cancer.
\newblock {\em Cancer prevention research\/}, {\bf 4}(8), 1149--1157.

\bibitem[Huang {\em et~al.}(2008)Huang, Sherman, and
  Lempicki]{huang2009systematic}
Huang, D.~W., Sherman, B.~T., and Lempicki, R.~A. (2008).
\newblock Systematic and integrative analysis of large gene lists using david
  bioinformatics resources.
\newblock {\em Nature Protocols\/}, {\bf 4}(1), 44--57.

\bibitem[Huerta {\em et~al.}(1998)Huerta, Salgado, Thieffry, and
  Collado-Vides]{Huerta01011998}
Huerta, A.~M., Salgado, H., Thieffry, D., and Collado-Vides, J. (1998).
\newblock Regulondb: A database on transcriptional regulation in escherichia
  coli.
\newblock {\em Nucleic Acids Research\/}, {\bf 26}(1), 55--59.

\bibitem[Ideker and Krogan(2012)Ideker and Krogan]{ideker2012differential}
Ideker, T. and Krogan, N.~J. (2012).
\newblock Differential network biology.
\newblock {\em Molecular Systems Biology\/}, {\bf 8}(1), 565.

\bibitem[Ideker {\em et~al.}(2011)Ideker, Dutkowski, and
  Hood]{ideker2011boosting}
Ideker, T., Dutkowski, J., and Hood, L. (2011).
\newblock Boosting signal-to-noise in complex biology: prior knowledge is
  power.
\newblock {\em Cell\/}, {\bf 144}(6), 860--863.

\bibitem[Joshi-Tope {\em et~al.}(2003)Joshi-Tope, Vastrik, Gopinath, Matthews,
  Schmidt, Gillespie, D'Eustachio, Jassal, Lewis, Wu, Birney, and
  Stein]{JOSHI-TOPE01012003}
Joshi-Tope, G., Vastrik, I., Gopinath, G., Matthews, L., Schmidt, E.,
  Gillespie, M., D'Eustachio, P., Jassal, B., Lewis, S., Wu, G., Birney, E.,
  and Stein, L. (2003).
\newblock The genome knowledgebase: A resource for biologists and
  bioinformaticists.
\newblock {\em Cold Spring Harbor Symposia on Quantitative Biology\/}, {\bf
  68}, 237--244.

\bibitem[Kanehisa and Goto(2000)Kanehisa and Goto]{Kanehisa01012000}
Kanehisa, M. and Goto, S. (2000).
\newblock Kegg: Kyoto encyclopedia of genes and genomes.
\newblock {\em Nucleic Acids Research\/}, {\bf 28}(1), 27--30.

\bibitem[Khatri {\em et~al.}(2012)Khatri, Sirota, and
  Butte]{10.1371/journal.pcbi.1002375}
Khatri, P., Sirota, M., and Butte, A.~J. (2012).
\newblock Ten years of pathway analysis: current approaches and outstanding
  challenges.
\newblock {\em PLoS Comput Biol\/}, {\bf 8}(2), e1002375.

\bibitem[Lauritzen(1996)Lauritzen]{lauritzen1996graphical}
Lauritzen, S.~L. (1996).
\newblock {\em Graphical Models\/}.
\newblock Oxford University Press.

\bibitem[Lindstrom and Bates(1988)Lindstrom and Bates]{Lindstrom:1988ri}
Lindstrom, M.~J. and Bates, D.~M. (1988).
\newblock Newton-raphson and em algorithms for linear mixed-effects models for
  repeated-measures data.
\newblock {\em Journal of the American Statistical Association\/}, {\bf
  83}(404), 1014--1022.

\bibitem[Meinshausen and B{\"u}hlmann(2006)Meinshausen and
  B{\"u}hlmann]{Meinshausen:2006rw}
Meinshausen, N. and B{\"u}hlmann, P. (2006).
\newblock High dimensional graphs and variable selection with the lasso.
\newblock {\em Annals of Statistics\/}, {\bf 34}(3), 1436--1462.

\bibitem[Mirsky(1975)Mirsky]{mirsky1975trace}
Mirsky, L. (1975).
\newblock A trace inequality of john von neumann.
\newblock {\em Monatshefte f{\"u}r Mathematik\/}, {\bf 79}(4), 303--306.

\bibitem[Morgan {\em et~al.}(2011)Morgan, Meyer, Miller, Sims, Cagnan,
  Faratian, Harrison, Millar, and Langdon]{morgan2011gnrh}
Morgan, K., Meyer, C., Miller, N., Sims, A.~H., Cagnan, I., Faratian, D.,
  Harrison, D.~J., Millar, R.~P., and Langdon, S.~P. (2011).
\newblock Gnrh receptor activation competes at a low level with growth
  signaling in stably transfected human breast cell lines.
\newblock {\em BMC cancer\/}, {\bf 11}(1), 476.

\bibitem[Narayan and Allen(2016)Narayan and Allen]{narayan2015mixed}
Narayan, M. and Allen, G.~I. (2016).
\newblock Mixed effects models to find differences in multi-subject functional
  connectivity.
\newblock {\em Frontiers in Neuroscience\/}, {\bf 10}(108).

\bibitem[Nishimura(2001)Nishimura]{nishimura2001biocarta}
Nishimura, D. (2001).
\newblock Biocarta.
\newblock {\em Biotech Software \& Internet Report: The Computer Software
  Journal for Scient\/}, {\bf 2}(3), 117--120.

\bibitem[Prill {\em et~al.}(2010)Prill, Marbach, Saez-Rodriguez, Sorger,
  Alexopoulos, Xue, Clarke, Altan-Bonnet, and Stolovitzky]{DREAM32010}
Prill, R.~J., Marbach, D., Saez-Rodriguez, J., Sorger, P.~K., Alexopoulos,
  L.~G., Xue, X., Clarke, N.~D., Altan-Bonnet, G., and Stolovitzky, G. (2010).
\newblock Towards a rigorous assessment of systems biology models: the dream3
  challenges.
\newblock {\em PloS one\/}, {\bf 5}(2), e9202.

\bibitem[Putluri {\em et~al.}(2011)Putluri, Shojaie, Vasu, Vareed, Nalluri,
  Putluri, Thangjam, Panzitt, Tallman, Butler, {\em et~al.}]{Putluri:2011ap}
Putluri, N., Shojaie, A., Vasu, V.~T., Vareed, S.~K., Nalluri, S., Putluri, V.,
  Thangjam, G.~S., Panzitt, K., Tallman, C.~T., Butler, C., {\em et~al.}
  (2011).
\newblock Metabolomic profiling reveals potential markers and bioprocesses
  altered in bladder cancer progression.
\newblock {\em Cancer Research\/}, {\bf 71}(24), 7376--7386.

\bibitem[Rothman {\em et~al.}(2008)Rothman, Bickel, Levina, and
  Zhu]{Rothman:2008ce}
Rothman, A.~J., Bickel, P.~J., Levina, E., and Zhu, J. (2008).
\newblock Sparse permutation invariant covariance estimation.
\newblock {\em Electronic Journal of Statistics\/}, {\bf 2}, 494--515.

\bibitem[Searle(1971)Searle]{searle1971linear}
Searle, S. (1971).
\newblock {\em Linear Models\/}.
\newblock John Wiley \& Sons.

\bibitem[Shojaie and Michailidis(2009)Shojaie and Michailidis]{Shojaie:2009lf}
Shojaie, A. and Michailidis, G. (2009).
\newblock Analysis of gene sets based on the underlying regulatory network.
\newblock {\em Journal of Computational Biology\/}, {\bf 16}(3), 407--426.

\bibitem[Shojaie and Michailidis(2010)Shojaie and Michailidis]{Shojaie:2010nz}
Shojaie, A. and Michailidis, G. (2010).
\newblock Network enrichment analysis in complex experiments.
\newblock {\em Statistical Applications in Genetics and Molecular Biology\/},
  {\bf 9}(1), Article 22.

\bibitem[Song {\em et~al.}(2011)Song, Rawal, Nemeth, and Haura]{song2011}
Song, L., Rawal, B., Nemeth, J.~A., and Haura, E.~B. (2011).
\newblock Jak1 activates stat3 activity in non-small--cell lung cancer cells
  and il-6 neutralizing antibodies can suppress jak1-stat3 signaling.
\newblock {\em Molecular cancer therapeutics\/}, {\bf 10}(3), 481--494.

\bibitem[Subramanian {\em et~al.}(2005)Subramanian, Tamayo, Mootha, Mukherjee,
  Ebert, Gillette, Paulovich, Pomeroy, Golub, Lander, {\em
  et~al.}]{subramanian2005}
Subramanian, A., Tamayo, P., Mootha, V.~K., Mukherjee, S., Ebert, B.~L.,
  Gillette, M.~A., Paulovich, A., Pomeroy, S.~L., Golub, T.~R., Lander, E.~S.,
  {\em et~al.} (2005).
\newblock Gene set enrichment analysis: a knowledge-based approach for
  interpreting genome-wide expression profiles.
\newblock {\em Proceedings of the National Academy of Sciences of the United
  States of America\/}, {\bf 102}(43), 15545--15550.

\bibitem[TCGA(2012)TCGA]{cancer2012comprehensive}
TCGA (2012).
\newblock Comprehensive molecular portraits of human breast tumours.
\newblock {\em Nature\/}, {\bf 490}(7418), 61--70.

\bibitem[Wermuth(1980)Wermuth]{wermuth1980linear}
Wermuth, N. (1980).
\newblock Linear recursive equations, covariance selection, and path analysis.
\newblock {\em Journal of the American Statistical Association\/}, {\bf
  75}(372), 963--972.

\bibitem[Wilson {\em et~al.}(2010)Wilson, Wang, Shen, McKenna, Lemieux, Cho,
  Koellhoffer, Pomeroy, Orkin, and Roberts]{wilson2010epigenetic}
Wilson, B.~G., Wang, X., Shen, X., McKenna, E.~S., Lemieux, M.~E., Cho, Y.-J.,
  Koellhoffer, E.~C., Pomeroy, S.~L., Orkin, S.~H., and Roberts, C.~W. (2010).
\newblock Epigenetic antagonism between polycomb and swi/snf complexes during
  oncogenic transformation.
\newblock {\em Cancer cell\/}, {\bf 18}(4), 316--328.

\bibitem[Zaki {\em et~al.}(2013)Zaki, Efimov, and Berengueres]{23688127}
Zaki, N., Efimov, D., and Berengueres, J. (2013).
\newblock Protein complex detection using interaction reliability assessment
  and weighted clustering coefficient.
\newblock {\em BMC Bioinformatics\/}, {\bf 14}(1), 163.

\bibitem[Zhou {\em et~al.}(2011)Zhou, Rutimann, Xu, and
  B{\"u}hlmann]{Zhou:2010on}
Zhou, S., Rutimann, P., Xu, M., and B{\"u}hlmann, P. (2011).
\newblock High-dimensional covariance estimation based on gaussian graphical
  models.
\newblock {\em Journal of Machine Learning Research\/}, {\bf 12}, 2975--3026.

\end{thebibliography}

\newpage

\newcommand{\mcf}{\mathcal{F}}
\theoremstyle{plain}
\newtheorem{lemma}{Lemma}
\setcounter{table}{0}
\renewcommand{\thetable}{A\arabic{table}}
\numberwithin{theorem}{section}
\numberwithin{lemma}{section}
\numberwithin{equation}{section}

\begin{center}{\bf \huge Supplementary materials to \\
``Network-based pathway enrichment analysis with incomplete network information"}
\end{center}
\appendix

\section{Theoretical Analysis and Proofs}\label{app:netgsa:pf}

We first introduce additional notation needed in the remainder. Define $\tilde{\Omega}_0 = \diag(\Omega_0) + \Omega_{0, E\cap \hat{E}}$, where $E$ and $\hat{E}$ are the true and the estimated edge set, respectively. By definition, $\tilde{\Omega}_0$ and $\Omega_0$ will be different at position $(i,i')$ only when the edge $(i,i')$ is falsely rejected. In the following, we first derive an upper bound for the size of $\hat{E}$ and $\|\tilde{\Omega}_0  - \Omega_0\|_F $. 
For the ease of presentation, we drop the superscript $i$ for sets $J_0$ and $J_1$ in the $i$th regression, but they should be understood as $J_0^i$ and $J_1^i$, respectively. 

The following lemma is needed in the proof of Theorem~\ref{coro_bias} below. 
\begin{lemma}\label{corr_event}
For $i=1, \ldots, p$, denote by $\M\xi^i = \bZ_i - \sum_{i'\ne i} \theta_{i'}^i \bZ_{i'}$, where $\btheta^i$ is the optimal prediction coefficient vector in the $i$th regression. Consider the event 
\[
\mcf_i := \left\{ \bZ: \frac{1}{m}\|\bZ_{-i}^T\M\xi^i\|_{\infty}\le \frac {c_1}{ 2} \sqrt{ \frac{\log (p-rp) }{ m \omega_{0,ii} } } \right\}
\]
with a constant $c_1>4$, where $\omega_{0,ii}$ is the $i$th diagonal element of the true inverse covariance matrix $\Omega_0$. Define the event $\mcf = \bigcap_{i=1}^p \mcf_i$.
Then
$\mpr(\mcf ) \ge 1-2p^{2-c_1^2/8} $.
\end{lemma}

The proof of Lemma~\ref{corr_event} will be provided shortly. Denote by $\Lambda_{\max}$ the maximal eigenvalue of $\bZ^T\bZ/m$. Conditioning on the event $\mcf$, we have the following results on controlling the size of $\hat{E}$ and the Frobenius norm of the deviance, $\|\tilde{\Omega}_0  - \Omega_0\|_F $.
\begin{theorem}\label{coro_bias}
Suppose the conditions in Theorem 2.2 are satisfied. Then on event $\mcf$, for appropriately chosen $\lambda$, we have
\begin{equation}\label{Ehat_bound}
|\hat{E}| \le \frac{64\Lambda_{\max}}{\kappa^2(s)} (1-r)S_0+ rS_0,
\end{equation}
and
\begin{equation}\label{A0_bias}
\|\tilde{\Omega}_0  - \Omega_0 \|_F \le c_3\sqrt{\frac{S_0\log (p-rp)} { m}}\le k_1 \phi_1,
\end{equation}
where $c_3 = 16c_1\sqrt{(1-r)}/ \kappa^2(2s) . $
\end{theorem}
\begin{remark}
The result indicates that the cardinality of the estimated edge set is upper bounded by a function of $r$, the percentage of the external information. The bound for $|\hat{E}|$ also depends on the restricted eigenvalue $\kappa(s)$, which is necessarily positive by the assumption that $\kappa(2s)>0$. Two extreme cases occur when (i) $r=0$, i.e. we do not observe anyexternal  information, thus reducing problem (2.4) to the original neighborhood selection in \cite{Meinshausen:2006rw}; (ii) $r=1$, i.e. the exact network topology is known and hence $\hat{E} = E$. On the other hand, the upper bound for $\|\tilde{\Omega}_0  - \Omega_0 \|_F$ decreases as $r$ increases, i.e. when more external information becomes available. 
However, since the coefficients also need to be estimated, this deviance always stays positive, even when $r=1$. 
\end{remark}

\begin{proof}[Proof of Theorem~\ref{coro_bias}]
Recall $\tilde{J} = V \backslash \{ J_1 \cup J_0 \cup \{i\}   \}$ is the set of indices for which there is no information available. Denote by ${\bf P}_{J_1} = \bZ_{J_1} (\bZ_{J_1} ^T\bZ_{J_1})^{-1} \bZ_{J_1}^T$ the projection onto the column space of $\bZ_{J_1}$.
It is easy to see that the problem (2.4) is equivalent to solving
\begin{equation}\label{opt_i_tilde}
\min_{\btheta_{\tilde{J}} } ~\frac{1}{ m} \|(\bIP-\bP_{J_1})\bZ_i-  \bZ_{\tilde{J}} \btheta_{\tilde{J} }\|_2^2 + 2\lambda \|\btheta_{\tilde{J}} \|_1.
\end{equation}
To bound $\hat E$ and $\|\tilde{\Omega}_0  - \Omega_0 \|_F$, it suffices to focus mainly on the set $\tilde{J}$, as false positive and negative errors will only occur on this set. 

Denote by $s_1^i$ and $s^i$, respectively, the number of known ones and the number of nonzero coordinates after excluding the known ones in the $i$th regression, and $s=\underset{i=1,\ldots, p}{\max} (s_1^i+s^i)$. 
If $\bZ$ satisfies the restricted eigenvalue condition in Assumption 2 with $\kappa(2s)>0$, then $ \bZ_{\tilde{J}}$ satisfies the same assumption with $\kappa(2s^i)\ge \kappa(2s)>0$ for $s^i \le s$. Moreover, $\kappa(s^i)\ge \kappa(s)\ge \kappa(2s)>0$.
Let $\hat{\btheta}^i_{\tilde{J}}$ be the lasso estimator in \eqref{opt_i_tilde} with
\begin{equation}\label{lambda}
\lambda = {c_1 } \sqrt{ \frac{\log (p-rp) }{ m \omega_{0,ii}} }
\end{equation} 
for $c_1>4$.
Conditioning on the event $\mcf$, we can invoke Theorem 7.2 of \cite{Bickel:2009vp} and obtain simultaneously for all $i$,
\begin{equation}\label{card_hat}
\|\hat{\btheta}^i_{\tilde{J}}\|_0 \le \frac{64 \Lambda_{\max} }{ \kappa^2(s^i)} s^i,
\end{equation}
and
\begin{equation}\label{norm_diff}
\| \hat{\btheta}^i_{\tilde{J}} - {\btheta}^{i}_{\tilde{J}} \|_2 \le  \frac{16c_1 }{ \omega_{0,ii} \kappa^2(2s^i)} \sqrt{\frac{s^i \log (p-rp) }{ m}}.
\end{equation}
Combining \eqref{card_hat} with the number of known edges $s^i_1$ as given in $J_1^i$, we get 
\[
|\hat{E}| \le \sum_{i=1}^p \{\|\hat{\btheta}_{\tilde{J}}^i\|_0 + |J_1^i| \} \le \frac{64\Lambda_{\max} }{ \kappa^2(s)} \sum_{i=1}^p  s^i+ \sum_{i=1}^p  s_1^i.
\]
The upper bound in \eqref{Ehat_bound} follows immediately, since by definition the number of known and unknown edges are $\sum_{i=1}^p  s_1^i = rS_0$ and $\sum_{i=1}^p  s^i = (1-r)S_0$, respectively. 

To bound $\|\tilde{\Omega}_0  - \Omega_0 \|_F $, recall that for every $i'\ne i$, $\omega_{0,ii'} =- \theta^i_{i'}\omega_{0,ii} $. Using the bound in \eqref{norm_diff}, we have
\begin{align*}
\|\tilde{\Omega}_0  - \Omega_0 \|_F^2 &= \sum_{i=1}^p \sum_{i' \in J(\theta^i) \cap J({\hat{\theta}^i})^c} (\theta^i_{i'} \omega_{0,ii})^2  = \sum_{i=1}^p \omega_{0,ii}^2\sum_{i' \in J(\theta^i) \cap J({\hat{\theta}^i})^c} |\theta^i_{i'} -\hat{\theta}^i_{i'} |^2 \\
& \le \sum_{i=1}^p \omega_{0,ii}^2 \| \btheta^i_{\tilde{J}} - \hat{\btheta}^i_{\tilde{J}}\|_2^2 \le \left\{ \frac{16c_1}{ \kappa^2(2s) } \right\}^2 \frac{(1-r)S_0\log (p-rp) }{ m}.
\end{align*}
The last inequality in \eqref{A0_bias} follows from condition (2.7) in Theorem 2.2.
\end{proof}

\begin{proof}[Proof of Lemma~\ref{corr_event}]
For every $i$, it is easy to verify that $\M\xi^i$ is normally distributed with mean $\M 0$ and variance $1/\omega_{0,ii} {\bf I}_m$. Define random variables $\Upsilon_{ii'} =  {(\omega_{0,ii}/m )}^{1/2} \bZ_{i'}^T \M\xi^i$ for $i'\ne i$. 
Then, $\bZ_{i'}^T \bZ_{i'}/m = 1$ implies that $\Upsilon_{ii'} \sim \MN(0,1)$. Let $\lambda$ be defined as in \eqref{lambda}.  
Using an elementary bound on the tails of Gaussian distributions,
\begin{align*}
\mpr(\mcf^c ) & \le \sum_{i=1}^p \sum_{i' \ne i} \mpr \left(\{|\bZ_{i'}^T \M\xi^i |/m> \lambda/2\}\right) \\
& \le \sum_{i=1}^p \sum_{i' \ne i}  \mpr \left(|\Upsilon_{ii'}| >{{(m \omega_{0,ii})}^{1/2}\lambda }/2 \right) \le \sum_{i=1}^p \sum_{i' \ne i} 2  \exp\left\{-{m \omega_{0,ii} \lambda^2/ 8 } \right\}\\
&  \le 2p (p-1) \exp \left\{ - {c_1^2\log (p-rp)/ 8} \right\} \le 2p^{2-c_1^2/8}.
\end{align*}
Therefore, $\mpr(\mcf ) \ge 1-2p^{2-c_1^2/8} $.
\end{proof}

With Lemma~\ref{corr_event} and Theorem~\ref{coro_bias}, we are ready to prove our main result in Theorem 2.2. The following proof is adapted from \cite{Zhou:2010on}.  

\begin{proof}[Proof of Theorem 2.2]
Consider $\hat{\Omega}$ defined in (2.5). 
It suffices to show that on the event $\mcf$
$$
\|\hat{\Omega} - \tilde{\Omega}_0 \|_F = O \left(\sqrt{ \frac{S_0\log (p-rp)} { m} }\right),
$$
since by triangle inequality and Theorem~\ref{coro_bias}, we can conclude
\begin{align*}
\|\hat{\Omega} - \Omega_0 \|_F \le \|\hat{\Omega} - \tilde{\Omega}_0 \|_F + \|\tilde{\Omega}_0  - \Omega_0 \|_F \le  O\left(\sqrt{ \frac{S_0\log (p-rp)} { m} }\right).
\end{align*}
Denote $\tilde{\Sigma}_0 = \tilde{\Omega}_0^{-1}$, which is positive definite since by Theorem~\ref{coro_bias},
\begin{align}
\phi_{\min}(\tilde{\Omega}_0) & \ge \phi_{\min}(\Omega_0) - \|\tilde{\Omega}_0  - \Omega_0 \|_2 \ge \phi_{\min}(\Omega_0) - \|\tilde{\Omega}_0  - \Omega_0 \|_F \ge \phi_1 - k_1 \phi_1 >0 \label{A0_min}.
\end{align}
The first inequality in \eqref{A0_min} comes from the fact that for any nonzero vector $\bdelta \in \mathbb{R}^p $, $\bdelta^T \tilde{\Omega}_0 \bdelta = \bdelta^T {\Omega}_0 \bdelta + \bdelta^T (\tilde{\Omega}_0 - {\Omega}_0) \bdelta \ge \phi_{\min}(\Omega_0) - \phi_{\max}(\tilde{\Omega}_0 - \Omega_0)$. 

Given $\tilde{\Omega}_0 \in \Sa \cap \mathcal{S}^p_{\hat{E}}$, define a new convex set:
$$
\mathcal{U}_m(\tilde{\Omega}_0) = \{\M B- \tilde{\Omega}_0  \mid \M B\in \mathcal{S}^p_{+} \cap \mathcal{S}^p_{\hat{E}}\} \subset \mathcal{S}^p_{\hat{E}}.
$$
Let
\begin{align*}
Q(\Omega) & =  \tr(\Omega\hat{\Sigma}) - \tr(\tilde{\Omega}_0\hat{\Sigma}) - \logdet \Omega +  \logdet \tilde{\Omega}_0 .
\end{align*}
Since the estimate $\hat{\Omega}$ minimizes $Q(\Omega)$, $\hat{\Delta} = \hat \Omega - \tilde{\Omega}_0$ minimizes $G(\Delta) = Q(\Delta + \tilde{\Omega}_0)$.

The main idea of this proof is as follows. For a sufficiently large $M>0$, consider sets
$$
\mathcal{T}_1 = \{ \Delta \in \mathcal{U}_m(\tilde{\Omega}_0) , \|\Delta\|_F = Mr_m \}, \quad \mathcal{T}_2 = \{ \Delta \in \mathcal{U}_m(\tilde{\Omega}_0) , \|\Delta\|_F \le Mr_m \},
$$
where 
$$
r_m = \sqrt{\frac{S_0\log (p-rp) }{ m }}.
$$
Note that $\mathcal{T}_1$ is non-empty. Indeed, consider $\M B_{\epsilon}=\epsilon\tilde{\Omega}_0$ for $\epsilon = Mr_m/\|\tilde{\Omega}_0\|_F$. Then $\M B_{\epsilon} = (1 + \epsilon)\tilde{\Omega}_0 - \tilde{\Omega}_0 \in \mathcal{U}_m(\tilde{\Omega}_0) $, hence $\M B_{\epsilon}  \in \mathcal{T}_1$.
Denote by $\bar{0}$ the matrix of all zero entries. It is clear that $G(\Delta) $ is convex, and
$
G(\hat{\Delta}) \le G(\bar{0}) =Q(\tilde{\Omega}_0) = 0.
$
Thus if we can show that $G(\Delta) >0$ for all $\Delta \in \mathcal{T}_1$, the minimizer $\hat{\Delta} $ must be inside $\mathcal{T}_2$ and hence $\|\hat{\Delta} \|_F \le Mr_m$. 
To see this, note that the convexity of $Q(\Omega)$ implies that 
$$
\inf_{\|\Delta \|_F = M r_m} Q(\tilde{\Omega}_0 + \Delta) > Q(\tilde{\Omega}_0) = 0.
$$
There exists therefore a local minimizer in the ball $\{ \tilde{\Omega}_0 + \Delta: \|\Delta \|_F \le M r_m\}$, or equivalently, for $\hat{\Delta} \in \mathcal{T}_2$, i.e. $\|\hat{\Delta} \|_F \le M r_m$. 

In the remainder of the proof, we focus on 
\begin{align}\label{qa}
G(\Delta) & = Q(\Delta + \tilde{\Omega}_0) = \tr(\Delta\hat{\Sigma})- \logdet(\Delta + \tilde{\Omega}_0)  +  \logdet \tilde{\Omega}_0.
\end{align}
Applying a Taylor expansion to $\logdet (\tilde{\Omega}_0 + \Delta)$ in \eqref{qa} gives
\begin{align}
& \logdet (\tilde{\Omega}_0 + \Delta) - \logdet \tilde{\Omega}_0 \notag \\
=& \frac{d }{dt} \logdet (\tilde{\Omega}_0 + t\Delta) \big |_{t=0} \Delta + \int_{0}^1 (1-t)  \frac{d^2}{ dt^2} \logdet (\tilde{\Omega}_0 + t\Delta)  dt  \notag \\
= &\tr (\Delta\tilde{\Sigma}_0) -  {\rm vec}(\Delta)^T \left\{ \int _{0}^1 (1-t) (\tilde{\Omega}_0 + t\Delta)^{-1} \otimes (\tilde{\Omega}_0 + t\Delta)^{-1} dt \right\} {\rm vec}(\Delta), \label{k1integral}
\end{align}
where 
${\rm vec}(\Delta)$ denotes the vectorized $\Delta$, and $\otimes$ is the Kronecker product. 
For $\Delta \in \mathcal{T}_1$, let $K_1$ be the integral term in \eqref{k1integral}, and define
\begin{align*}
K_2 & = \tr\left\{\Delta(\hat{\Sigma} - \Sigma_0)\right\}, \quad K_3 = \tr\left\{\Delta(\tilde{\Sigma}_0 - \Sigma_0)\right\}.
\end{align*}
We can then write
\begin{align*}
G(\Delta) & =  K_1 + \tr(\Delta\hat{\Sigma})- \tr (\Delta\tilde{\Sigma}_0) =  K_1 + K_2 - K_3.
\end{align*}
Next, we bound each of the terms $K_1, K_2$ and $K_3$ to find a lower bound for $G(\Delta)$. 

First consider $K_2$. Since the diagonal elements of $\hat{\Sigma}$ and $\Sigma_0$ are the same after scaling, 
\begin{align*}
|K_2| \le |\sum_{i\ne {i'}}(\hat{\Sigma}_{ii'} - \Sigma_{0,ii'}) \Delta_{ii'}|.
\end{align*}
By Lemma A.3 of \cite{Bickel:2008fk}, there exists a positive constant $c_2$ depending on $\phi_{\max}(\Sigma_0)$ such that 
\[
\max_{i \ne {i'}} |\hat{\Sigma}_{ii'} - \Sigma_{0,ii'} | \le c_2 \sqrt{\frac{\log (p-rp) } {m}},
\]
with probability tending to 1. 
Let $\Delta^+ = \diag(\Delta)$ be the diagonal matrix with the same diagonal as $\Delta$, and write $\Delta^- = \Delta - \Delta^+$. 
Then, $K_2$ is bounded by
\begin{align}\label{K2_bound}
|K_2| &\le c_2  \sqrt{\frac{\log (p-rp) } {m}} \|\Delta^-\|_1.
\end{align}
For $K_3$, we can use the upper bound for $\|\tilde{\Omega}_0 - \Omega_0 \|_F$ in \eqref{A0_bias}, and the lower bound for $\phi_{\min}(\tilde{\Omega}_0)$ in \eqref{A0_min}, to write, 
\begin{align}
|K_3|  & \le \|\Delta\|_F \| \tilde{\Sigma}_0 - \Sigma_0\|_F \le \|\Delta\|_F \frac{ \|\tilde{\Omega}_0 - \Omega_0 \|_F }{ \phi_{\min}(\tilde{\Omega}_0) \phi_{\min}(\Omega_0) } \label{K3_bound_1}\\
& \le \|\Delta\|_F \frac{c_3 \{S_0\log (p-rp)/m\} ^{1/2}} { (1-k_1) \phi^2_1}. \label{K3_bound_2}
\end{align}
The second inequality in \eqref{K3_bound_1} comes from the rotation invariant property of Frobenius norm, i.e.
\begin{align*}
\| \St -\S \|_F &= \|\S (\Omega_0 - \tilde \Omega_0 )\St\|_F \le \phi_{\max} (\S)  \| \Omega_0 - \At \|_F \phi_{\max} (\St).
\end{align*}
Using \eqref{A0_bias}, we can also obtain an upper bound for the maximum eigenvalue of $\At$:
\begin{align*}
\phi_{\max}(\tilde{\Omega}_0) &\le \phi_{\max}(\Omega_0) + \|\tilde{\Omega}_0  - \Omega_0 \|_2 \le \phi_{\max}(\Omega_0) + \|\tilde{\Omega}_0  - \Omega_0 \|_F \le \frac{1}{ \phi_2} + k_1 \phi_1. 
\end{align*}
Since $r_m \rightarrow 0$, there exists a sufficiently large $k_2>0$ such that for $\Delta \in \mathcal{T}_1$,
$$
\|\Delta\|_2 \le \|\Delta\|_F = Mr_m < \frac{1}{ \phi_2} k_2.
$$ 
Following \cite[][Page 502, proof of Theorem 1]{Rothman:2008ce}, a lower bound for $K_1$ can be found as
\begin{align}
K_1 & \ge \|\Delta\|_F^2 /\{2( \phi_{\max}(\At) + \|\Delta\|_2)^2 \} \notag\\
& \ge  \|\Delta\|_F^2  / \{2 \left({1/ \phi_2} + k_1 \phi_1 + {k_2 /\phi_2}\right)^2 \} = \frac{\phi_2^2 }{ 2(1+k_1\phi_1\phi_2 +k_2)^2}  \|\Delta\|_F^2 \label{K1_bound}.
\end{align}
Combining \eqref{K2_bound}, \eqref{K3_bound_2} and \eqref{K1_bound},
\begin{align*}
G(\Delta) & \ge \frac{\phi_2^2 }{ 2(1+k_1 \phi_1\phi_2+k_2)^2}  \|\Delta\|_F^2 -c_2  \sqrt{\frac{\log (p-rp) } {m}} \|\Delta^-\|_1 \\
& -  \frac{c_3 }{ (1-k_1) \phi^2_1}  \sqrt{\frac{S_0\log (p-rp) } {m}} \|\Delta\|_F .
\end{align*}
For $\Delta \in \mathcal{T}_1$, applying Cauchy-Schwarz inequality yields
$$
\|\Delta^-\|_1 \le \sqrt{|\hat{E}|} \cdot \|\Delta^-\|_F.
$$
We thus have
\begin{align*}
G(\Delta) & \ge \frac{\phi_2^2 }{ 2(1+k_1 \phi_1\phi_2+k_2)^2}  \|\Delta\|_F^2 -c_2  \sqrt{\frac{|\hat{E}| \log (p-rp) } {m}}\|\Delta^-\|_F  \\
& -  \frac{c_3 }{ (1-k_1) \phi^2_1}  \sqrt{\frac{S_0\log (p-rp) } {m}} \|\Delta\|_F \\
& \ge  \|\Delta\|_F^2 \left\{ \frac{\phi_2^2 }{ 2(1+k_1 \phi_1\phi_2+k_2)^2}  - \frac{c_2 }{ M}\sqrt{\frac{|\hat{E}|}{S_0}}- \frac{c_3 }{ M(1-k_1) \phi^2_1} \right\}>0,
\end{align*}
for $M$ sufficiently large.
\end{proof}

\begin{proof}[Proof of Corollary 1]
Under the assumptions in Theorem 2.2, we have
\[
\| \Delta_{ \Omega_0} \|_2=\| \hat \Omega  - \Omega_0 \|_2 = O_{\mpr} \left(\sqrt{ \frac{S_0\log (p-rp)} { m} }\right)= o_{\mpr}(1).
\]
The partial correlation matrix corresponding to $\hat \Omega$ can be written as
\begin{align*}
{\hat \bA}  & = \bIP - \hat \bD^{-1/2} {\hat \Omega} \hat \bD^{-1/2} =\bA_0  + \bD_0^{-1/2} \Omega_0 \bD_0^{-1/2} -  (\hat \bD)^{-1/2} {\hat \Omega}\hat \bD^{-1/2}  = \bA_0 + \Delta_{\bA_0},
\end{align*} 
where 
\begin{align}
\Delta_{\bA_0} & =  \bD_0^{-1/2} \Omega_0 \bD_0^{-1/2} -  (\hat \bD)^{-1/2} {\hat \Omega}\hat \bD^{-1/2} \notag \\
&= \bD_0^{-1/2} (\Omega_0 - \hat \Omega)\bD_0^{-1/2}  + \bD_0^{-1/2} \hat  \Omega \big( \bD_0^{-1/2}  - \hat \bD^{-1/2}  \big) +  \big( \bD_0^{-1/2}  - \hat \bD^{-1/2}  \big)   \hat  \Omega \hat \bD^{-1/2}. \label{e:deltaa0}
\end{align}
Next we show that each of the summands on the right hand side of \eqref{e:deltaa0} has $\ell_2$ norm $o_{\mpr}(1)$ and conclude thus $\|\Delta_{\bA_0}\|_2 = o_{\mpr}(1)$. 

By Assumption 1, the diagonal entries of $\Omega_0$ satisfy $\omega_{0,ii} \ge \phi_{\min}(\Omega_0) \ge \phi_1$ for all $i=1,\ldots, p$. Thus, $\|\bD_0^{-1/2}\|_2 =\max_{i} \omega_{0,ii}^{-1/2}\le \phi_1^{-1/2}$. It follows that
\begin{align*}
\|  \bD_0^{-1/2} (\Omega_0 - \hat \Omega)\bD_0^{-1/2}\|_2 & \le \| \bD_0^{-1/2}\|_2 ^2 \| \Omega_0 - \hat \Omega\|_2 = o_{\mpr}(1).
\end{align*}
For the remaining two terms, first notice that $\|\bD_0 - \hat \bD\|_2 \le \|\bD_0 - \hat \bD\|_F\le \|\Omega_0 - \hat \Omega\|_F = o_{\mpr}(1)$. Therefore,
\begin{align*}
\|\bD_0^{-1/2} - \hat \bD^{-1/2}\|_2 & = \max_{i=1,\ldots, p} |  \omega_{0,ii}^{-1/2} -  \hat \omega_{ii}^{-1/2} | =  \max_{i=1,\ldots, p} \left| \frac{ \omega_{0,ii}^{1/2} -  \hat \omega_{ii}^{1/2}} {\omega_{0,ii}^{1/2} \hat \omega_{ii}^{1/2}} \right| \\
&=  \max_{i=1,\ldots, p} \left| \frac{ \omega_{0,ii} -  \hat \omega_{ii} } {\omega_{0,ii}^{1/2} \hat \omega_{ii}^{1/2} ( \omega_{0,ii}^{1/2} +  \hat \omega_{ii}^{1/2}) } \right|  \le \phi_{1}^{-1}(\phi_1- o_{\mpr}(1))^{-1/2} \|\bD_0 - \hat \bD\|_2,
\end{align*}
where the last inequality comes from that fact that
\[ 
\min_i | \hat{\omega}_{ii}| = \min_i | \hat{\omega}_{ii} - \omega_{0,ii} + \omega_{0,ii}  | \ge \min_i |\omega_{0,ii} | - \max_i | \hat{\omega}_{ii} - \omega_{0,ii} | \ge \phi_1 - o_{\mpr}(1).
\]
Hence, $\|\bD_0^{-1/2} - \hat \bD^{-1/2}\|_2=o_{\mpr}(1)$. Note further, 
\[ 
\| \hat \Omega\|_2 = \|  \hat \Omega - \Omega_0 + \Omega_0 \|_2 \le \|\Omega_0\|_2 +
\|  \hat \Omega - \Omega_0\|_2 = \|\Omega_0\|_2 +o_{\mpr}(1)
\]
is bounded above.  It follows thus,
\begin{align*}
\| \bD_0^{-1/2} \hat  \Omega \big( \bD_0^{-1/2}  - \hat \bD^{-1/2}  \big) \|_2 & \le \|\bD_0^{-1/2}\|_2 \|\hat  \Omega\|_2 \|\bD_0^{-1/2}  - \hat \bD^{-1/2} \|_2 =o_{\mpr}(1), \\
\|  \big( \bD_0^{-1/2}  - \hat \bD^{-1/2}  \big)   \hat  \Omega \hat \bD^{-1/2} \|_2 & \le \|  \bD_0^{-1/2}  - \hat \bD^{-1/2} \|_2 \| \hat  \Omega \|_2 \|\hat \bD^{-1/2} \|_2 = o_{\mpr}(1).
\end{align*}
This completes the proof. 

\end{proof}

The following proof of Theorem 3.1 adapts from that of Theorem 2.1 in \cite{Shojaie:2010nz}.

\begin{proof}[Proof of Theorem 3.1]
Consider the special case where the row vector $\bfb = {\bf 1}^T$, i.e. the whole network is tested as one pathway. The general case when $\bfb \neq {\bf 1}^T$ follows from a similar argument.  

For the partial correlation $\bA_0^{(k)}\ (k=1,2)$ defined in Section 3.2, it holds that $\Lambda^{(k)} (\Lambda^{(k)} )^T = (\bIP - \bA_0^{(k)})^{-1} = \sum_{t=0}^{\infty} (\bA_0^{(k)})^t$. 
Hence 
\begin{align*}
\hat \Lambda^{(k)} (\hat \Lambda^{(k)} )^T =  \sum_{t=0}^{\infty} ({\hat \bA}^{(k)})^t 
&=  \sum_{t=0}^{\infty} (\bA_0^{(k)})^t  + \sum_{t=1}^{\infty} \sum_{u=1}^{t} {t \choose u} (\bA_0^{(k)})^{t-u} (\Delta_{\bA_0^{(k)}})^{u}\\
& = \Lambda^{(k)} (\Lambda^{(k)} )^T +  \Delta_{\Lambda^{(k)}}.
\end{align*}
For $\hat{\bA}^{(k)}$ defined under the assumptions in Theorem 2.2 and 3.1, we have $\|\Delta_{\bA_0^{(k)}}\|_2 = o_{\mpr}(1)$ by Corollary 1. Thus, $\|\Delta_{\Lambda^{(k)}}\|_2= o_{\mpr}(1)$.

Using results from \cite{Shojaie:2010nz}, the test statistic in (3.11) can be written as
\begin{align*}
TS = \frac{\bfb (\bar{\bY}^{(2)} - \bar{\bY}^{(1)})}{\sqrt{\hat{\sigma}_{\bgamma}^2\left[\bfb \left\{ \frac{1}{n_1} \hat \Lambda^{(1)} (\hat \Lambda^{(1)} )^T+ \frac{1}{n_2}\hat \Lambda^{(2)} (\hat \Lambda^{(2)} )^T  \right\}\bfb^T\right] + \hat \sigma^2_{\beps} \left(\frac{1}{n_1} + \frac{1}{n_2} \right) \bfb \bfb^T}},
\end{align*}
where $\bar{\bY}^{(k)}$ is the mean expression of genes in the experimental condition $k$. 
\cite{Shojaie:2010nz} show that $TS$ is an asymptotically most powerful unbiased test for (3.10) when the correct network information is provided. Therefore, to establish the result in Theorem 3.1, it suffices to show that the denominator of $TS$ is a consistent estimator. 

In the following, we first consider the log-likelihood $l_F (\vth; \hat \Lambda)$ based on the estimated networks $\hat \Lambda = (\hat \Lambda^{(1)}, \hat \Lambda^{(2)})$ and correct variance components $\vth = (\sigma^2_{\bgamma}, \sigma^2_{\beps})$. We then establish that the maximum likelihood estimator $\hat \vth_{\hat \Lambda} \rightarrow_{\mpr}  \vth$ as $\hat \Lambda^{(k)}(\hat \Lambda^{(k)} )^T \rightarrow_{\mpr} \Lambda^{(k)}(\Lambda^{(k)} )^T$ for both $k$. Hence the denominator of $TS$ is consistent and $TS$ is an asymptotically most powerful unbiased test for (3.10). 

Let $\hat \bW^{(k)} =  \sigma_{\bgamma}^2  \hat \Lambda^{(k)} (\hat \Lambda^{(k)} )^T + \sigma^2_{\beps} \bIP$ for $k=1, 2$. Up to a constant, the negative log-likelihood 
\[
l_F (\vth; \hat \Lambda) =\frac{n_1}{2n} l (\vth; \hat \Lambda^{(1)}) + \frac{n_2}{2n} l (\vth; \hat \Lambda^{(2)}) 
\] with 
\begin{align*}
 l (\vth; \hat \Lambda^{(1)})& =  \logdet(\hat \bW^{(1)}) + \frac{1}{n_1} \sum_{j=1}^{n_1} \bR_j^T (\hat \bW^{(1)})^{-1} \bR_j, \\
 l (\vth; \hat \Lambda^{(2)})& =  \logdet(\hat \bW^{(2)}) + \frac{1}{n_2} \sum_{j=1+n_1}^{n} \bR_j^T (\hat \bW^{(2)})^{-1} \bR_j,
\end{align*}
where $\bR_j = \bY_j^{(1)} - \bar \bY^{(1)}\ (j=1, \ldots, n_1)$ and $\bR_j = \bY_j^{(2)} - \bar \bY^{(2)}\ (j=1+n_1, \ldots, n)$. We treat $ l (\vth; \hat \Lambda^{(1)})$ first. In particular, we can approximate $l (\vth; \hat \Lambda^{(1)})$ using its one-term Taylor expansion around $ \bW^{(1)}$
\begin{equation*}
l(\vth; \hat \Lambda^{(1)}) =l (\vth; \Lambda^{(1)}) + \tr\big\{ \nabla_{\bW^{(1)}} l (\vth; \Lambda^{(1)})^T  \Delta_{\bW^{(1)}}\big\} + o(\|\Delta_{\bW^{(1)}}\|_2^2 ),
\end{equation*}
where $ \nabla_{\bW^{(1)}} l (\vth; \Lambda^{(1)})$ is the gradient of $l (\vth; \Lambda^{(1)})$ with respect to $\bW^{(1)}$ and
\[
 \nabla_{\bW^{(1)}} l (\vth; \Lambda^{(1)}) = (\bW^{(1)})^{-1} - n_1^{-1}\sum_{j=1}^{n_1} (\bW^{(1)})^{-1}  \bR_j \bR_j^T (\bW^{(1)})^{-1} .
\]
Let $\Gamma =  \Delta_{\bW^{(1)}}/\|  \Delta_{\bW^{(1)}}\|_2$ and denote 
\begin{align*}
g(\vth) &=  \tr\big\{ \nabla_{\bW^{(1)}} l (\vth; \Lambda^{(1)})^T \Gamma\big\} = \tr\big\{ (\bW^{(1)})^{-1}\Gamma\big\} - n_1^{-1} \sum_{j=1}^{n_1} \bR_j^T (\bW^{(1)})^{-1} \Gamma  (\bW^{(1)})^{-1} \bR_j.
\end{align*} 
then
\begin{align*}
l(\vth; \hat \Lambda^{(1)}) =l (\vth; \Lambda^{(1)}) +  g(\vth) \|  \Delta_{\bW^{(1)}}\|_2 + o(\|  \Delta_{\bW^{(1)}}\|_2^2).
\end{align*} 
Using von Neumann's trace inequality \citep{mirsky1975trace}, we can bound the first term in $g(\vth)$ by
\begin{align*}
\big|\tr\big\{ (\bW^{(1)})^{-1}\Gamma\big\} \big| & \le \sum_{i=1}^p \vs_{[i]} ((\bW^{(1)})^{-1}) \vs_{[i]} (\Gamma) \\
& \le p \vs_{[1]} \big(( \sigma_{\bgamma}^2 \Lambda^{(1)} (\Lambda^{(1)} )^T + \sigma^2_{\beps} \bIP)^{-1}\big) \vs_{[1]} (\Gamma) \\
& = p  \frac{1}{\phi_{\min} ( \sigma_{\bgamma}^2 \Lambda^{(1)} (\Lambda^{(1)} )^T + \sigma^2_{\beps} \bIP)}  \vs_{[1]} (\Gamma),
\end{align*}
where $\vs_{[i]} (\bA)$ denotes the $i$th largest singular value of $\bA$. By construction, $ \vs_{[1]} (\Gamma)=1$ and $\phi_{\min} ( \sigma_{\bgamma}^2 \Lambda^{(1)} (\Lambda^{(1)} )^T + \sigma^2_{\beps} \bIP) \ge \sigma^2_{\beps}$. Hence $
|\tr\{ (\bW^{(1)})^{-1}\Gamma\} | \le p/\sigma^2_{\beps}.
$
On the other hand, with probability tending to 1, 
\begin{align*}
n_1^{-1} \sum_{j=1}^{n_1} \bR_j^T (\bW^{(1)})^{-1} \Gamma  (\bW^{(1)})^{-1} \bR_j & \le \|(\bW^{(1)})^{-1} \Gamma  (\bW^{(1)})^{-1} \|_2 n_1^{-1} \sum_{j=1}^{n_1} \bR_j^T \bR_j\\
& \le \|(\bW^{(1)})^{-1} \|_2^2 \| \Gamma\|_2 n_1^{-1} \sum_{j=1}^{n_1} \bR_j^T \bR_j = \sigma^{-4}_{\beps} \mbe(\| \bR_j\|_2^2),
\end{align*}
where the last step follows from the strong law of large numbers. This implies that $g(\vth)$ is bounded for nontrivial $\sigma^2_{\beps}$. Note also $ \Delta_{\bW^{(1)}} = \hat \bW^{(1)}  - \bW^{(1)} = \sigma_{\bgamma}^2 \{ \hat \Lambda^{(1)} (\hat \Lambda^{(1)} )^T -  \Lambda^{(1)} ( \Lambda^{(1)} )^T\} = \sigma_{\bgamma}^2 \Delta_{\Lambda^{(k)}} $. Hence $g(\vth) \|  \Delta_{\bW^{(1)}}\|_2 = g(\vth) \sigma_{\bgamma}^2 \|\Delta_{\Lambda^{(k)}}\|_2 = o_{\mpr}(1)$. Therefore
$
l(\vth; \hat \Lambda^{(1)}) =l (\vth; \Lambda^{(1)})  + o_{\mpr}(1),
$
and similarly one can show that  $l(\vth; \hat \Lambda^{(2)}) =l (\vth; \Lambda^{(2)})  + o_{\mpr}(1)$. They together imply that 
\[
l_F(\vth; \hat \Lambda) =l_F(\vth; \Lambda)  + o_{\mpr}(1).
\]

Now conditioning on the event $\{ l_F(\vth; \hat \Lambda) =l_F(\vth; \Lambda)\}$, the estimate of the variance components is $\hat{\vth} = \argmin_{\vth} l_F(\vth; \Lambda)$. Since $ l_F(\vth; \Lambda)$ is convex with respect to $\vth$,  M-estimation results in \cite{haberman1989concavity} imply that $\mpr(\hat \vth = \vth) = 1$ and hence $\hat \vth \rightarrow_{\mpr}  \vth$ as $\hat \Lambda^{(k)} (\hat \Lambda^{(k)})^T \rightarrow_{\mpr}  \Lambda^{(k)} (\Lambda^{(k)})^T$ for both $k$. It follows immediately that the denominator of the test statistic $TS$ is a consistent estimator as $\hat \Lambda^{(k)} (\hat \Lambda^{(k)})^T \rightarrow_{\mpr}  \Lambda^{(k)} (\Lambda^{(k)})^T$ for both $k$. This concludes the proof. 
\end{proof}
%%%%%%%%%%%%%%%%%%%%

%\section{Derivation for Newton's Method}
\section{Efficient Estimation of Model Parameters}\label{app:netgsa:newton}
In this section, we present in details the strategy used to scale up the NetGSA algorithm for large scale networks as well as necessary derivations. 

As pointed out in Section 2.2.1 of the main text, inference in NetGSA requires estimation of the mean parameters $\bmu^{(1)}$ and $\bmu^{(2)}$ and variance components ${\sigma}_{\bgamma}^2$ and ${\sigma}_{\beps}^2$. After rearranging the data $\mathcal{D}$ to be a $N \times 1$ vector $\bY$, we can write the model using the matrix notation as
\begin{equation}\label{TSmodel:2}
  \bY = \Psi \bbeta + \Pi \bgamma + \beps,
\end{equation}
where the design matrix
\[
\Pi = {\rm bdiag} (\Lambda^{(1)} , \ldots, \Lambda^{(1)} , \Lambda^{(2)}, \ldots, \Lambda^{(2)}  ) \in \bR^{N \times N},
\]
and 
\[
\Psi = 
\begin{pmatrix}
\Lambda^{(1)} & \\
\vdots && \\
\Lambda^{(1)} &\\
& \Lambda^{(2)} \\
 &\vdots\\
&\Lambda^{(2)}  \\
\end{pmatrix}
\in \bR^{N \times 2p}.
\]
The variance of $\bY$, i.e. $\bW =  \sigma_{\beps}^2{\bf I}_{N} + \sigma_{\bgamma} ^2 \Pi \Pi '$.
%is thus block diagonal with the $j$-th block being $\Var(\bY_j)= \Sigma^{(k)} $ for $j= \sum_{t=1}^{k-1} n_{t} + 1, \ldots,  \sum_{t=1}^{k} n_{t}$. 
%\subsection{Parameter Estimation}
%A similar procedure as used in the two-sample case can be adopted to estimate the unknown parameters for the mixed linear model in \eqref{TSmodel}. In particular, the profile likelihood method introduced in the previous section can be used to estimate the variance components. 
The mean $\bbeta$ can be estimated via the maximum likelihood as
\[
\hat \bbeta = (\Psi' \hat \bW^{-1} \Psi)^{-1} \Psi'\hat \bW^{-1} \bY,
\]
where $\hat \bW$ is defined using the estimated variances. The variances are often estimated via the maximum likelihood or restricted maximum likelihood using the profile likelihood. Thus, one can use an iterative algorithm to jointly estimate $\bbeta$ and the variance components.

However, estimation of the variance components is computationally demanding for large networks. 
To ensure stability, the earlier version of the NetGSA considered profiling out one of the variance components and implemented an algorithm from \cite{byrd1995limited}, which uses a limited-memory modification of the Broyden--Fletcher--Goldfarb--Shanno quasi-Newton method to optimize the profile log-likelihood. However, the above implementation has a few issues. 
The first issue is its high computational cost due to the inefficient evaluation of matrix inverses and determinants. 
Moreover, the algorithm from \cite{byrd1995limited} requires finite values of the objective function within the supplied box constraints, which is often not satisfied, even after the constraints are adjusted to be within a small range of the optimal estimate. This is particularly the case when the underlying networks are large. To extend the applicability of the NetGSA, we consider using Newton's method for estimating the variance parameters based on the profile log-likelihood to improve the computational stability. In particular, we make the following two key improvements for implementation of Newton's method.

First, it is clear that 
$\Var(\bY_j^{(k)}) =\sigma^2_{\beps}\left\{\bIP + \tau \Lambda^{(k)} (\Lambda ^{(k)})^T\right\}= \sigma^2_{\beps} \Sigma^{(k)}$, where $\tau = \sigma^2_{\bgamma} / \sigma^2_{\beps}$.
Since the profile log-likelihood as well as its gradient and Hessian matrix with respect to $\tau$ all depend on $\Sigma^{(k)}\ (k=1, 2)$ and their inverses, 
we choose to invert from their Cholesky decompositions $\sk = \bU^T\bU,$
where $\bU$ is an upper triangular matrix. The inversion of the triangular matrices results in significant speedup and the inverses of the original matrices can then be computed as
$
(\sk)^{-1} = (\bU^{-1}) (\bU^{-1})^T.
$
In the meantime, we also simplify the calculation of the determinant of $\sk$ since $\det (\sk) = \det (\bU)^2$, which is necessary for evaluating the profile log-likelihood. 

Second, the quality of the starting point as well as step sizes will both affect convergence of Newton's method. To select a good starting point, we use a method-of-moment-type estimate of the variance components. Specifically, denote the residuals $\bR_j = \bY^{(k)}_j - \Lambda^{(k)} \hat{\bmu}^{(k)} $ for $j = 1, \ldots, n$, where $\hat{\bmu}^{(k)} $ is the estimate of ${\bmu}^{(k)}$. Assume that there is a single variance $\sigma^2_{\beps}$ that applies to all $\beps_j \ (j = 1, \ldots, n)$ and variances of $\bgamma_j$ are different. The variance of $\bR_j$ can be decomposed as $(\sigma^2_{\bgamma})_j + \sigma^2_{\beps}$. We then take the minimum of $\Var(\bR_j)$ as the estimate of $\sigma^2_{\beps}$ and average of the remaining variances as the estimate of $\sigma^2_{\bgamma}$. Their ratio is used as the initial value for $\tau$. The approximation runs very fast and does not add much computational cost to the method. To find the appropriate step sizes, we use backtracking line search as described in \cite[][page 464]{Boyd:2004gs}.

With the above two modifications, Newton's method can then be implemented to optimize the profile log-likelihood and returns an estimate of $\tau$. Estimates of $\hat{\sigma}_{\bgamma}^2$ and $ \hat{\sigma}_{\beps}^2$ follow immediately. The implementation of Newton's method requires the gradient and the Hessian of the objective function, i.e., the profile log-likelihood. Next we provide details on how to calculate these quantities from the profile log-likelihood when profiling out $\sigma_{\beps}$, based on the general framework introduced in \cite{Lindstrom:1988ri}. The derivation follows similarly when profiling out $\sigma_{\bgamma}$.

Let $N=np$ be the total number of observations for all genes. Recall that for $k=1, 2$, $\Sigma^{(k)} = \bIP + \tau \Lambda^{(k)} (\Lambda ^{(k)})^T$ with $\tau = \sigma^2_{\bgamma} / \sigma^2_{\beps}$. The residuals $\bR_j = \bY^{(k)}_j - \Lambda^{(k)} \hat{\bmu}^{(k)} $ for $j = 1, \ldots, n$, where $\hat{\bmu}^{(k)} $ is the estimate of ${\bmu}^{(k)}$. 
Given the observations $\bY_1, \ldots, \bY_n$ (with the first $n_1$ samples from condition 1 and the remaining $n_2=n-n_1$ samples from condition 2), the nonconstant part of the ``full'' log-likelihood $l_F$ is
\begin{align*}
l_F(\sigma_{\beps}, \tau \mid \bY_1, \ldots, \bY_n)  = &-\frac{1}{ 2} \left\{ n_1 \logdet(\sigma^2_{\beps} \Sigma^{(1)}) +  n_2 \logdet (\sigma^2_{\beps} \Sigma^{(2)})  \right\}  \\
&  - \frac{1}{ 2} \sigma^{-2}_{\beps} \left\{ \sum_{j=1}^{n_1} \bR^T_j (\Sigma^{(1)})^{-1} \bR_j + \sum_{j=n_1+1}^{n} \bR^T_j (\Sigma^{(2)})^{-1} \bR_j\right\}.
\end{align*}
Similarly, the nonconstant part of the log-likelihood using the restricted maximum likelihood is
\begin{align*}
l_R(\sigma_{\beps}, \tau\mid \bY_1, \ldots, \bY_n) = & l_F(\sigma_{\beps}, \tau \mid \bY_1, \ldots, \bY_n)\\
& - \frac{1}{ 2} \logdet \left\{n_1\sigma_{\beps}^{-2} (\Lambda^{(1)})^T (\Sigma^{(1)})^{-1} \Lambda^{(1)} + n_2\sigma_{\beps}^{-2} (\Lambda^{(2)})^T (\Sigma^{(2)})^{-1} \Lambda^{(2)}\right\}.\label{reml}
\end{align*}
We first solve for $\sigma^2_{\beps}$ as a function of $\tau$. The maximum likelihood estimate of $\sigma^2_{\beps}$ is 
\begin{equation}\label{sigmahat:ml}
\hat{\sigma}^2_{\beps} = \frac{1}{ N} \left\{ \sum_{j=1}^{n_1} \bR^T_j (\Sigma^{(1)})^{-1} \bR_j + \sum_{j=n_1+1}^{n} \bR^T_j (\Sigma^{(2)})^{-1} \bR_j\right\},
\end{equation}
whereas its restricted maximum likelihood estimate is
\begin{equation}\label{sigmahat:rml}
\hat{\sigma}^2_{\beps} = \frac{1}{ N - 2p} \left\{ \sum_{j=1}^{n_1} \bR^T_j (\Sigma^{(1)})^{-1} \bR_j + \sum_{j=n_1+1}^{n} \bR^T_j (\Sigma^{(2)})^{-1} \bR_j\right\}.
\end{equation}
Substituting $\sigma^2_{\beps}$ with its corresponding estimate, we obtain the profile log-likelihood 
\begin{align}
p_F(\tau\mid \bY_1, \ldots, \bY_n) & = - \frac{1}{ 2} (n_1\logdet \Sigma^{(1)} + n_2 \logdet \Sigma^{(2)}) \notag \\
 &\quad - \frac{1}{ 2}N \log \left\{ \sum_{j=1}^{n_1} \bR^T_j (\Sigma^{(1)})^{-1} \bR_j + \sum_{j=n_1+1}^{n} \bR^T_j (\Sigma^{(2)})^{-1} \bR_j\right\},
\end{align} for maximum likelihood and
 \begin{align}
p_R(\tau \mid \bY_1, \ldots, \bY_n) &= -  \frac{1}{2} (n_1\logdet \Sigma^{(1)} + n_2 \logdet \Sigma^{(2)})\notag   \\
&\quad - \frac{1}{2} (N-2p) \log \left\{ \sum_{j=1}^{n_1} \bR^T_j (\Sigma^{(1)})^{-1} \bR_j + \sum_{j=n_1+1}^{n} \bR^T_j (\Sigma^{(2)})^{-1} \bR_j\right\} \notag \\
& \quad - \frac{1}{2} \logdet \left\{  n_1 (\Lambda^{(1)})^T (\Sigma^{(1)})^{-1} \Lambda^{(1)} + n_2(\Lambda^{(2)})^T (\Sigma^{(2)})^{-1} \Lambda^{(2)} \right\},
\end{align}
 for restricted maximum likelihood.

As $\sk \ (k=1, 2)$ are the only terms that depend on $\tau$, we first look at the derivatives of
$\logdet \sk$, $\bR^T_j (\sk )^{-1} \bR_j$, and $\logdet(\M H)$ 
with respect to $\tau$, where $\M H = n_1 \M H^{(1)} + n_2 \M H^{(2)}$ and $\hk = (\lk)^T(\sk )^{-1} \lk$ for $k=1, 2$. Let 
\[
\M B^{(k)} = (\sk ) ^{-1} \frac{d \sk}{ d\tau} (\sk  )  ^{-1}.
\]  
Then
\[
\frac{d \logdet (\sk )  }{ d \tau}  = \tr \left\{(\sk )^{-1} \frac{d \sk}{ d\tau} \right\},
\]
\[  
\frac{d^2 \logdet (\sk ) }{ d \tau^2 }  =  \tr \left\{ -(\M B^{(k)})^T\frac{d \sk}{ d\tau} + (\sk )^{-1} \frac{d^2  \sk}{ d \tau^2 }\right\},
\]
\[
\frac{d ~ \bR^T_j ( \sk)^{-1} \bR_j }{ d \tau}  = -  \bR^T_j \M B^{(k)} \bR_j, \quad \frac{d^2 ~ \bR^T_j ( \sk )^{-1} \bR_j  }{ d \tau^2}   = - \bR^T_j   \frac{d \M B^{(k)} }{ d\tau} \bR_j, 
\]
\[
\frac{d   \logdet ( \M H) }{ d \tau}  = -  \tr  \left\{ \M H^{-1} \sum_{k=1,2} n_k(\Lambda^{(k)})^T \M B^{(k)} \Lambda^{(k)} \right\},
\]
and
\begin{align*}
\frac {d^2  \logdet(\M H ) }{ d \tau^2}  
 = & - \tr  \left\{ \M H^{-1} \sum_{k=1,2} n_k (\lk)^T\M B^{(k)} \lk  \times  \M H^{-1}  \sum_{k=1,2} n_k  (\lk)^T\M B^{(k)} \lk  \right\} \\
&- \tr \left\{ \M H^{-1}  \sum_{k=1,2} n_k  (\lk)^T \frac{d\M B^{(k)}  }{ d\tau} \lk  \right\},
\end{align*}
where
\begin{align*}
\frac{d \M B^{(k)} }{ d\tau} & = - (\sk )^{-1} \left\{2  \frac{d \sk}{ d\tau} (\sk )^{-1} \frac{d \sk}{ d\tau} -  \frac{d^2  \sk}{ d \tau^2 }\right\} (\sk )^{-1}.
\end{align*}

\noindent Given the covariance $\sk \ (k=1, 2)$ defined in Section 3.1, we can further simplify the above derivatives and obtain
\[
\frac{d \logdet \sk}{ d \tau}  = \tr  \left\{ \hk\right\},  \quad  \frac{d^2 \logdet\sk}{ d \tau^2 }   =  -\tr  \left\{ \hk\hk\right\},
\]
\[
\frac{d ~ \bR^T_j (\sk )^{-1} \bR_j }{ d \tau}  = -  \bR^T_j (\sk ) ^{-1}  \lk   (\lk )^T (\sk )^{-1} \bR_j,
\]
\[
\frac{d^2 ~ \bR^T_j (\sk )^{-1} \bR_j  }{ d \tau^2}   = 2 \bR^T_j (\sk ) ^{-1} \lk \hk( \lk )^T (\sk )^ {-1} \bR_j,
\]
\[
 \frac {d   \logdet (\M H) }{ d \tau} = -  \tr  \left\{ \M H^{-1}  \sum_{k=1,2} n_k  \hk \hk \right\}, 
\]
\[  
\frac{d^2  \logdet (\M H) }{ d \tau^2}   =-  \tr  \left\{  \M H^{-1} \sum_{k=1,2} n_k \hk\hk\right\} + 2 \tr  \left\{  \M H^{-1} \sum_{k=1,2} n_k \hk\hk \hk\right\}.
\]

With the above quantities, one can then calculate the gradient and Hessian of the profile log-likelihood $p_R$ for restricted maximum likelihood and use Newton's method to obtain an estimate of $\tau$. Estimate of $\hat{\sigma}^2_{\beps}$ is calculated from  \eqref{sigmahat:rml}, and $\hat{\sigma}^2_{\bgamma} =\hat{\tau} \hat{\sigma}^2_{\beps}$. Estimation with maximum likelihood follows similarly by applying Newton's method to $p_F$ and utilizing \eqref{sigmahat:ml}.

\section{Additional Simulation Results}\label{app:netgsa:sim}

%In this section, we first present additional details for the two simulations discussed in the main paper. 
To benchmark the performance of the proposed network estimation procedure as well as NetGSA, we first revisit the two simulation experiments presented in Section 3 of the main paper and report the Type I error (or the observed false discovery proportion) when the null hypothesis is true. In addition, we consider two other simulation experiments and refer to them as the third and fourth settings, following the earlier two settings in the main paper. The simulations in this section are also discussed when comparing the run time of NetGSA with different variance estimation algorithms in Section 3 of the main paper. 

\subsection{Simulation Studies 1 and 2}

\subsubsection{Powers}
We have shown the estimated powers in Tables 2 and 3 in the main paper for the two experiments based on adjusted false discovery rate (FDR) cutoffs. For completeness, we present here the estimated powers in Tables \ref{power:100:q1} and \ref{power:160:q1} when the FDR cutoff is $q^*=0.05$. Due to the use of different FDR cutoffs, one expects to see higher powers for the columns corresponding to 0.2, 0.8, 0.2(m) and 0.8(m), and slightly lower powers for E and \GSA-c in Tables \ref{power:100:q1} and \ref{power:160:q1} compared to, respectively, Tables 2 and 3 in the main paper. In both Table \ref{power:100:q1} and Table \ref{power:160:q1}, we still observe the following: NetGSA with the exact networks does a very good job in recovering the true powers for each pathway; NetGSA with more external structural information generally reports powers that are closer to the true power; further NetGSA is robust to misspecification in external structural information. Further, for pathway 1 that has neither mean nor structural changes, we note that the powers are sometimes greater than 0.05 when NetGSA with estimated network information is applied. This is partly due to the network estimation error. 

\begin{table}[ht]
\centering
\caption{{Powers with false discovery rate cutoff $q^*=0.05$ in experiment 1. 0.2/0.8 refer to NetGSA with 20\%/80\% external information; E refers to NetGSA with the exact networks; T refers to the true power; \GSA-c/\GSA-s refer to Gene Set Analysis with/without randomization of the genes in 1000 permutations, respectively;  0.2(m)/0.8(m)  refer to NetGSA with 20\%/80\% misspecified external information.
}}{
\begin{tabular}{ccccccccc}
\hline\hline
\multicolumn{1}{c}{}& \multicolumn{8}{c}{$p=100$}  \\
Pathway & 0.2 & 0.8 & E & T & \GSA-s & \GSA-c &0.2(m) & 0.8(m)\\ 
\hline
  1 & 0.05 & 0.10 & 0.03 & 0.06 & 0.17 & 0.01 & 0.05 & 0.08 \\ 
  2 & 0.18 & 0.13 & 0.03 & 0.06 & 0.09 & 0.00 & 0.16 & 0.16 \\ 
  3 & 0.50 & 0.48 & 0.30 & 0.46 & 0.36 & 0.00 & 0.50 & 0.63 \\ 
  4 & 0.49 & 0.29 & 0.02 & 0.07 & 0.25 & 0.04 & 0.44 & 0.29 \\ 
  5 & 0.94 & 0.98 & 0.89 & 0.97 & 0.97 & 0.00 & 0.95 & 0.95 \\ 
  6 & 0.46 & 0.49 & 0.20 & 0.26 & 0.36 & 0.00 & 0.49 & 0.41 \\ 
  7 & 0.84 & 0.90 & 0.94 & 0.99 & 0.98 & 0.04 & 0.82 & 0.92 \\ 
  8 & 0.54 & 0.68 & 0.42 & 0.57 & 0.87 & 0.00 & 0.56 & 0.61 
\\ \hline
\end{tabular}}\label{power:100:q1}
\end{table}

\begin{table}[ht]
\centering
\caption{{Powers with false discovery rate cutoff $q^*=0.05$ in experiment 2. 0.2/0.8 refer to NetGSA with 20\%/80\% external information; E refers to NetGSA with the exact networks; T refers to the true power; \GSA-c/\GSA-s refer to Gene Set Analysis with/without randomization of the genes in 1000 permutations, respectively;  0.2(m)/0.8(m)  refer to NetGSA with 20\%/80\% misspecified external information.
}}{
\begin{tabular}{ccccccccc}
\hline\hline
\multicolumn{1}{c}{}& \multicolumn{8}{c}{$p=160$}  \\
Pathway & 0.2 & 0.8 & E & T & \GSA-s & \GSA-c &0.2(m) & 0.8(m)\\ 
\hline
  1 & 0.10 & 0.11 & 0.02 & 0.05 & 0.07 & 0.01 & 0.10 & 0.11 \\ 
  2 & 0.52 & 0.60 & 0.15 & 0.36 & 0.51 & 0.00 & 0.53 & 0.60 \\ 
  3 & 0.96 & 1.00 & 0.95 & 0.99 & 1.00 & 0.00 & 0.97 & 1.00 \\ 
  4 & 0.98 & 1.00 & 1.00 & 1.00 & 1.00 & 0.09 & 0.98 & 1.00 \\ 
  5 & 0.38 & 0.34 & 0.02 & 0.11 & 0.10 & 0.03 & 0.41 & 0.36 \\ 
  6 & 0.46 & 0.35 & 0.01 & 0.07 & 0.24 & 0.01 & 0.46 & 0.34 \\ 
  7 & 0.78 & 0.83 & 0.89 & 0.92 & 0.99 & 0.00 & 0.78 & 0.82 \\ 
  8 & 0.92 & 0.99 & 1.00 & 1.00 & 1.00 & 0.02 & 0.91 & 0.98
\\ \hline
\end{tabular}}\label{power:160:q1}
\end{table}

\subsubsection{Type I errors}

To validate the type I error when the null hypothesis is true, we use the same null setup as presented in Section 3 of the main paper for both experiment 1 and 2. The network structure and node mean expressions under the alternative are set to be the same as in the null case. We use $n_1=n_2=25$ samples for each condition in experiment 1 and $n_1=n_2=40$ in experiment 2 for pathway enrichment analysis. When the underlying networks are not available, we estimate the networks based on external information ranging from 0\%, 20\%, 80\% to 100\% and 100 observations generated from the true network. Scenarios with misspecified structural information are also considered. In the following, we present type I errors based on both the adjusted FDR cutoffs and uniform FDR cutoff at $q^*=0.05$, where the former corresponds to using $q^*=0.01$ for cases 0.2, 0.8, 0.2(m) and 0.8(m), 0.05 for GSA-s and 0.10 for E and GSA-c.

Table~\ref{p100:null:q1} and \ref{p160:null:q1} present the type I errors based on adjusted FDR cutoffs evaluated over 100 replications for experiment 1 and 2, respectively. As expected, the type I errors when all true parameters are plugged in the NetGSA model are 0.05 for all subnetworks in both experiments. When the exact networks are known, one only estimates the variance components in the NetGSA model and observes small false discovery proportions. When the exact networks are not available such that one estimates the partial correlations as well as the variance components, the type I errors are generally greater than $q^*$; in particular, the type I errors get worse as the amount of external information decreases. This is likely due to the small sample sizes for estimating the networks. In general, one benefits from having more external structural information and/or more observations for recovering the underlying networks when using NetGSA. In comparison, both \GSA-c and \GSA-s have type I errors smaller than 0.05. 

\begin{table}[ht]
\centering
\caption{Type I error when the null hypothesis is true in experiment 1. False discovery rate cutoffs are $q^*=0.01$ for cases 0.2, 0.8, 0.2(m) and 0.8(m), 0.05 for \GSA-s and 0.10 for E and \GSA-c.}
\begin{tabular}{ccccccccccc}
  \hline\hline
Pathway & 0.2 & 0.8 & E & T & \GSA-s & \GSA-c & 0.2(m) & 0.8(m) \\ 
  \hline
  1 & 0.00 & 0.00 & 0.00 & 0.05 & 0.02 & 0.03 & 0.00 & 0.01 \\ 
  2 & 0.06 & 0.06 & 0.01 & 0.05 & 0.03 & 0.05 & 0.07 & 0.02 \\ 
  3 & 0.29 & 0.16 & 0.02 & 0.05 & 0.01 & 0.02 & 0.28 & 0.14 \\ 
  4 & 0.16 & 0.12 & 0.00 & 0.05 & 0.03 & 0.05 & 0.17 & 0.11 \\ 
  5 & 0.02 & 0.01 & 0.01 & 0.05 & 0.00 & 0.03 & 0.03 & 0.01 \\ 
  6 & 0.24 & 0.12 & 0.00 & 0.05 & 0.01 & 0.02 & 0.23 & 0.12 \\ 
  7 & 0.18 & 0.13 & 0.03 & 0.05 & 0.03 & 0.05 & 0.19 & 0.14 \\ 
  8 & 0.11 & 0.10 & 0.00 & 0.05 & 0.02 & 0.02 & 0.09 & 0.09 \\
   \hline
\end{tabular}\label{p100:null:q1}
\end{table}

\begin{table}[ht]
\centering
\caption{Type I error when the null hypothesis is true in experiment 2. False discovery rate cutoffs are $q^*=0.01$ for cases 0.2, 0.8, 0.2(m) and 0.8(m), 0.05 for \GSA-s and 0.10 for E and \GSA-c.}
\begin{tabular}{ccccccccccc}
  \hline\hline
Pathway & 0.2 & 0.8 & E & T & \GSA-s & \GSA-c & 0.2(m) & 0.8(m) \\ 
  \hline  
  1 & 0.03 & 0.03 & 0.00 & 0.05 & 0.01 & 0.03 & 0.03 & 0.03 \\ 
  2 & 0.08 & 0.05 & 0.00 & 0.05 & 0.01 & 0.05 & 0.08 & 0.05 \\ 
  3 & 0.17 & 0.09 & 0.00 & 0.05 & 0.01 & 0.01 & 0.17 & 0.08 \\ 
  4 & 0.34 & 0.25 & 0.01 & 0.05 & 0.03 & 0.03 & 0.35 & 0.25 \\ 
  5 & 0.25 & 0.09 & 0.00 & 0.05 & 0.01 & 0.02 & 0.27 & 0.11 \\ 
  6 & 0.32 & 0.18 & 0.00 & 0.05 & 0.02 & 0.04 & 0.33 & 0.18 \\ 
  7 & 0.23 & 0.13 & 0.00 & 0.05 & 0.00 & 0.04 & 0.28 & 0.13 \\ 
  8 & 0.26 & 0.12 & 0.00 & 0.05 & 0.02 & 0.01 & 0.29 & 0.11 \\
   \hline
\end{tabular}\label{p160:null:q1}
\end{table}

As a comparison, Table~\ref{p100:null} and \ref{p160:null} present the type I errors based on the uniform FDR cutoffs 0.05 evaluated over 100 replications for experiment 1 and 2, respectively. The reported false discovery proportions for NetGSA with estimated networks including columns 0.2, 0.8, 0.2(m) and 0.8(m) are generally higher than the corresponding columns in Table \ref{p100:null:q1} and \ref{p160:null:q1}, especially pathways 2-4 and 6-8.

\begin{table}[ht]
\centering
\caption{Type I error when the null hypothesis is true in experiment 1. False discovery rate cutoff is $q^*=0.05$.}
\begin{tabular}{ccccccccccc}
  \hline\hline
Pathway & 0.2 & 0.8 & E & T & \GSA-s & \GSA-c & 0.2(m) & 0.8(m) \\ 
  \hline
  1 & 0.03 & 0.04 & 0.00 & 0.05 & 0.01 & 0.01 & 0.04 & 0.03 \\ 
  2 & 0.18 & 0.18 & 0.01 & 0.05 & 0.04 & 0.02 & 0.18 & 0.17 \\ 
  3 & 0.32 & 0.25 & 0.00 & 0.05 & 0.02 & 0.00 & 0.36 & 0.25 \\ 
  4 & 0.39 & 0.29 & 0.01 & 0.05 & 0.03 & 0.00 & 0.39 & 0.32 \\ 
  5 & 0.09 & 0.04 & 0.00 & 0.05 & 0.02 & 0.05 & 0.11 & 0.07 \\ 
  6 & 0.30 & 0.19 & 0.00 & 0.05 & 0.04 & 0.03 & 0.32 & 0.21 \\ 
  7 & 0.39 & 0.29 & 0.01 & 0.05 & 0.04 & 0.02 & 0.39 & 0.32 \\ 
  8 & 0.29 & 0.19 & 0.02 & 0.05 & 0.04 & 0.03 & 0.27 & 0.21 \\
   \hline
\end{tabular}\label{p100:null}
\end{table}

\begin{table}[ht]
\centering
\caption{Type I error when the null hypothesis is true in experiment 2. False discovery rate cutoff is $q^*=0.05$.}
\begin{tabular}{ccccccccccc}
  \hline\hline
Pathway & 0.2 & 0.8  & E & T & \GSA-s & \GSA-c & 0.2m & 0.8m \\ 
  \hline
  1 & 0.15 & 0.14 & 0.00 & 0.05 & 0.01 & 0.03 & 0.13 & 0.15 \\ 
  2 & 0.14 & 0.12 & 0.00 & 0.05 & 0.00 & 0.01 & 0.15 & 0.12 \\ 
  3 & 0.25 & 0.17 & 0.00 & 0.05 & 0.01 & 0.01 & 0.26 & 0.17 \\ 
  4 & 0.33 & 0.26 & 0.00 & 0.05 & 0.04 & 0.01 & 0.32 & 0.28 \\ 
  5 & 0.46 & 0.32 & 0.00 & 0.05 & 0.01 & 0.02 & 0.42 & 0.32 \\ 
  6 & 0.43 & 0.26 & 0.00 & 0.05 & 0.02 & 0.03 & 0.44 & 0.26 \\ 
  7 & 0.40 & 0.30 & 0.00 & 0.05 & 0.02 & 0.02 & 0.40 & 0.32 \\ 
  8 & 0.47 & 0.29 & 0.00 & 0.05 & 0.01 & 0.01 & 0.46 & 0.29 \\
   \hline
\end{tabular}\label{p160:null}
\end{table}

It is important to make a distinction between the samples used for enrichment analysis and those for network estimation. If one has access to a large number of observations that can only be used for network estimation as well as some external structural information, then NetGSA can leverage both resources to achieve more reliable enrichment testing. However, methods like GSA are unable to take advantage of such rich external information.

\subsection{Simulation Studies 3 and 4}

\subsubsection{The setup}
Our third simulation experiment considers an undirected network with $p=160$ and a design similar to the second experiment. However, in this case, each of the 8 subnetworks has a denser structure and there are more interactions between subnetworks. Specifically, there are 70 edges connecting the 20 nodes in each subnetwork under the null. 
There is 30\% chance of an interaction between four randomly selected nodes from each subnetwork to four randomly selected nodes from every other subnetwork. Under the alternative, there is an increase of 0.5 in mean values for varying proportions of nodes (0\%, 30\% and 50\%) for subnetworks 1-3 and 5-7. For subnetworks 4 and 8, 70\% of the nodes have mean values decreased by 0.5. Moreover, 13\% of the edges in subnetworks 5-8 under the alternative are different from their null counterpart. 

The fourth experiment uses an undirected network of size $p=400$ and illustrates the scalability of the proposed method using the new optimization algorithm. 
The network consists of 20 subnetworks, each corresponding to a pathway with 20 genes. 
The probability of an interaction between the hub node in one subnetwork and two randomly selected nodes from another subnetwork is 0.4. 
Under the null, all subnetworks have the same topology generated from a scale-free random graph such that there are 37 edges linking the 20 nodes; all the nodes have mean expression values 1. 
Under the alternative, subnetworks 1-6 and 11-16 keep the same mean expression values, but 20\%, 40\%, 60\% and 80\% of nodes in subnetworks 7-10 and 17-20 have 0.5 unit increase in their mean values, respectively. 
In addition, subnetworks 11-20 under the alternative all have 39 edges and their structure differs from their null equivalent by 30\%. 
This experiment is also of interest because we created a setting where there are enough subnetworks in order for the permutation based Gene Set Analysis~\citep{Efron:2007vz} to calibrate the number of permutations required. 

In both experiments, we also included scenarios where a proportion of the supplied structural information is incorrectly specified. This is to check whether NetGSA is robust to model misspecification. In particular, about 50\% (20\%) of the supplied edges are actually not present in the true model for the case $r=0.2$ ($r=0.8$). 

\subsubsection{Network estimation}
Table~\ref{p160:400:net} presents the deviance measures for estimating the networks with 100 replicates and sample sizes of $m=500$ for $p=160$ and $m=400$ for $p=400$, when varying levels of external information are available. In both experiments, we see performance improvement in Matthews correlation coefficient and Frobenius norm loss as the correctly specified structural information of the networks $r$ increases ($r=0.2, 0.8$ corresponding to 20\% and 80\% total information, respectively). When there exists misspecified edges in the external information (denoted by 0.2(m) and 0.8(m)), we used two tuning parameters for network estimation, one for controlling the overall sparsity of the network and the other for correcting the misspecified edges. The optimal tuning parameters were selected over a grid of values using \bic. It can be seen that the performance of network estimation is not compromised by much after properly selected tuning parameters.

\begin{table}[ht]
\centering
\caption{Deviance measures for network estimation in experiment 3 and 4. FPR(\%), false positive rate in percentage; FNR(\%), false negative rate in percentage; MCC, Matthews correlation coefficient; Fnorm, Frobenius norm loss.}
{
\resizebox{0.85\linewidth}{!}{% 
\begin{tabular}{llccccccccc}
\hline\hline
\multicolumn{2}{c}{} &\multicolumn{4}{c}{$p=160$} & &\multicolumn{4}{c}{$p=400$} \\
 &$r$ & FPR(\%) & FNR(\%) & MCC & Fnorm& & FPR(\%) & FNR(\%) & MCC & Fnorm \\
 \hline  
\multirow{5}{*}{Null} 
& 0.0 & 8.20 & 12.38 & 0.59 & 0.58 &   &4.19 & 14.58 & 0.44 & 0.50  \\
& 0.2 & 7.21 & 14.88 & 0.60 & 0.58 &   &3.89 & 12.29 & 0.46 & 0.48  \\
& 0.8 & 3.04 & 5.47   & 0.80 & 0.47  &   & 1.88 & 3.87   & 0.65 & 0.40 \\
& 0.2(m) &7.48 & 12.83 & 0.60 & 0.57&  & 3.89 &12.44 & 0.46 & 0.49  \\
& 0.8(m) &3.07 &   8.39 & 0.78 & 0.49 &   & 1.88 & 4.23  & 0.64 & 0.40 \\
& & & & & &  & & & \\
\multirow{3}{*}{Alternative} 
& 0.0 &  8.16 & 11.62 & 0.59 & 0.57  &   &4.25 & 14.70 & 0.44 & 0.50  \\
& 0.2 &  7.15 & 14.41 & 0.60 & 0.57  &   &3.96 & 12.38 & 0.46 & 0.48  \\
& 0.8 &  3.02 &   5.24 & 0.80 & 0.46  &   & 1.95 & 3.75 & 0.64 & 0.40 \\
& 0.2(m) &7.42 & 12.60& 0.60 & 0.56  &   & 3.97 & 12.57 & 0.46 & 0.48  \\
& 0.8(m) &3.03 &  8.17 & 0.78 & 0.48  &   & 1.95 & 4.13 & 0.64 & 0.40 \\
\hline
\end{tabular}}
\label{p160:400:net}
}%
\end{table}

\subsubsection{Powers}
Table~\ref{power:160:dense} shows the estimated powers after correcting for false discovery rate in the third experiment with $p=160$. 
When the exact networks are known, NetGSA estimated powers match very well with the true powers. In the case of unknown networks, we see consistent recovery of high powers for subnetworks 3, 4, 6, 7 and 8 using NetGSA even with only 20\% external information. This suggests that, with large enough samples for network estimation, a small amount of external knowledge is sufficient for making reliable inference using the network-based method.  
Interestingly, \GSA-c identifies only subnetworks 3 and 4 as significantly differential with high power, whereas \GSA-s returns relatively high power for subnetworks 3 and 6 but surprisingly low power for subnetwork 8. 
One possible reason for this pattern is that the busy interactions between subnetworks and the negative mean changes in subnetworks 4 and 8 affected the ability of \GSA~to properly recognize the correct differential behavior. 
The last two columns in Table~\ref{power:160:dense} show the estimated powers from NetGSA when the external information is misspecified. For both cases ($r=20\%$ and $r=80\%$), the results bear high similarity to those in the first two columns, which suggests that NetGSA is robust to model misspecification.

\begin{table}[ht]
\centering
\caption{{Powers in experiment 3. 0.2/0.8 refer to NetGSA with 20\%/80\% external information; E refers to NetGSA with the exact networks; T refers to the true power; \GSA-c/\GSA-s refer to Gene Set Analysis with/without randomization of the genes in 1000 permutations, respectively;  0.2(m)/0.8(m)  refer to NetGSA with 20\%/80\% misspecified external information. False discovery rate cutoffs are $q^*=0.01$ for 0.2, 0.8, 0.2(m), 0.8(m) and \GSA-s, 0.10 for E and \GSA-c.
}}{
\begin{tabular}{ccccccccc}
\hline\hline
\multicolumn{1}{c}{}& \multicolumn{8}{c}{$p=160$}  \\
Pathway & 0.2 & 0.8 & E & T & \GSA-s & \GSA-c &0.2(m) & 0.8(m)\\ 
\hline
  1 & 0.00 & 0.00 & 0.05 & 0.05 & 0.06 & 0.04 & 0.00 & 0.00 \\ 
  2 & 0.49 & 0.50 & 0.77 & 0.71 & 0.59 & 0.04 & 0.49 & 0.52 \\ 
  3 & 0.75 & 0.71 & 0.91 & 0.90 & 0.97 & 0.76 & 0.78 & 0.72 \\ 
  4 & 0.90 & 0.88 & 1.00 & 0.99 & 0.59 & 0.97 & 0.89 & 0.87 \\ 
  5 & 0.18 & 0.12 & 0.07 & 0.05 & 0.48 & 0.03 & 0.17 & 0.13 \\ 
  6 & 0.47 & 0.44 & 0.66 & 0.68 & 0.78 & 0.27 & 0.48 & 0.44 \\ 
  7 & 0.68 & 0.62 & 0.96 & 0.95 & 0.64 & 0.04 & 0.69 & 0.60 \\ 
  8 & 0.76 & 0.77 & 0.99 & 1.00 & 0.14 & 0.50 & 0.77 & 0.75 \\
\hline
\end{tabular}}\label{power:160:dense}
\end{table}

\begin{table}[ht]
\centering
\caption{{Powers in experiment 3. 0.2/0.8 refer to NetGSA with 20\%/80\% external information; E refers to NetGSA with the exact networks; T refers to the true power; \GSA-c/\GSA-s refer to Gene Set Analysis with/without randomization of the genes in 1000 permutations, respectively;  0.2(m)/0.8(m) refer to NetGSA with 20\%/80\% misspecified external information. False discovery rate cutoffs are $q^*=0.05$.
}}{
\begin{tabular}{ccccccccc}
\hline\hline
\multicolumn{1}{c}{}& \multicolumn{8}{c}{$p=160$}  \\
Pathway & 0.2 & 0.8 & E & T & \GSA-s & \GSA-c &0.2(m) & 0.8(m)\\ 
\hline
  1 & 0.02 & 0.02 & 0.01 & 0.05 & 0.02 & 0.00 & 0.02 & 0.02 \\ 
  2 & 0.70 & 0.72 & 0.67 & 0.71 & 0.65 & 0.02 & 0.72 & 0.75 \\ 
  3 & 0.81 & 0.78 & 0.87 & 0.90 & 0.97 & 0.39 & 0.80 & 0.80 \\ 
  4 & 0.95 & 0.92 & 0.99 & 0.99 & 0.53 & 0.79 & 0.94 & 0.92 \\ 
  5 & 0.30 & 0.32 & 0.03 & 0.05 & 0.44 & 0.00 & 0.27 & 0.31 \\ 
  6 & 0.56 & 0.65 & 0.63 & 0.68 & 0.88 & 0.11 & 0.57 & 0.66 \\ 
  7 & 0.78 & 0.74 & 0.93 & 0.95 & 0.72 & 0.06 & 0.78 & 0.76 \\ 
  8 & 0.90 & 0.87 & 1.00 & 1.00 & 0.12 & 0.27 & 0.88 & 0.88 
\\ \hline
\end{tabular}}\label{power:160:dense:q1}
\end{table}

The estimated powers after correcting for false discovery rate in the fourth experiment are shown separately in Table~\ref{power:400}. 
When the exact networks with the correct edge weights are known, we again see that NetGSA estimated powers match the true powers closely, with very low powers for subnetworks 1-6 which have no changes in neither mean expressions nor structures, high powers for subnetworks 8-10 which have significant changes in mean expression values, low powers for subnetworks 11-16 that have changes in structures and very high powers for pathways 17-20 with changes in both. 
When there is 20\% external information on the underlying network topology, NetGSA's powers for subnetworks 8-10 and 18-20 are close to true powers. However, NetGSA overestimates the powers for subnetworks 11-16. This is due to the small sample size ($m=400$) for estimating the underlying networks. When the external information is slightly misspecified, the last two columns indicate that NetGSA still returns valid powers that are comparable to those obtained with correctly specified structural information. 
In comparison, \GSA-s believes almost all subnetworks except 1-6 are significantly differential. On the other hand, when testing against the competitive null, the results from \GSA-c suggest that only subnetworks 10, 19 and 20 are significantly differential. The conflicting results from \GSA~with or without randomization of the genes also raise concerns as to which version to choose in practice.

\begin{table}[!t]
\centering
\caption{{Powers in experiment 4. 0.2/0.8 refer to NetGSA with 20\%/80\% external information; E refers to NetGSA with the exact networks; T refers to the true power; \GSA-c/\GSA-s refer to Gene Set Analysis with/without randomization of the genes in 1000 permutations, respectively;  0.2(m)/0.8(m)  refer to NetGSA with 20\%/80\% misspecified external information. False discovery rate cutoffs are $q^*=0.01$ for 0.2, 0.8, 0.2(m) and 0.8(m), 0.05 for \GSA-s and 0.10 for E and \GSA-c.
}}{
\begin{tabular}{ccccccccc}
\hline\hline
\multicolumn{1}{c}{}& \multicolumn{8}{c}{$p=400$}  \\
Pathway & 0.2 & 0.8 & E & T & \GSA-s & \GSA-c &0.2(m) & 0.8(m) \\ 
\hline
  1 & 0.02 & 0.02 & 0.05 & 0.05 & 0.09 & 0.03 & 0.02 & 0.02 \\ 
  2 & 0.14 & 0.12 & 0.06 & 0.05 & 0.13 & 0.03 & 0.15 & 0.12 \\ 
  3 & 0.19 & 0.18 & 0.05 & 0.05 & 0.11 & 0.01 & 0.18 & 0.18 \\ 
  4 & 0.29 & 0.24 & 0.03 & 0.05 & 0.15 & 0.04 & 0.30 & 0.25 \\ 
  5 & 0.32 & 0.29 & 0.05 & 0.05 & 0.15 & 0.02 & 0.32 & 0.29 \\ 
  6 & 0.46 & 0.44 & 0.05 & 0.05 & 0.18 & 0.03 & 0.44 & 0.43 \\ 
  7 & 0.56 & 0.50 & 0.85 & 0.83 & 0.93 & 0.00 & 0.58 & 0.52 \\ 
  8 & 0.75 & 0.73 & 1.00 & 1.00 & 1.00 & 0.00 & 0.75 & 0.73 \\ 
  9 & 0.92 & 0.89 & 1.00 & 1.00 & 1.00 & 0.19 & 0.93 & 0.89 \\ 
  10 & 0.98 & 0.99 & 1.00 & 1.00 & 1.00 & 1.00 & 0.98 & 0.99 \\ 
  11 & 0.49 & 0.46 & 0.09 & 0.06 & 1.00 & 0.00 & 0.47 & 0.48 \\ 
  12 & 0.57 & 0.60 & 0.07 & 0.07 & 1.00 & 0.00 & 0.54 & 0.59 \\ 
  13 & 0.59 & 0.57 & 0.07 & 0.05 & 1.00 & 0.05 & 0.59 & 0.57 \\ 
  14 & 0.58 & 0.63 & 0.12 & 0.07 & 0.99 & 0.03 & 0.59 & 0.63 \\ 
  15 & 0.58 & 0.66 & 0.07 & 0.07 & 0.99 & 0.02 & 0.57 & 0.66 \\ 
  16 & 0.60 & 0.50 & 0.08 & 0.07 & 1.00 & 0.03 & 0.57 & 0.51 \\ 
  17 & 0.65 & 0.68 & 0.90 & 0.89 & 1.00 & 0.02 & 0.63 & 0.68 \\ 
  18 & 0.73 & 0.81 & 1.00 & 1.00 & 1.00 & 0.34 & 0.74 & 0.80 \\ 
  19 & 0.85 & 0.85 & 1.00 & 1.00 & 1.00 & 1.00 & 0.85 & 0.86 \\ 
  20 & 0.86 & 0.88 & 1.00 & 1.00 & 1.00 & 1.00 & 0.86 & 0.87 
\\ \hline
\end{tabular}}\label{power:400}
\end{table}

\begin{table}[!t]
\centering
\caption{{Powers in experiment 4. 0.2/0.8 refer to NetGSA with 20\%/80\% external information; E refers to NetGSA with the exact networks; T refers to the true power; \GSA-c/\GSA-s refer to Gene Set Analysis with/without randomization of the genes in 1000 permutations, respectively;  0.2(m)/0.8(m)  refer to NetGSA with 20\%/80\% misspecified external information. False discovery rate cutoffs are $q^*=0.05$.
}}{
\begin{tabular}{ccccccccc}
\hline\hline
\multicolumn{1}{c}{}& \multicolumn{8}{c}{$p=400$}  \\
Pathway & 0.2 & 0.8 & E & T & \GSA-s & \GSA-c &0.2(m) & 0.8(m) \\ 
\hline
  1 & 0.10 & 0.11 & 0.04 & 0.05 & 0.10 & 0.03 & 0.10 & 0.11 \\ 
  2 & 0.25 & 0.20 & 0.04 & 0.05 & 0.12 & 0.03 & 0.25 & 0.20 \\ 
  3 & 0.37 & 0.34 & 0.02 & 0.05 & 0.07 & 0.00 & 0.36 & 0.31 \\ 
  4 & 0.36 & 0.38 & 0.05 & 0.05 & 0.19 & 0.03 & 0.36 & 0.38 \\ 
  5 & 0.57 & 0.51 & 0.03 & 0.05 & 0.08 & 0.01 & 0.58 & 0.51 \\ 
  6 & 0.48 & 0.45 & 0.05 & 0.05 & 0.11 & 0.01 & 0.48 & 0.45 \\ 
  7 & 0.58 & 0.58 & 0.69 & 0.83 & 0.83 & 0.00 & 0.56 & 0.59 \\ 
  8 & 0.84 & 0.82 & 1.00 & 1.00 & 1.00 & 0.00 & 0.85 & 0.83 \\ 
  9 & 0.94 & 0.93 & 1.00 & 1.00 & 1.00 & 0.11 & 0.93 & 0.93 \\ 
  10 & 0.98 & 0.97 & 1.00 & 1.00 & 1.00 & 0.91 & 0.98 & 0.97 \\ 
  11 & 0.65 & 0.66 & 0.01 & 0.06 & 1.00 & 0.00 & 0.67 & 0.64 \\ 
  12 & 0.64 & 0.68 & 0.06 & 0.07 & 1.00 & 0.01 & 0.64 & 0.67 \\ 
  13 & 0.63 & 0.58 & 0.00 & 0.05 & 1.00 & 0.03 & 0.63 & 0.58 \\ 
  14 & 0.71 & 0.72 & 0.04 & 0.07 & 0.99 & 0.00 & 0.69 & 0.72 \\ 
  15 & 0.67 & 0.69 & 0.03 & 0.07 & 1.00 & 0.02 & 0.66 & 0.69 \\ 
  16 & 0.68 & 0.61 & 0.02 & 0.07 & 1.00 & 0.00 & 0.69 & 0.62 \\ 
  17 & 0.68 & 0.68 & 0.76 & 0.89 & 1.00 & 0.00 & 0.68 & 0.69 \\ 
  18 & 0.74 & 0.78 & 0.99 & 1.00 & 1.00 & 0.08 & 0.78 & 0.78 \\ 
  19 & 0.93 & 0.94 & 1.00 & 1.00 & 1.00 & 0.92 & 0.91 & 0.94 \\ 
  20 & 0.90 & 0.92 & 1.00 & 1.00 & 1.00 & 0.99 & 0.92 & 0.92 
\\ \hline
\end{tabular}}\label{power:400:q1}
\end{table}

\subsubsection{Type I errors}

Finally, we also look at the scenarios where the null hypothesis is true for experiment 3 and 4. Again we present type I errors obtained based on both the adjusted FDR cutoffs and the uniform FDR cutoffs at 0.05, where the adjusted FDR cutoffs are $q^*=0.01$ for 0.2, 0.8, 0.2(m) and 0.8(m), 0.05 for \GSA-s and 0.10 for E and \GSA-c. The results can be found in Tables \ref{p160d:null:q1}, \ref{p160d:null}, \ref{p400:null:q1} and \ref{p400:null}.

Since the sample size used for network estimation in experiment 3 is sufficiently large, we observe very good control of type I errors in Table \ref{p160d:null:q1} for NetGSA, even with estimated networks. In Table \ref{p400:null:q1}, type I errors are high for some pathways, which is again due to the small sample size for estimating $400\times 400$ networks. 

\begin{table}[ht]
\centering
\caption{Type I error when the null hypothesis is true in experiment 3. False discovery rate cutoffs are $q^*=0.01$ for 0.2, 0.8, 0.2(m) and 0.8(m), 0.05 for \GSA-s and 0.10 for E and \GSA-c.}
\begin{tabular}{ccccccccccc}
  \hline\hline
Pathway & 0.2 & 0.8  & E & T & \GSA-s & \GSA-c & 0.2m & 0.8m \\ 
  \hline
  1 & 0.00 & 0.00 & 0.01 & 0.05 & 0.03 & 0.02 & 0.00 & 0.00 \\ 
  2 & 0.01 & 0.00 & 0.00 & 0.05 & 0.02 & 0.04 & 0.01 & 0.00 \\ 
  3 & 0.04 & 0.05 & 0.00 & 0.05 & 0.01 & 0.04 & 0.04 & 0.03 \\ 
  4 & 0.07 & 0.07 & 0.01 & 0.05 & 0.02 & 0.04 & 0.09 & 0.08 \\ 
  5 & 0.11 & 0.11 & 0.01 & 0.05 & 0.02 & 0.04 & 0.11 & 0.11 \\ 
  6 & 0.09 & 0.11 & 0.02 & 0.05 & 0.04 & 0.05 & 0.07 & 0.10 \\ 
  7 & 0.13 & 0.12 & 0.00 & 0.05 & 0.03 & 0.03 & 0.12 & 0.11 \\ 
  8 & 0.13 & 0.10 & 0.00 & 0.05 & 0.02 & 0.03 & 0.14 & 0.11 \\ 
   \hline
\end{tabular}\label{p160d:null:q1}
\end{table}

\begin{table}[ht]
\centering
\caption{Type I error when the null hypothesis is true in experiment 3. False discovery rate cutoff is $q^*=0.05$.}
\begin{tabular}{ccccccccccc}
  \hline\hline
Pathway & 0.2 & 0.8  & E & T & \GSA-s & \GSA-c & 0.2m & 0.8m \\ 
  \hline
  1 & 0.05 & 0.03 & 0.00 & 0.05 & 0.04 & 0.01 & 0.05 & 0.03 \\ 
  2 & 0.09 & 0.08 & 0.00 & 0.05 & 0.03 & 0.04 & 0.10 & 0.09 \\ 
  3 & 0.18 & 0.17 & 0.00 & 0.05 & 0.05 & 0.02 & 0.18 & 0.17 \\ 
  4 & 0.16 & 0.15 & 0.01 & 0.05 & 0.02 & 0.01 & 0.15 & 0.13 \\ 
  5 & 0.23 & 0.18 & 0.00 & 0.05 & 0.07 & 0.02 & 0.21 & 0.18 \\ 
  6 & 0.21 & 0.17 & 0.01 & 0.05 & 0.05 & 0.02 & 0.18 & 0.19 \\ 
  7 & 0.18 & 0.16 & 0.01 & 0.05 & 0.02 & 0.01 & 0.17 & 0.16 \\ 
  8 & 0.25 & 0.19 & 0.01 & 0.05 & 0.02 & 0.05 & 0.22 & 0.19 \\
   \hline
\end{tabular}\label{p160d:null}
\end{table}

\begin{table}[ht]
\centering
\caption{Type I error when the null hypothesis is true in experiment 4. False discovery rate cutoffs are $q^*=0.01$ for 0.2, 0.8, 0.2(m) and 0.8(m), 0.05 for \GSA-s and 0.10 for E and \GSA-c.}
\begin{tabular}{ccccccccccc}
  \hline\hline
Pathway & 0.2 & 0.8  & E & T & \GSA-s & \GSA-c & 0.2m & 0.8m \\ 
  \hline
  1 & 0.03 & 0.03 & 0.02 & 0.05 & 0.03 & 0.02 & 0.03 & 0.03 \\ 
  2 & 0.21 & 0.18 & 0.00 & 0.05 & 0.02 & 0.00 & 0.21 & 0.18 \\ 
  3 & 0.24 & 0.26 & 0.00 & 0.05 & 0.04 & 0.01 & 0.25 & 0.26 \\ 
  4 & 0.25 & 0.29 & 0.02 & 0.05 & 0.03 & 0.01 & 0.26 & 0.28 \\ 
  5 & 0.34 & 0.35 & 0.02 & 0.05 & 0.01 & 0.01 & 0.35 & 0.34 \\ 
  6 & 0.41 & 0.40 & 0.01 & 0.05 & 0.03 & 0.02 & 0.42 & 0.41 \\ 
  7 & 0.37 & 0.40 & 0.01 & 0.05 & 0.02 & 0.00 & 0.40 & 0.39 \\ 
  8 & 0.43 & 0.40 & 0.02 & 0.05 & 0.02 & 0.02 & 0.41 & 0.39 \\ 
  9 & 0.48 & 0.52 & 0.01 & 0.05 & 0.02 & 0.01 & 0.50 & 0.52 \\ 
  10 & 0.45 & 0.52 & 0.01 & 0.05 & 0.02 & 0.02 & 0.44 & 0.51 \\ 
  11 & 0.45 & 0.47 & 0.00 & 0.05 & 0.03 & 0.00 & 0.45 & 0.48 \\ 
  12 & 0.50 & 0.50 & 0.03 & 0.05 & 0.02 & 0.06 & 0.51 & 0.49 \\ 
  13 & 0.55 & 0.51 & 0.00 & 0.05 & 0.04 & 0.01 & 0.55 & 0.52 \\ 
  14 & 0.63 & 0.65 & 0.01 & 0.05 & 0.03 & 0.00 & 0.60 & 0.65 \\ 
  15 & 0.52 & 0.50 & 0.02 & 0.05 & 0.02 & 0.02 & 0.52 & 0.50 \\ 
  16 & 0.56 & 0.54 & 0.02 & 0.05 & 0.02 & 0.04 & 0.57 & 0.52 \\ 
  17 & 0.63 & 0.67 & 0.02 & 0.05 & 0.03 & 0.00 & 0.64 & 0.66 \\ 
  18 & 0.58 & 0.55 & 0.01 & 0.05 & 0.03 & 0.02 & 0.57 & 0.57 \\ 
  19 & 0.59 & 0.59 & 0.01 & 0.05 & 0.02 & 0.04 & 0.58 & 0.60 \\ 
  20 & 0.53 & 0.51 & 0.01 & 0.05 & 0.01 & 0.03 & 0.55 & 0.52 \\ 
   \hline
\end{tabular}\label{p400:null:q1}
\end{table}

\begin{table}[ht]
\centering
\caption{Type I error when the null hypothesis is true in experiment 4. False discovery rate cutoff is $q^*=0.05$.}
\begin{tabular}{ccccccccccc}
  \hline\hline
Pathway & 0.2 & 0.8 & E & T & \GSA-s & \GSA-c & 0.2m & 0.8m \\ 
  \hline
  1 & 0.03 & 0.04 & 0.00 & 0.05 & 0.02 & 0.00 & 0.03 & 0.03 \\ 
  2 & 0.27 & 0.27 & 0.00 & 0.05 & 0.04 & 0.00 & 0.27 & 0.26 \\ 
  3 & 0.36 & 0.36 & 0.00 & 0.05 & 0.04 & 0.00 & 0.35 & 0.35 \\ 
  4 & 0.39 & 0.40 & 0.01 & 0.05 & 0.04 & 0.02 & 0.40 & 0.41 \\ 
  5 & 0.51 & 0.49 & 0.01 & 0.05 & 0.04 & 0.00 & 0.52 & 0.51 \\ 
  6 & 0.59 & 0.59 & 0.00 & 0.05 & 0.02 & 0.00 & 0.61 & 0.58 \\ 
  7 & 0.56 & 0.54 & 0.00 & 0.05 & 0.02 & 0.02 & 0.55 & 0.53 \\ 
  8 & 0.61 & 0.60 & 0.01 & 0.05 & 0.04 & 0.04 & 0.58 & 0.58 \\ 
  9 & 0.59 & 0.60 & 0.00 & 0.05 & 0.02 & 0.02 & 0.59 & 0.59 \\ 
  10 & 0.61 & 0.59 & 0.00 & 0.05 & 0.04 & 0.00 & 0.58 & 0.62 \\ 
  11 & 0.60 & 0.58 & 0.00 & 0.05 & 0.04 & 0.01 & 0.61 & 0.58 \\ 
  12 & 0.63 & 0.60 & 0.00 & 0.05 & 0.03 & 0.00 & 0.60 & 0.60 \\ 
  13 & 0.64 & 0.64 & 0.01 & 0.05 & 0.02 & 0.02 & 0.62 & 0.67 \\ 
  14 & 0.56 & 0.67 & 0.03 & 0.05 & 0.02 & 0.03 & 0.56 & 0.68 \\ 
  15 & 0.73 & 0.76 & 0.00 & 0.05 & 0.02 & 0.00 & 0.71 & 0.77 \\ 
  16 & 0.64 & 0.60 & 0.00 & 0.05 & 0.02 & 0.01 & 0.63 & 0.63 \\ 
  17 & 0.68 & 0.64 & 0.00 & 0.05 & 0.02 & 0.01 & 0.64 & 0.62 \\ 
  18 & 0.64 & 0.64 & 0.00 & 0.05 & 0.04 & 0.00 & 0.58 & 0.63 \\ 
  19 & 0.70 & 0.60 & 0.00 & 0.05 & 0.02 & 0.01 & 0.66 & 0.60 \\ 
  20 & 0.72 & 0.73 & 0.00 & 0.05 & 0.04 & 0.00 & 0.71 & 0.73 \\ 
   \hline
\end{tabular}\label{p400:null}
\end{table}

\section{Additional Results on Metabolomics and Genomics}

Table~\ref{metabs:app}, \ref{microarray:app} and \ref{TCGA:app} present the full list of pathways used in each of the studies and their corresponding false discovery rate corrected $p$-values, respectively.

\begin{table}[!t]
\centering
\caption{$p$-values after false discovery rate correction for all pathways in the metabolomics data}
{
\resizebox{0.8\linewidth}{!}{% 
\begin{tabular}{lrrr}
\hline\hline
Pathway & NetGSA & \GSA-s &  \GSA-c   \\
\hline
  Tryptophan metabolism  & $3e^{-5}$ & 0.00 & 1.00 \\ 
  beta-Alanine metabolism & $3e^{-5}$ & 0.00 & 1.00 \\ 
  Aminoacyl-tRNA biosynthesis & $2e^{-4}$ & 0.00 & 1.00 \\ 
  ABC transporters         & $4e^{-4}$ & 0.00 & 1.00 \\ 
  Fatty acid biosynthesis & $2e^{-3}$ & 1.00 & 1.00 \\ 
  Pyrimidine metabolism  & $2e^{-3}$ & 0.00 & 1.00 \\ 
  Phenylalanine metabolism & $4e^{-3}$ & 0.00 & 1.00 \\ 
  Pantothenate and CoA biosynthesis & 0.01 & 0.00 & 1.00 \\ 
  Phenylalanine, tyrosine and tryptophan biosynthesis & 0.02 & 1.00 & 1.00 \\ 
  Caffeine metabolism & 0.04 & 0.15 & 1.00  \\ 
  Glycine, serine and threonine metabolism & 0.15 & $4e^{-3}$ & 1.00 \\ 
  Lysine biosynthesis & 0.19 & 1.00 & 1.00 \\ 
  Methionine metabolism & 0.20 & 1.00 & 1.00 \\ 
  Histidine metabolism & 0.26 & 0.00 & 0.42 \\ 
  Propanoate metabolism  & 0.34 & 0.04 & 1.00 \\ 
  Arginine and proline metabolism & 0.39 & 0.06 & 1.00 \\ 
  Glutathione metabolism & 0.43 & 0.12 & 1.00 \\ 
  Arginine biosynthesis    & 0.47 & 0.01 & 1.00 \\ 
  Alanine and aspartate metabolism & 0.57 & 1.00 & 1.00 \\ 
  Valine, leucine and isoleucine biosynthesis & 0.61 & 1.00 & 1.00 \\ 
  Purine metabolism & 1.00 & 0.03 & 1.00 \\ 
  Glutamate metabolism & 1.00 & 1.00 & 1.00 \\ 
  Tyrosine metabolism & 1.00 & 1.00 & 1.00 \\ 
  Cyanoamino acid metabolism & 1.00 & 1.00 & 1.00 \\ 
  Nitrogen metabolism  & 1.00 & 0.43 & 1.00 \\ 
  Tropane, piperidine and pyridine alkaloid biosynthesis  & 1.00 & 0.02 & 1.00 \\ 
  Neuroactive ligand-receptor interaction & 1.00 & 0.02 & 1.00 \\ 
\hline
\end{tabular}}\label{metabs:app}
}%
\end{table}

\begin{table}[!t]
\centering
\caption{$p$-values after false discovery rate correction for all pathways in the Lung cancer data}
{
\resizebox{0.75\linewidth}{!}{% 
\begin{tabular}{lrrr}
\hline\hline
Pathway & NetGSA & \GSA-s &  \GSA-c   \\
\hline
   Jak-STAT signaling pathway   & 0.18  & 0.31 & 1.00 \\
   p53 signaling pathway       & 0.22  & 0.68 & 1.00\\
   Wnt signaling pathway & 0.28 & 0.61 & 1.00 \\ 
   mTOR signaling pathway & 0.42 & 0.46 & 1.00 \\ 
   Glutathione metabolism & 0.42 & 1.00 & 1.00 \\ 
   Purine metabolism & 0.49 & 0.46 & 1.00 \\ 
   Cysteine and methionine metabolism & 0.49 & 0.46 & 1.00 \\ 
   ErbB signaling pathway & 0.74 & 0.07 & 1.00 \\ 
   Chemokine signaling pathway & 0.74 & 0.61 & 1.00 \\ 
   MAPK signaling pathway & 0.77 & 0.61 & 1.00 \\ 
   Pentose phosphate pathway & 0.82 & 1.00 & 1.00 \\ 
   Pyrimidine metabolism & 0.83 & 0.46 & 1.00 \\ 
   Cell cycle & 0.87 & 0.80 & 1.00 \\ 
   Glycolysis / Gluconeogenesis & 1.00 & 0.98 & 1.00 \\ 
   Citrate cycle (TCA cycle) & 1.00 & 1.00 & 1.00 \\ 
   Fructose and mannose metabolism & 1.00 & 1.00 & 1.00 \\ 
   Galactose metabolism & 1.00 & 1.00 & 1.00 \\ 
   Fatty acid metabolism & 1.00 & 1.00 & 1.00 \\ 
   Oxidative phosphorylation & 1.00 & 1.00 & 1.00 \\ 
   Alanine, aspartate and glutamate metabolism & 1.00 & 1.00 & 1.00 \\ 
   Valine, leucine and isoleucine degradation & 1.00 & 1.00 & 1.00 \\ 
   Lysine degradation & 1.00 & 1.00 & 1.00 \\ 
   Arginine and proline metabolism & 1.00 & 1.00 & 1.00 \\ 
   Histidine metabolism & 1.00 & 1.00 & 1.00 \\ 
   Tyrosine metabolism & 1.00 & 1.00 & 1.00 \\ 
   Tryptophan metabolism & 1.00 & 1.00 & 1.00 \\ 
   beta-Alanine metabolism & 1.00 & 1.00 & 1.00 \\ 
   Starch and sucrose metabolism & 1.00 & 0.61 & 1.00 \\ 
   Amino sugar and nucleotide sugar metabolism & 1.00 & 1.00 & 1.00 \\ 
   PPAR signaling pathway & 1.00 & 1.00 & 1.00 \\ 
   Calcium signaling pathway & 1.00 & 1.00 & 1.00 \\ 
   Phosphatidylinositol signaling system & 1.00 & 1.00 & 1.00 \\ 
   Notch signaling pathway & 1.00 & 0.68 & 1.00 \\ 
   Hedgehog signaling pathway & 1.00 & 1.00 & 1.00 \\ 
   TGF-beta signaling pathway & 1.00 & 1.00 & 1.00 \\ 
   VEGF signaling pathway & 1.00 & 0.98 & 1.00 \\ 
   Toll-like receptor signaling pathway & 1.00 & 0.98 & 1.00 \\ 
   NOD-like receptor signaling pathway & 1.00 & 0.61 & 1.00 \\ 
   RIG-I-like receptor signaling pathway & 1.00 & 1.00 & 1.00 \\ 
   T cell receptor signaling pathway & 1.00 & 0.46 & 1.00 \\ 
   B cell receptor signaling pathway & 1.00 & 0.61 & 1.00 \\ 
   Fc epsilon RI signaling pathway & 1.00 & 0.98 & 1.00 \\ 
   Neurotrophin signaling pathway & 1.00 & 0.46 & 1.00 \\ 
   Insulin signaling pathway & 1.00 & 0.46 & 1.00 \\ 
   GnRH signaling pathway & 1.00 & 1.00 & 1.00 \\ 
   Adipocytokine signaling pathway & 1.00 & 1.00 & 1.00 \\ 
   Epithelial cell signaling in Helicobacter pylori infection & 1.00 & 1.00 & 1.00 \\ 
   \hline
\end{tabular}}\label{microarray:app}
}%
\end{table}

\begin{table}[!t]
\centering
\caption{$p$-values after false discovery rate correction for all pathways in the TCGA data}
{
\resizebox{0.8\linewidth}{!}{% 
\begin{tabular}{lrrr}
\hline\hline
Pathway & NetGSA & \GSA-s &  \GSA-c   \\
\hline
  Epithelial cell signaling in Helicobacter pylori infection & $5e^{-95}$ & 0.00 & 1.00 \\ 
  Cell cycle & $2e^{-47}$ & 0.00 & 1.00 \\ 
  Galactose metabolism & $3e^{-31}$ & 0.00 & 1.00 \\ 
  Glutathione metabolism & $1e^{-27}$ & 0.00 & 1.00 \\ 
  NOD-like receptor signaling pathway & $1e^{-24}$ & 0.00 & 1.00 \\ 
  Pyrimidine metabolism & $4e^{-23}$ & 0.00 & 1.00 \\ 
  Cysteine and methionine metabolism & $1e^{-22}$ & 0.00 & 1.00 \\ 
  Starch and sucrose metabolism & $1e^{-18}$ & 0.00 & 1.00 \\ 
  Toll-like receptor signaling pathway & $1e^{-18}$ & 0.00 & 1.00 \\ 
  Glycolysis / Gluconeogenesis &$3e^{-17}$ & 0.00 & 1.00 \\ 
  Jak-STAT signaling pathway & $9e^{-15}$ & 0.00 & 1.00 \\ 
  Chemokine signaling pathway & $3e^{-14}$ & 0.00 & 1.00 \\ 
  ErbB signaling pathway & $7e^{-13}$ & 0.00 & 1.00 \\ 
  p53 signaling pathway & $7e^{-12}$ & 0.00 & 1.00 \\ 
  Hedgehog signaling pathway & $5e^{-10}$ & 0.00 & 1.00 \\ 
  beta-Alanine metabolism & $1e^{-7}$ & 0.00 & 1.00 \\ 
  Fc epsilon RI signaling pathway & $5e^{-7}$ & 0.00 & 1.00 \\ 
  Fructose and mannose metabolism & $2e^{-6}$ & 0.00 & 1.00 \\ 
  Pentose phosphate pathway & $2e^{-6}$ & 0.00 & 1.00 \\ 
  PPAR signaling pathway & $5e^{-6}$ & 0.00 & 1.00 \\ 
  Adipocytokine signaling pathway & $4e^{-5}$ & 0.00 & 1.00 \\ 
  Purine metabolism & $6e^{-5}$ & 0.00 & 1.00 \\ 
  Valine, leucine and isoleucine degradation & $5e^{-4}$ & $1e^{-3}$ & 1.00 \\ 
  GnRH signaling pathway & $2e^{-3}$ & 0.00 & 1.00 \\ 
  TGF-beta signaling pathway & $3e^{-3}$ & 0.00 & 1.00\\
  Neurotrophin signaling pathway & 0.02 & 0.00 & 1.00 \\ 
  Fatty acid metabolism & 0.03 & 0.01 & 1.00 \\ 
  Oxidative phosphorylation & 0.04 & 0.00 & 1.00 \\ 
  Lysine degradation & 0.04 & 0.00 & 1.00 \\ 
  Arginine and proline metabolism & 0.06 & 0.00 & 1.00 \\ 
  VEGF signaling pathway & 0.07 & 0.00 & 1.00 \\ 
  mTOR signaling pathway & 0.08 & 0.00 & 1.00 \\ 
  Glycine serine and threonine metabolism & 0.10 & 0.00 & 1.00 \\ 
  Phosphatidylinositol signaling system & 0.17 & 0.00 & 1.00 \\ 
  Notch signaling pathway & 0.65 & 0.00 & 1.00 \\ 
  MAPK signaling pathway & 0.82 & 0.00 & 1.00 \\ 
  Citrate cycle (TCA cycle) & 1.00 & 0.00 & 1.00 \\ 
  Tryptophan metabolism & 1.00 & 0.00 & 1.00 \\ 
  Amino sugar and nucleotide sugar metabolism & 1.00 & 0.00 & 1.00 \\ 
  Calcium signaling pathway & 1.00 & 0.00 & 1.00 \\ 
  Wnt signaling pathway & 1.00 & 0.00 & 1.00 \\ 
  RIG-I-like receptor signaling pathway & 1.00 & 0.00 & 1.00 \\ 
  T cell receptor signaling pathway & 1.00 & 0.00 & 1.00 \\ 
  B cell receptor signaling pathway & 1.00 & 0.00 & 1.00 \\ 
  Insulin signaling pathway & 1.00 & 0.00 & 1.00 \\ 
\hline
\end{tabular}}\label{TCGA:app}
}%
\end{table}

\end{document}